\input harvmac.tex
\input psfig
\input epsf
\noblackbox
\def\figin{\epsfcheck\figin}\def\figins{\epsfcheck\figins}
\def\epsfcheck{\ifx\epsfbox\UnDeFiNeD
\message{(NO epsf.tex, FIGURES WILL BE IGNORED)}
\gdef\figin##1{\vskip2in}\gdef\figins##1{\hskip.5in}
\else\message{(FIGURES WILL BE INCLUDED)}%
\gdef\figin##1{##1}\gdef\figins##1{##1}\fi}
\def\DefWarn#1{}
\def\figinsert{\goodbreak\midinsert}
\def\ifig#1#2#3{\DefWarn#1\xdef#1{fig.~\the\figno}
\writedef{#1\leftbracket fig.\noexpand~\the\figno}%
\figinsert\figin{\centerline{#3}}\medskip\centerline{\vbox{\baselineskip12pt
\advance\hsize by -1truein\noindent\footnotefont{\bf Fig.~\the\figno } \it#2}}
\bigskip\endinsert\global\advance\figno by1}


\def\encadremath#1{\vbox{\hrule\hbox{\vrule\kern8pt\vbox{\kern8pt
 \hbox{$\displaystyle #1$}\kern8pt}
 \kern8pt\vrule}\hrule}}
 %
 %
 

 \font\cmss=cmss10
 \font\cmsss=cmss10 at 7pt
 \def\rlx{\relax\leavevmode}
 \def\inbar{\vrule height1.5ex width.4pt depth0pt}
 \def\IC{\relax\,\hbox{$\inbar\kern-.3em{\rm C}$}}
 \def\IN{\relax{\rm I\kern-.18em N}}
 \def\IP{\relax{\rm I\kern-.18em P}}

\def\ZZ{\rlx\leavevmode\ifmmode\mathchoice{\hbox{\cmss Z\kern-.4em Z}}
  {\hbox{\cmss Z\kern-.4em Z}}{\lower.9pt\hbox{\cmsss Z\kern-.36em Z}}
  {\lower1.2pt\hbox{\cmsss Z\kern-.36em Z}}\else{\cmss Z\kern-.4em Z}\fi}
 \def\IZ{\relax\ifmmode\mathchoice
 {\hbox{\cmss Z\kern-.4em Z}}{\hbox{\cmss Z\kern-.4em Z}}
 {\lower.9pt\hbox{\cmsss Z\kern-.4em Z}}
 {\lower1.2pt\hbox{\cmsss Z\kern-.4em Z}}\else{\cmss Z\kern-.4em Z}\fi}
 \def\IZ{\relax\ifmmode\mathchoice
 {\hbox{\cmss Z\kern-.4em Z}}{\hbox{\cmss Z\kern-.4em Z}}
 {\lower.9pt\hbox{\cmsss Z\kern-.4em Z}}
 {\lower1.2pt\hbox{\cmsss Z\kern-.4em Z}}\else{\cmss Z\kern-.4em Z}\fi}

 \def\narrowplus{\kern -.04truein + \kern -.03truein}
 \def\narrowminus{- \kern -.04truein}
 \def\narrowminussub{\kern -.02truein - \kern -.01truein}

 \def\frac#1#2{{#1\over #2}}

 \def\mp{m_{{\rm P}}}

 \def\IZ{\relax\ifmmode\mathchoice
 {\hbox{\cmss Z\kern-.4em Z}}{\hbox{\cmss Z\kern-.4em Z}}
 {\lower.9pt\hbox{\cmsss Z\kern-.4em Z}}
 {\lower1.2pt\hbox{\cmsss Z\kern-.4em Z}}\else{\cmss Z\kern-.4em Z}\fi}
 \def\IB{\relax{\rm I\kern-.18em B}}
 \def\IC{{\relax\hbox{$\inbar\kern-.3em{\rm C}$}}}
 \def\Ic{{\relax\hbox{$\inbar\kern-.22em{\rm c}$}}}
 \def\ID{\relax{\rm I\kern-.18em D}}
 \def\IE{\relax{\rm I\kern-.18em E}}
 \def\IF{\relax{\rm I\kern-.18em F}}
 \def\IG{\relax\hbox{$\inbar\kern-.3em{\rm G}$}}
 \def\IGa{\relax\hbox{${\rm I}\kern-.18em\Gamma$}}
 \def\IH{\relax{\rm I\kern-.18em H}}
 \def\II{\relax{\rm I\kern-.18em I}}
 \def\IK{\relax{\rm I\kern-.18em K}}
 \def\IP{\relax{\rm I\kern-.18em P}}
\def\IR{\relax{\rm I\kern-.18em P}}

 \font\cmss=cmss10 \font\cmsss=cmss10 at 7pt
 \def\IR{\relax{\rm I\kern-.18em R}}

 %

 %
 %
 \def\eqnn#1{\xdef
#1{(\secsym\the\meqno)}\writedef{#1\leftbracket#1}%
 \global\advance\meqno by1\wrlabeL#1}
 \def\eqna#1{\xdef
#1##1{\hbox{$(\secsym\the\meqno##1)$}}

\writedef{#1\numbersign1\leftbracket#1{\numbersign1}}%
 \global\advance\meqno by1\wrlabeL{#1$\{\}$}}
 \def\eqn#1#2{\xdef
#1{(\secsym\the\meqno)}\writedef{#1\leftbracket#1}%
 \global\advance\meqno by1$$#2\eqno#1\eqlabeL#1$$}

\newdimen\tableauside\tableauside=1.0ex
\newdimen\tableaurule\tableaurule=0.4pt
\newdimen\tableaustep
\def\phantomhrule#1{\hbox{\vbox to0pt{\hrule height\tableaurule width#1\vss}}}
\def\phantomvrule#1{\vbox{\hbox to0pt{\vrule width\tableaurule height#1\hss}}}
\def\sqr{\vbox{%
  \phantomhrule\tableaustep
  \hbox{\phantomvrule\tableaustep\kern\tableaustep\phantomvrule\tableaustep}%
  \hbox{\vbox{\phantomhrule\tableauside}\kern-\tableaurule}}}
\def\squares#1{\hbox{\count0=#1\noindent\loop\sqr
  \advance\count0 by-1 \ifnum\count0>0\repeat}}
\def\tableau#1{\vcenter{\offinterlineskip
  \tableaustep=\tableauside\advance\tableaustep by-\tableaurule
  \kern\normallineskip\hbox
    {\kern\normallineskip\vbox
      {\gettableau#1 0 }%
     \kern\normallineskip\kern\tableaurule}%
  \kern\normallineskip\kern\tableaurule}}
\def\gettableau#1 {\ifnum#1=0\let\next=\null\else
  \squares{#1}\let\next=\gettableau\fi\next}

\tableauside=1.0ex
\tableaurule=0.4pt

\def\IE{\relax{\rm I\kern-.18em E}}
\def\IP{\relax{\rm I\kern-.18em P}}

\lref\gvmone{R. Gopakumar and C. Vafa, ``M-theory and topological
strings, I," hep-th/9809187. }
\lref\zasgr{T. Graber and E. Zaslow, ``Open-string Gromov-Witten invariants:
calculations and a mirror `theorem','' hep-th/0109075.}
\lref\wittcs{E. Witten, ``Chern-Simons gauge theory as
a string theory,'' hep-th/9207094, in {\it The Floer memorial volume},
H. Hofer, C.H. Taubes, A. Weinstein and E. Zehner, eds.,
Birkh\"auser 1995, p. 637.}
\lref\gv{R. Gopakumar and C. Vafa, ``On the gauge theory/geometry
correspondence," hep-th/9811131, Adv. Theor. Math. Phys. {\bf 3} (1999)
1415.}
\lref\lm{J.M.F. Labastida and M. Mari\~no, ``Polynomial invariants
for torus knots and topological strings,''  hep-th/0004196,
Commun. Math. Phys. {\bf 217} (2001) 423.}
\lref\gvtwo{R. Gopakumar and C. Vafa, ``M-theory and topological
strings, II," hep-th/9812127. }
\lref\ov{H. Ooguri and C. Vafa, ``Knot invariants and topological
strings," hep-th/9912123, Nucl. Phys. {\bf B 577} (2000) 419.}
\lref\jones{E. Witten, ``Quantum field theory and the Jones polynomial,"
Commun. Math. Phys. {\bf 121} (1989) 351.}
\lref\lmv{J.M.F. Labastida, M. Mari\~no and C. Vafa, ``Knots, links and
branes at large $N$,'' hep-th/0010102, JHEP {\bf 0011} (2000) 007.}
\lref\dfg{E. Diaconescu, B. Florea and A. Grassi,
``Geometric transitions and open string instantons,'' hep-th/0205234.}
\lref\mv{M. Mari\~no and C. Vafa, ``Framed knots at large $N$,''
hep-th/0108064.}
\lref\kr{R.K. Kaul and P. Ramadevi, ``Three-manifold invariants from
Chern-Simons field theory with arbitrary semi-simple gauge groups,''
hep-th/0005096, Commun.\ Math.\ Phys.\  {\bf 217} (2001) 295.}
\lref\ilr{J.M. Isidro, J.M.F. Labastida and A.V. Ramallo,
``Polynomials for torus links from Chern-Simons gauge theory,''
hep-th/9210124,
Nucl. Phys. {\bf B 398} (1993) 187.}
\lref\macdonald{I.G. Macdonald, {\it Symmetric functions and Hall
polynomials}, 2nd edition, Oxford University Press, 1995.}
\lref\ml{H.R. Morton and S.G. Lukac, ``The HOMFLY polynomial of the
decorated Hopf link,'' math.GT/0108011.}
\lref\lukac{S.G. Lukac, ``HOMFLY skeins and the Hopf link,'' Ph.D. Thesis,
June 2001, in http://www.liv.ac.uk/~su14/knotgroup.html}
\lref\bp{J. Bryan and R. Pandharipande, ``BPS states of curves in
Calabi-Yau threefolds,'' math.AG/0009025, Geom. Topolo. {\bf 5} (2001)
287.}
\lref\lmqa{J.~M.~F.~Labastida and M.~Mari\~no,
``A new point of view in the theory of knot and link invariants,''
 math.QA/0104180, J. Knot Theory and Its Ramifications {\bf 11} (2002)
173.}
\lref\kz{A. Klemm and E. Zaslow,
``Local mirror symmetry at higher genus,'' hep-th/9906046, in {\it Winter
School on Mirror Symmetry, Vector bundles and Lagrangian
Submanifolds}, p. 183, American Mathematical Society 2001.}
\lref\kkv{S.~Katz, A.~Klemm and C.~Vafa,
``M-theory, topological strings and spinning black holes,'' hep-th/9910181,
Adv.\ Theor.\ Math.\ Phys.\  {\bf 3} (1999) 1445.}
\lref\geomen{S.~Katz, A.~Klemm and C.~Vafa,
``Geometric engineering of quantum field theories,'' hep-th/9609239,
Nucl.\ Phys.\ {\bf B 497} (1997) 173.}
\lref\AVG{M.~Aganagic and C.~Vafa,
``G(2) manifolds, mirror symmetry and geometric engineering,'' hep-th/0110171.}
\lref\ckyz{T.~M.~Chiang, A.~Klemm, S.~T.~Yau and E.~Zaslow,
``Local mirror symmetry: Calculations and interpretations,''
hep-th/9903053,
Adv.\ Theor.\ Math.\ Phys.\  {\bf 3} (1999) 495. }
\lref\guada{E. Guadagnini, ``The universal link polynomial,"
Int. J. Mod. Phys. {\bf A 7} (1992) 877;  {\it The link invariants of the
Chern-Simons field theory,} Walter de Gruyter, 1993.}
\lref\proof{H. Ooguri and C. Vafa, ``Worldsheet derivation of a large $N$
duality,'' hep-th/0205297.}
\lref\AK{
M.~Aganagic, A.~Karch, D.~Lust and A.~Miemiec,
``Mirror symmetries for brane configurations and branes at singularities,''
Nucl.\ Phys.\ B {\bf 569}, 277 (2000)
[arXiv:hep-th/9903093].}
\lref\AV{M. Aganagic and C. Vafa, ``Mirror symmetry, D-branes and
counting holomorphic discs,'' hep-th/0012041.}
\lref\AKV{M. Aganagic,
A. Klemm and C. Vafa, ``Disk instantons, mirror symmetry and the duality
web,'' hep-th/0105045, Z.\ Naturforsch.\ {\bf A 57} (2002) 1.}
\lref\kl{S. Katz and M. Liu, ``Enumerative geometry of stable
maps with Lagrangian boundary conditions and multiple covers of the disc,''
math.AG/0103074, Adv.\ Theor.\ Math.\ Phys.\  {\bf 5} (2002) 1.}
\lref\sv{S. Sinha and C. Vafa, ``SO and Sp Chern-Simons at large $N$,''
hep-th/0012136.}
\lref\rs{P. Ramadevi and T. Sarkar, ``On link invariants and
topological string amplitudes,'' hep-th/0009188, Nucl. Phys. {\bf B 600}
(2001) 487.}
\lref\FI{F.~Cachazo, K.~A.~Intriligator and C.~Vafa,
``A large $N$ duality via a geometric transition,'' hep-th/0103067,
Nucl.\ Phys.\  {\bf B 603} (2001) 3.}
\lref\FII{F.~Cachazo, S.~Katz and C.~Vafa,
``Geometric transitions and N = 1 quiver theories,'' hep-th/0108120.}
\lref\FIII{F.~Cachazo, B.~Fiol, K.~A.~Intriligator, S.~Katz and C.~Vafa,
``A geometric unification of dualities,'' hep-th/0110028,
Nucl.\ Phys.\ {\bf B 628} (2002) 3.}
\lref\aw{M.~Atiyah and E.~Witten,
``M-theory dynamics on a manifold of G(2) holonomy,'' hep-th/0107177.}
\lref\Acharya{B. Acharya,
``On realising ${\cal N} = 1$ super Yang-Mills
in M theory,'' hep-th/0011089.}
\lref\gflop{M.~Aganagic and C.~Vafa,
``Mirror symmetry and a G(2) flop,'' hep-th/0105225.}
\lref\aha{O.~Aharony and A.~Hanany,
``Branes, superpotentials and superconformal fixed points,''
hep-th/9704170, Nucl.\ Phys.\  {\bf B 504} (1997) 239.
O.~Aharony, A.~Hanany and B.~Kol,
``Webs of (p,q) 5-branes, five dimensional field theories and grid
diagrams,'' hep-th/9710116,
JHEP {\bf 9801} (1998) 002.}
\lref\vafaleung{N.~C.~Leung and C.~Vafa,
``Branes and toric geometry,'' hep-th/9711013,
Adv.\ Theor.\ Math.\ Phys.\  {\bf 2} (1998) 91.}
\lref\sft{E.~Witten, ``Noncommutative geometry and string field theory,''
Nucl.\ Phys.\  {\bf B 268} (1986) 253.}
\lref\HV{K.~Hori and C.~Vafa,``Mirror symmetry,'' hep-th/0002222.}
\lref\phases{E.~Witten,``Phases of N = 2 theories in two dimensions,''
Nucl.\ Phys.\ B {\bf 403}, 159 (1993), [arXiv:hep-th/9301042].}
\lref\klebstra{I.~R.~Klebanov and M.~J.~Strassler,
``Supergravity and a confining gauge theory: Duality cascades and
chiSB-resolution of naked singularities,'' hep-th/0007191,
JHEP {\bf 0008} (2000) 052.}
\lref\lllr{J.M.F.~Labastida, P.~M.~Llatas and A.~V.~Ramallo,
``Knot operators in Chern-Simons gauge theory,''
Nucl.\ Phys.\ {\bf B 348} (1991) 651.}
\lref\DD{M.~Bershadsky, C.~Vafa and V.~Sadov,
``D-strings on D-manifolds,'' hep-th/9510225,
Nucl.\ Phys.\  {\bf B 463} (1996) 398.}
\lref\kon{M.~Kontsevich, ``Enumeration of rational curves via torus
actions,'' hep-th/9405035, in {\it The moduli space of curves}, p. 335,
Birkh\"auser, 1995.}
\lref\gp{T. Graber and R. Pandharipande, ``Localization of virtual
classes,'' alg-geom/9708001, Invent. Math. {\bf 135} (1999) 487.}
\lref\konts{M. Kontsevich, ``Intersection theory on the moduli space of
curves and the matrix Airy function," Commun. Math. Phys. {\bf 147} (1992)
1.}
\lref\faber{C. Faber, ``Algorithms for computing intersection numbers
of curves, with an application to the class of the locus of Jacobians,"
alg-geom/9706006, in {\it New trends in algebraic geometry},
Cambridge Univ. Press, 1999.}
\lref\gmm{E. Guadagnini, M. Martellini and M. Mintchev, ``Wilson
lines in Chern-Simons theory and link invariants,'' Nucl. Phys.
{\bf B 330} (1990) 575.}
\lref\qgt{E.~Witten, ``On quantum gauge theories in two-dimensions,''
Commun.\ Math.\ Phys.\  {\bf 141} (1991) 153.}
\lref\cmr{S.~Cordes, G.~W.~Moore and S.~Ramgoolam,
``Lectures on 2-d Yang-Mills theory, equivariant cohomology
and topological field theories,'' hep-th/9411210,
Nucl.\ Phys.\ Proc.\ Suppl.\  {\bf 41} (1995) 184.}
\lref\kmv{S.~Katz, P.~Mayr and C.~Vafa,
``Mirror symmetry and exact solution of 4D ${\cal N} = 2$ gauge theories. I,''
hep-th/9706110, Adv.\ Theor.\ Math.\ Phys.\  {\bf 1} (1998) 53.}
\lref\verlinde{E. Verlinde,
``Fusion rules and modular transformations
in 2-D conformal field theory,''
Nucl.\ Phys.\ {\bf B 300} (1988) 360.}
\lref\BCOV{
M.~Bershadsky, S.~Cecotti, H.~Ooguri and C.~Vafa,
``Kodaira-Spencer theory of gravity and exact results for quantum string
amplitudes,'' hep-th/9309140,
Commun.\ Math.\ Phys.\  {\bf 165} (1994) 311.}
\lref\vaaug{C. Vafa, ``Superstrings and topological strings at large $N$,''
hep-th/0008142, J.\ Math.\ Phys.\  {\bf 42} (2001) 2798.}
\lref\bhv{B.~Acharya, M.~Aganagic, K.~Hori and C.~Vafa,
``Orientifolds, mirror symmetry and superpotentials,'' hep-th/0202208.}
\lref\cv{F.~Cachazo and C.~Vafa, ``${\cal N}=1$ and ${\cal N} = 2$
geometry from fluxes,'' hep-th/0206017.}
\lref\kaul{R.~K.~Kaul, ``Chern-Simons theory, knot invariants, vertex
models and three-manifold  invariants,'' hep-th/9804122, in {\it
Frontiers of field theory, quantum gravity and strings}, p. 45, Nova
Science, 1999.}
\lref\AMV{M.~Atiyah, J.~M.~Maldacena and C.~Vafa,
``An M-theory flop as a large N duality,'' hep-th/0011256,
J.\ Math.\ Phys.\  {\bf 42} (2001) 3209.}
\lref\gtso{J.~D.~Edelstein, K.~Oh and R.~Tatar,
``Orientifold, geometric transition and large N duality for
SO/Sp gauge  theories,'' hep-th/0104037,
JHEP {\bf 0105} (2001) 009.}
\lref\dot{K.~Dasgupta, K.~Oh and R.~Tatar,
``Geometric transition, large N dualities and MQCD dynamics,''
hep-th/0105066, Nucl.\ Phys.\  {\bf B 610} (2001) 331; ``Open/closed string
dualities and Seiberg
duality from geometric  transitions in M-theory,'' hep-th/0106040;
``Geometric transition versus cascading solution,'' hep-th/0110050,
JHEP {\bf 0201} (2002) 031.}
\lref\fo{H.~Fuji and Y.~Ookouchi,
``Confining phase superpotentials for SO/Sp gauge theories via  geometric
transition,'' hep-th/0205301.}
\lref\mp{D.~R.~Morrison and M.~Ronen Plesser,
``Summing the instantons: Quantum cohomology and mirror symmetry in toric
varieties,'' hep-th/9412236, Nucl.\ Phys.\ {\bf B 440} (1995) 279.}
\lref\aklm{M.~Aganagic, A.~Karch, D.~Lust and A.~Miemiec,
``Mirror symmetries for brane configurations and branes at singularities,''
hep-th/9903093, Nucl.\ Phys.\ B {\bf 569} (2000) 277.}
\lref\newp{J.~M.~Labastida and M.~Mari\~no,
``A new point of view in the theory of
knot and link invariants,'' math.QA/0104180, J. Knot Theory
Ramifications {\bf 11} (2002) 173.}
\lref\phases{
E.~Witten, ``Phases of ${\cal N} = 2$ theories in two dimensions,''
hep-th/9301042, Nucl.\ Phys.\  {\bf B 403} (1993) 159.}
\lref\gkp{A.~Giveon, A.~Kehagias and H.~Partouche,
``Geometric transitions, brane dynamics and gauge theories,''
hep-th/0110115, JHEP {\bf 0112} (2001) 021.}

\Title
{\vbox{
 \baselineskip12pt
\hbox{HUTP-02/A024}
\hbox{hep-th/0206164}\hbox{}\hbox{}
}}
 {\vbox{
 \centerline{All Loop Topological String Amplitudes}
\centerline{}
 \centerline{From Chern-Simons Theory}
 }}
\centerline{Mina Aganagic, Marcos Mari\~no and Cumrun Vafa
}
 \bigskip\centerline{Jefferson Physical Laboratory}
 \centerline{Harvard University}
\centerline{Cambridge, MA 02138, USA}
 \smallskip
 \vskip .3in \centerline{\bf Abstract}
\medskip
 {We demonstrate the equivalence
 of all loop closed topological string amplitudes on toric
 local Calabi-Yau threefolds with computations of certain
knot invariants for Chern-Simons theory.  We use this equivalence
to compute the topological string amplitudes in certain cases to very
high degree and to all genera.  In particular we
explicitly compute the topological
string amplitudes for $\IP^2$ up to degree $12$
and $\IP^1\times \IP^1$ up to total degree $10$ to all genera.
This also leads to certain novel large $N$ dualities in the
context of ordinary superstrings, involving duals of type II
superstrings on local Calabi-Yau three-folds without any fluxes.}
\smallskip \Date{June 2002}

\newsec{Introduction}

In \gv\ it was conjectured that $U(N)$ Chern-Simons theory on
${\bf S}^3$, which describes
topological A-model of $N$ D-branes
on $X=T^{*}{\bf S}^3$, is dual at large $N$
to topological closed string theory on
$X^{t}={\cal O}(-1)\oplus {\cal O}(-1)\rightarrow {\IP}^1$.
There it was shown that the 't Hooft expansion of Chern-Simons free
energy agrees with topological string amplitudes on $X^{t}$ to all genera.
The conjecture was further tested in \ov, where computations of
certain Wilson loop observables in Chern-Simons theory were
shown to match the corresponding quantities on $X^{t}$.
Various aspects of the duality were
studied in \refs{\lm,\rs,\lmv,\newp,\AKV,\mv,\sv,\bhv} from different points
of view. The topological string duality was embedded in the superstring
theory in \vaaug.
In \refs{\AMV} the target space derivation of the superstring duality
of \vaaug\ was found by lifting up to M-theory \refs{\AMV,\Acharya}.
This was further studied in \refs{\gflop, \aw}, and
 also in a related context in
\refs{\klebstra,\FI,\FII,\FIII,\gtso,\dot,\fo,
\gkp}.
Recently, \proof\ gave a world-sheet proof of the topological
string duality based on some earlier ideas in \gv.

In \AVG\ a large class of new large $N$ dualities was proposed
which generalize the conjecture of \gv\ to
more general backgrounds, employing the
philosophy of \gv\ that the large
$N$ dualities are geometric transitions.
On the open string side, replacing $T^*{\bf S}^3$
with a more general Calabi-Yau manifold $X$ led one to
incorporate large open string instantons whose contributions
deform Chern-Simons theory \refs{\wittcs\AVG\dfg}.
In the spirit of 't Hooft's original large
$N$ conjecture, the holes in open string
Riemann surfaces fill up at large $N$, and the complicated open
string instanton sums that arise in a
general Calabi-Yau $X$ get related to a complicated
structure of instantons on the dual closed string side.
Some important aspects of how this works
were clarified in \dfg.
For one of the examples of \AVG, where both sides of the duality
are explicitly computable to all orders \dfg\
verifies the correspondence at the level of the partition
functions. However, in a general setting,
the descriptions of the theory in terms of open and closed strings
are at the same level of complexity, and the duality
was not easy to check (beyond the leading disk amplitude).

In this paper, by combining all of the ideas above
together with several new technical ingredients, we show that Chern-Simons
theory with product gauge groups and
topological matter in bifundamental representations computes
all loop topological string amplitudes on non-compact toric
Calabi-Yau manifolds. Namely, it is shown that open string duals of
a certain class of local toric Calabi-Yau manifolds involve
D-branes on chains of Lagrangian submanifolds
that are coupled only via annuli. In terms of Chern-Simons
theory this is related to computations of appropriate combinations
of Wilson loop observables associated with knots that are
the boundaries of the annuli.
The duality is local in the sense
that, as in \gv, the three-manifolds wrapped
by D-branes get replaced by $\IP^1$'s in the dual. However in this
case, open string theories build very complicated closed string geometries:
in fact any noncompact toric Calabi-Yau manifold arises in some limit of
this.

The paper is organized as follows.
In section 2 we review the relevant geometries for open
and closed strings that are related by large $N$ duality.
In section 3, we discuss the physics
of open string theories, and explain why the model simplifies
dramatically using a deformation argument.
 In section 4
we explain what is the relevant Chern-Simons computation in
terms of three-manifolds glued with annuli.
In section 5 we propose the large $N$ dualities
and we argue that the results of \proof\ should be applicable to
derive them.
In section 6 we discuss the relation between
the predictions of this
duality to localization in the
A-model closed string computation.
In section 7 we present explicit evaluations of the
amplitudes and provide predictions for
the integer invariants for some examples including
$\IP^2$ and $\IP^1\times \IP^1$, and we
show that they agree with the
known results when they are available \geomen\kz\ckyz\kkv.
In section 8 we consider embedding of this in the superstring
context. Results of previous sections
give open string duals of closed string geometries with no RR flux.
Moreover, we show that some local geometries in IIB
string theory have dual description in terms of gauge theory alone.

The work in section 7.2. was done in collaboration with
P. Ramadevi, to whom we are very grateful.  Also,
our work has some overlap with the
work of \ref\diagr{D.-E. Diaconescu, B. Florea and A. Grassi,
``Geometric transitions, del Pezzo surfaces and open string
instantons,'' to appear.}, and we thank the authors
for discussing their work prior to publication.
In particular we learned of their result that only a
limited number of holomorphic curves
contributes to the amplitudes
 before we found the general argument
presented in section 3.  The argument
discussed in \diagr\  (in the context
of $dP_2$) uses localization principle, whereas
our argument that only annuli
contribute for all toric 3-folds is based on complex structure
deformation invariance.

\newsec{Geometry}
\subsec{Open String Geometry: $T^2$ Fibrations and Their Degenerations}
In this paper, the relevant  Calabi-Yau manifolds are non-compact and
admit a description
as a special Lagrangian $T^2 \times \IR$ fibration over $\IR^3$.
The $T^2$ fibers
degenerate over loci in the base.  The geometry of the  manifold is encoded in
the one dimensional graphs in $\IR^3$ that correspond to the discriminant of
the fibration.
A very familiar example of a Calabi-Yau manifold of this type is
$X=T^*{\bf S}^3$.
The  complex structure of $X$ is given by
\eqn\ts{xy=z,\quad uv =z+\mu,}
The two-torus is visible in the above equation
as it is generated by two $U(1)$ isometries of
$X$ acting as
$$x,y,u,v \rightarrow xe^{i\alpha},ye^{-i\alpha},ue^{i\beta},ve^{-i\beta}.$$
The $\alpha$ and $\beta$ actions above can be taken to generate
the $(1,0)$ and $(0,1)$ cycle of the $T^2$.

The local type of the singularity has a
$T^2$ fiber that degenerates to ${\bf S}^1$ by collapsing
one of its one-cycles. In the equation above,
the $U(1)_{\alpha}$ action fixes $x=0=y$ and therefore fails to
generate a circle there.
In the total space, the locus where this happens, i.e. the $x=0=y=z$
subspace of $X$, is another cylinder $uv=\mu$. The projection to
the base space forgets the circle of this cylinder
and is a line in $\IR^3$.

Such a geometry locally looks like a Taub-Nut (TN) space times a cylinder
${\bf C^* }= \IR \times {\bf S}^1$.
Here, the TN space itself is thought of as a cylinder $xy={\rm const}$
which is fibered over the $z$ plane and which degenerates at $z=0$.
Analogous considerations apply to the
$U(1)_{\beta}$ action. The locus of degenerate fibers in the base $\IR^3$
of the deformed conifold is given in the figure below. In this and
similar figures below, two of the
directions of the base are the axes of the two cylinders, and the third
direction represents the real axis of the $z-$plane.

\ifig\onethreesphere{The figure depicts the  discriminant locus of the $T^2\times \IR$ fibration
in the base $\IR^3$.  The $\alpha$ and $\beta$ cycles of the $T^2$
degenerate over lines $z=0$, $z=-\mu$.}
{\epsfxsize4.0truein\epsfbox{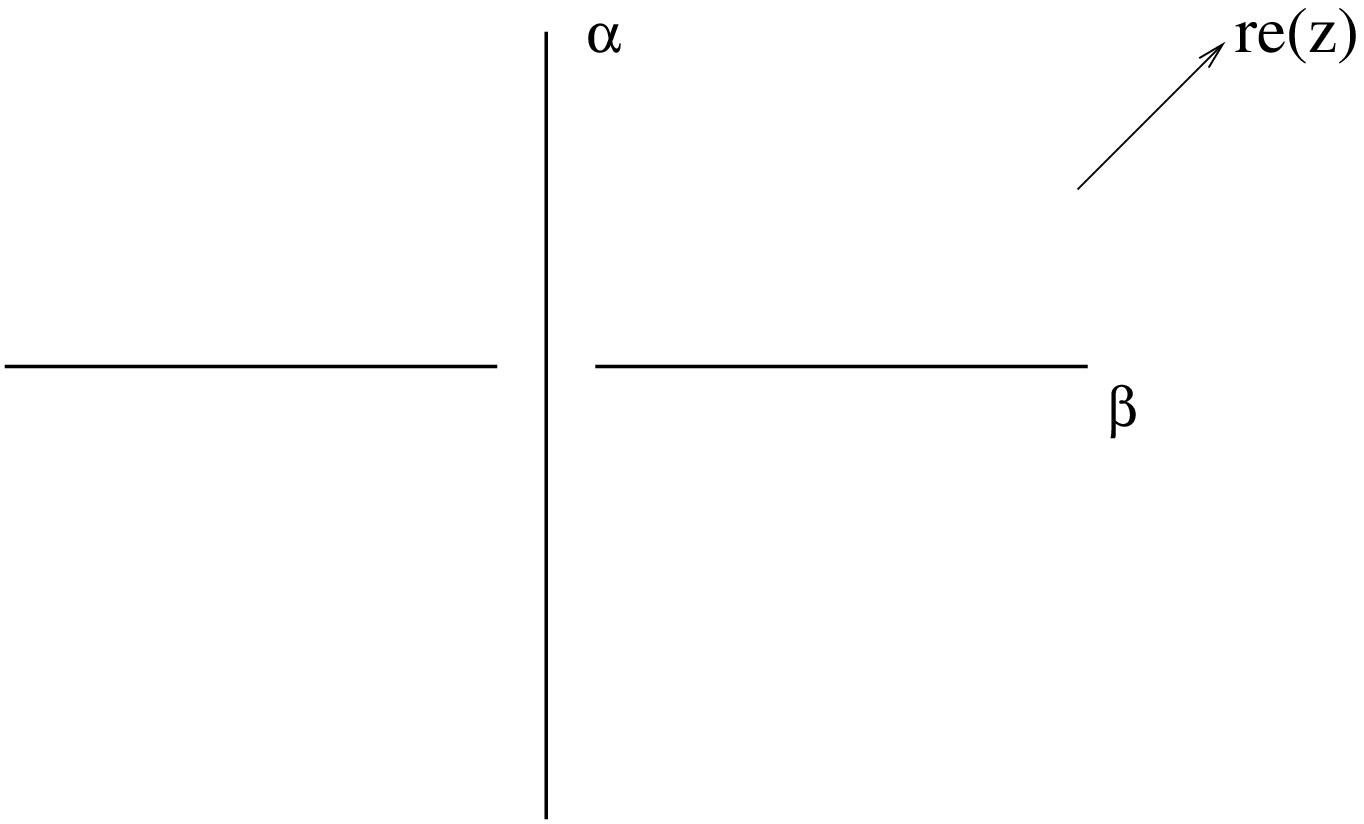}}

In general, any $(p,q)$ cycle  of the $T^2$ can degenerate in
this way. As long as the degenerate loci do not intersect,
the local geometry is that of Taub-Nut space, as
an ${\rm SL}(2,{\bf Z})$ transformation on the
$T^2$ fiber can be used to relate
it to the degenerations discussed above.
In what follows, it will be important that the orientation of the
locus where the $T^2$ fiber degenerates in the base $\IR^3$
is correlated with the $(p,q)$ type of degenerating cycle.
We have seen an example of this above in the case of $T^{*}{\bf S}^3$,
where the ${\alpha}$ and $\beta$ cycles degenerated
along orthogonal directions in the base in \onethreesphere.
The origin of this is the fact that the Calabi-Yau manifold
is a complex manifold and the fibration is special Lagrangian.

We will not go into detail here in this language as it is cumbersome for
physicists, and explained in the literature (see for example \AVG)\foot{ To
give an idea of the more general situation,
let $\hat x_{\alpha,\beta}$ be the single valued holomorphic coordinates
on  ${\bf C}^*\times  {\bf C}^*$, and let $z$ be a
coordinate on $\IR^2$.
If a $(p,q)$ cycle of the  $T^2$ degenerates at  a point in $z$,
then the fixed point locus which is invariant under
$\hat{x}_{\alpha}\rightarrow \hat{x}_{\alpha}{\rm e}^{i
p \theta} , \hat{x}_{\beta}\rightarrow \hat{x}_{\beta}{\rm e}^{i q \theta}$.
In terms of periodic variables  $\hat{x}_{\alpha,\beta}=\exp(x_{\alpha,\beta})$
we can write the degeneration locus as
by
$$q x_{\alpha} - p x_{\beta}= \,{\rm const.}$$}, especially because
there is a string theory duality that
provides excellent intuition about the geometry, which we would like to
explain instead.

\ifig\threesphere{The degeneration locus of the $T^2$
fibration in the base specifies the Calabi-Yau geometry.
The orientation of the lines are related to the
$(p,q)$ type of the 1-cycle that degenerates over it.
In the type IIB language, this corresponds to different $(p,q)$ fivebranes.}
{\epsfxsize 3.0truein\epsfbox{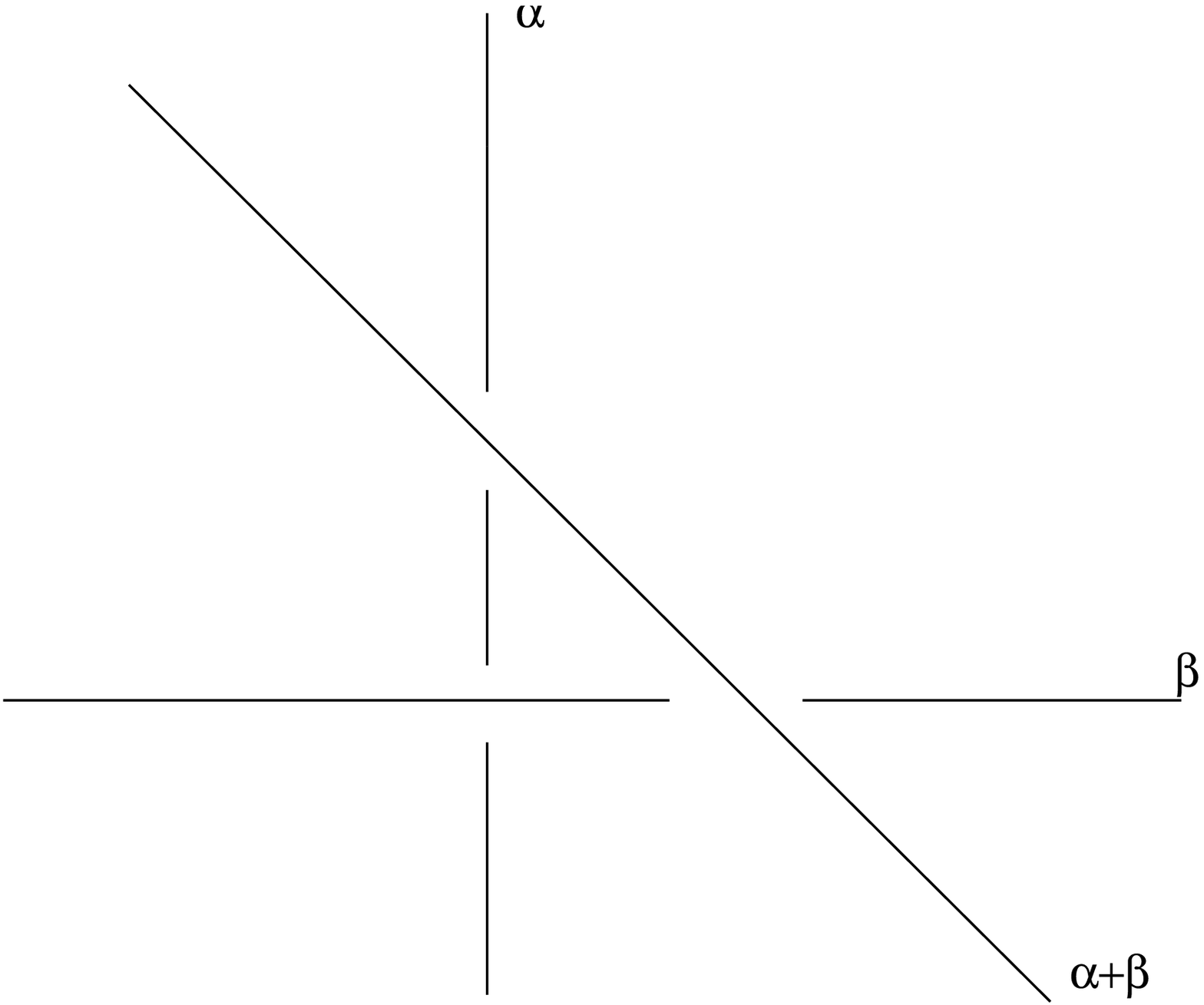}}

\subsec{Relation to $(p,q)$ Fivebranes in IIB}
In this section we connect the description
of Calabi-Yau geometry by a duality to the web of $(p,q)$
fivebranes \aha. This will be
helpful for us for an intuitive
picture of holomorphic curves in the geometry.  The connection
was derived in \vafaleung\ and we will now review it.

Recall that M-theory on $T^2$
is related to type IIB string theory
on ${\bf S}^1$. Since
the Calabi-Yau manifolds we have been considering are $T^2$
fibered over $B=\IR^4$, we can relate geometric M theory compactification
on Calabi-Yau manifold $X$ to type IIB on flat space on $B\times {\bf S}^1$.
However, due to the fact that
$T^2$ is not fibered trivially, this is not related
to the vacuum type IIB compactification.

The local type of singularity,
as we have seen above, is the Taub Nut space $TN_{p,q}$,
where the $(p,q)$ label denotes
which cycle of the $T^2$ corresponds to the ${\bf S}^1$ of the
Taub-Nut geometry. Under the duality, this local
degeneration of $X$ is mapped to the $(p,q)$
five-brane that wraps the discriminant locus in the base space $B$,
and lives on a point on the ${\bf S}^1$.
The fact that the $(p,q)$ type of the five brane
is correlated with its orientation in the base is a consequence
of the BPS condition. More precisely,
configurations of five branes that preserve supersymmetry and
$4+1$-dimensional Lorentz invariance are pointlike in a fixed
$\IR^2$ subspace of the base that we called the $z$ plane above.
In the two remaining directions of the base,
the five branes are lines
where the equation of the $(p,q)$ five brane is
$px_{\alpha}+qx_{\beta}={\rm const}$.

\subsec{Geometric Transitions}

Consider a pair of lines in the base space over which
two one-cycles of the $T^2$ degenerate.
Any path in the base space ending on the two lines, together with
the $T^2$ fiber over it, gives rise to a closed three-manifold in the
total space. This is because a cycle of the $T^2$
degenerates over the start and the end point of the path,
so the three-manifold has no boundaries. If the two lines intersect
in the base space, the three-cycle obtained in this way
can be shrunken to a point. If they don't,
it generates a homology class in $H_3(X,{\bf Z})$.
Let $n$ be the number of five-branes.
If the five branes are in generic positions and do not intersect,
the manifold is smooth, and it is easy to see that the
the dimension of third homology is $b_3(X)=n-1$.

In the superstring context, among Lagrangian three-cycles in the Calabi-Yau
manifold, special Lagrangian three-cycles are of
particular interest as they are  supersymmetric, i.e. D-branes wrapped on
them preserve some supersymmetry of the theory. These cycles
are volume-minimizing and
project to paths of shortest length in the base (the Lagrangian
condition can always be satisfied with some choice of
symplectic form on $X$). In the non-compact situation we are discussing, the
meaning of this is particularly transparent in the
IIB string theory as the five-branes live in $\IR^4$ with flat metric.
The number of supersymmetric cycles, for five-branes in generic
positions, is easily counted by doing the projection of
 the base $\IR^3\rightarrow \IR^2$
that suppresses the $z$-direction and counts the number of
intersections. Generically there will be $n(n-1)/2$ such intersection
points (unless some $(p,q)$ 5-branes are of the same type).

The Calabi-Yau manifolds we have been discussing have geometric transitions
where three-cycles in geometry shrink and the resulting singularity
is smoothed to a manifold $X^{t}$ of different topology.
This was explained in some detail in \AVG.
In the examples we will be studying
in this paper, the local geometry of the singularity will
be $T^*{\bf S}^3$, so the geometric transition in question involves an
${\bf S}^3$ shrinking and a $\IP^1$ growing.
The geometric transitions do not spoil the fact that the manifold
is $T^2$ fibered, however they do change the locus of singular fibers.
After the transition that shrinks all the three-cycles (and
these always exist in the family of $X$'s we consider), the resulting
manifolds are toric varieties.
Toric varieties admit a group of $U(1)$ isometries whose
rank is the complex dimension of the manifold.
In our case this is $U(1)^3$, and the symmetry enhancement
comes from the fact that the transition which gets rid of all the three-cycles
requires all the loci of singular fibers to coincide in the $z$-plane,
and the extra $U(1)$ is the group of rotations about this point.
While the reader might get an impression from the above discussion
that the manifold after transition gains new cycles
only in $H_2(X^{t},{\bf Z})$,
this is in fact not the case. In fact, in the generic case the number
of shrinking minimal three-cycles is larger than the number
of classes in $H_3(X)$. Then, since not
all three-cycles are independent in homology, there are
four-chains with boundaries on some of them corresponding to the relations
which they satisfy. After the transition, the four-chains
close off because their boundaries shrink. As a consequence,
the dual geometry does involve compact cycles in $H_4(X, {\bf Z})$.
\ifig\threetrans{This shows the geometric transition of the
Calabi-Yau in the previous figure.
In the leftmost geometry
there are three minimal 3-cycles.
The lengths of the
dashed lines are proportional to their sizes.
The intermediate geometry
is singular, and the figure on the right is the base of the smooth
toric Calabi-Yau after the transition. This Calabi-Yau
is related to $\IB_3$ by
flopping three $\IP^1$'s.}
{\epsfxsize 5.5truein\epsfbox{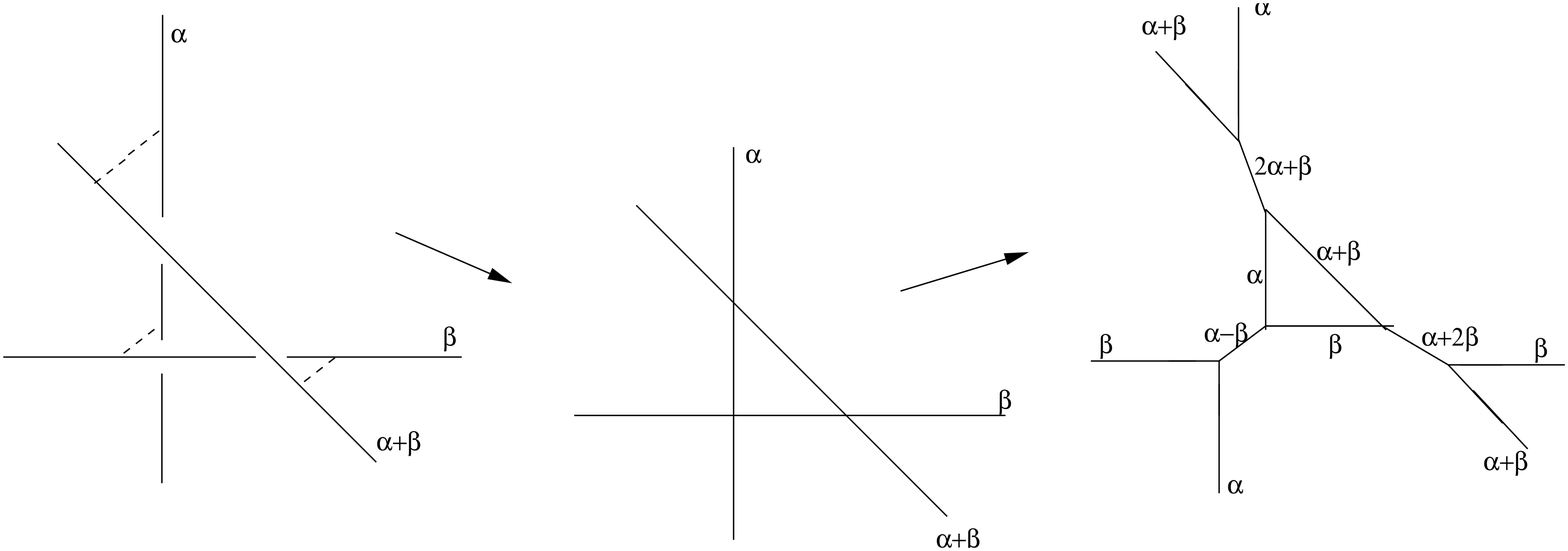}}

In the language of $(p,q)$ five branes, the geometric
transition corresponds to a phase transition in the five-dimensional
theory. Namely, the configuration of
intersecting $(1,0)$ and $(0,1)$ five-branes is a
phase transition point: the Higgs phase with five-branes
separated in the $z$ plane\foot{Recall that this is the complex structure
modulus of the geometry. The five dimensional hyper-multiplet contains
a compact scalar from the period of the $C$-field through the ${\bf S}^3$
(or the positions of the five-branes on the ${\bf S}^1$ in type IIB).
It also has a non-compact scalar from the ``period'' of the $C$-field through
the non-compact three-cycle dual to ${\bf S}^3$ or
the ``Wilson line on $\IR$'' for the five-brane. Topology-changing
transitions of toric manifolds are discussed in \aklm.}
meets the Coulomb phase, where
a piece of $(1,1)$ brane resolves the singularity.
In the geometry, there is a $T^2$ fiber whose $(1,1)$ cycle degenerates
over this interval, and the cylinder is capped off to a $\IP^1$
by all the cycles of the $T^2$ degenerating over the boundaries of the
interval. The singularity can also be resolved
with a $(1,-1)$ brane, which corresponds to the flopped $\IP^1$.

\subsec{Geometry of Holomorphic Curves}

Calabi-Yau manifolds generally come with families of
embedded curves. In the topological A-model
only holomorphic curves are relevant, as
the A-model string amplitudes localize on them.
In the presence of D-branes wrapping Lagrangian submanifolds
$M_i$ in $X$, we must also consider holomorphic curves with boundaries on
the $M_i$'s.

Holomorphic curves have a very simple description in the toric
base, or equivalently, in the $(p,q)$ five brane language.
Let us first consider closed string geometries, the family of Calabi-Yau
manifolds we have called $X^t$ above. In this case, it can be shown that
all the compact holomorphic curves in a non-compact toric
Calabi-Yau manifold wrap a 1-cycle in the $T^2$
fiber direction.
Holomorphic curves project to lines in the toric base, and locally 
the direction of the curve in the base is correlated with its direction in
the fiber. For the compact curves, the direction in the fiber is the $(p,q)$ 
1-cycle of the $T^2$.
This is most transparent in the $(p,q)$ five-brane language.
\ifig\dpthree{The figure on the left depicts a genus one holomorphic curve
with three holes ending on three minimal three-cycles. The figure on
the right is after the transition, and also depicts a genus one curve, but
without boundaries.}
{\epsfxsize 4.5truein\epsfbox{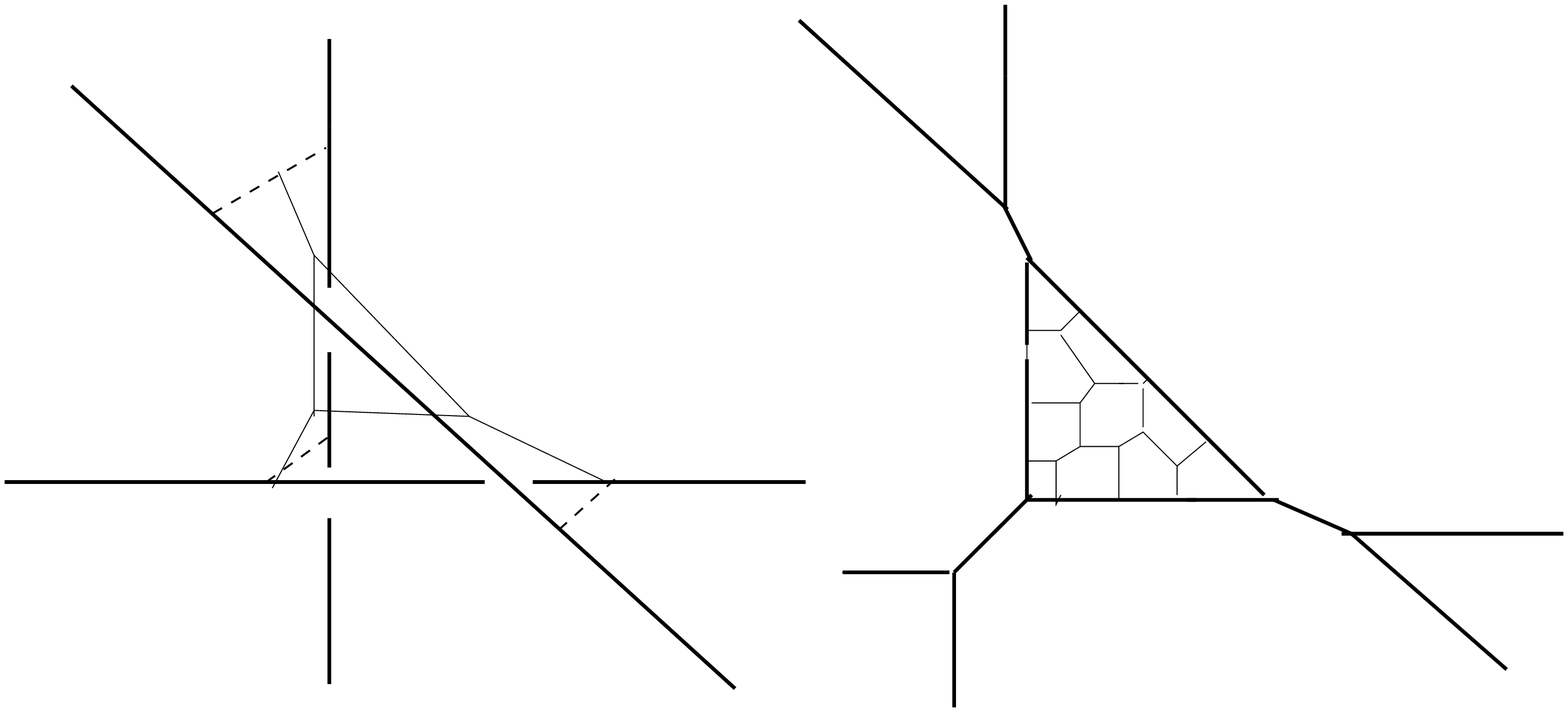}}

Namely, consider an M-theory membrane, wrapping a holomorphic curve on $X^t$.
By M-theory/type IIB duality, a membrane wrapping a $(p,q)$ cycle
of the $T^2$ is dual to a $(p,q)$ string, therefore membranes on
holomorphic curves in $X^t$ that are along the $T^2$ in the fiber
are dual to webs of $(p,q)$ strings in
type IIB string theory that are BPS. Moreover, the compact curves
are related to webs ending on $(p,q)$ five-branes
\aha.
An example of such a curve is given in the
right portion of \dpthree.

Much of the same considerations are clearly true in $X$
as well. There is however an important distinction from the point of view
of topological string amplitudes. Namely, in a generic situation,
there are no compact holomorphic curves.
This is easy to see in the $(p,q)$ five-brane
picture because a condition for the web to be
supersymmetric is that it lives in the plane parallel to
the web of the five branes, and is therefore pointlike in
the $z$ plane (there is an additional condition
that fixes the orientation of the $(p,q)$ string
depending on its charge, which comes from the balance of tensions,
by requiring it to be orthogonal to a $(p,q)$ five brane.
We refer the reader to \aha\ for
a detailed discussion). However, a given $(p,q)$ string can only end on the
five brane of the same charge, and so for five-branes at generic
locations in the $z$ plane, the string webs are never compact.
This situation changes if strings can end elsewhere.
For example, if there are M-theory five-branes wrapped on Lagrangian cycles
in $X$, membranes can end on them. The M-theory five-brane is replaced with
a D3-brane in IIB string theory ending on the various $(p,q)$
five-branes. There are then compact string webs ending on the D3-branes,
corresponding to holomorphic curves with boundaries. In relation to
large $N$ dualities in superstring context it is more natural to consider
type IIA string on $X$ instead, with D6-branes on the $M_i$'s.
This is related, via duality of IIA/M-theory on ${\bf S}^1$,
to IIB string theory with Kaluza-Klein monopoles
ending on the $(p,q)$ five brane web \AKV.
Namely, the D6 branes in Lagrangian submanifolds of $X$ lift
to M-theory on a $G_2$ holonomy manifold. To obtain this manifold,
we need to consider an extra ${\bf S}^1$ which is
fibered over the corresponding CY. This is related to IIB on
$B \times {\bf S}^1$ where we
exchange the 11-th circle with the $T^2$ that fibers $X$.
What used to be the 11-th circle is now fibered nontrivially over $B$.
In particular, the circle vanishes
over a 2-dimensional subspace of the type IIB 5-dimensional geometry.
It vanishes along the line in $B$ ending on the $(p,q)$
five branes as well as on the ${\bf S}^1$
which is dual to the $T^2$ of M-theory.
This line in $B$ corresponds to the line in the base of $X$ over which
there is the three-manifold that the D6 brane wraps. The
IIB 5-brane web is in a background of ALF
geometry dictated by the location of the Lagrangian submanifold in $X$.

\subsec{Geometry of Three-Cycles}
In this paper, we will
wrap D-branes on the minimal three-manifolds in $X$.
The physics of the D-branes depends on both $X$
and the three-manifold $M$ it is wrapped on, so we
will describe the geometries of the latter.

In our context, $M$ is obtained by pinching the cycles of the $T^2$ fibers
over the endpoints of an interval in the base.
Clearly, if the same cycle vanishes at both ends,
the topology of the three-manifold is ${\bf S}^2 \times {\bf S}^1$, as
there is a cycle of the $T^2$ that never vanishes on $M$.
An example where the manifold is ${\bf S}^3$ arises
in the familiar context of $T^*{\bf S}^3$.
This ${\bf S}^3$ comes from a $(1,0)$ cycle of the $T^2$ vanishing
at one end, and
$(0,1)$ cycle vanishing on the other. To see that this is an ${\bf S}^3$,
note that at $x=\bar y$ and $u=-\bar v$
the equation \ts\ defining $T^*{\bf S}^3$ becomes
$$|x|^2+|u|^2 = \mu,$$
and $\mu$ is real and positive, so this is a three-sphere.
In view of the discussion above, we can regard this as a real interval,
together with the $(1,0)$ one-cycle that corresponds to the
phase of $x=\bar y$,
degenerating at the end with $x=0$, and the $(0,1)$ cycle
that is the phase of $u=-\bar v$ degenerating over the $u=0$ endpoint
\foot{This is in fact the minimal ${\bf S}^3$ in $X$, as it is a
fixed point set of the real involution on \ts\ given by
$x\rightarrow \bar y$ and $u\rightarrow - \bar v$.}.

More generally, we have the following.
For our current purpose, by an ${\rm SL}(2,{\bf Z})$
transformation of the $T^2$ we can make $(1,0)$ be the
vanishing cycle over one of the boundaries, and let $(q,p)$
be the cycle that vanish over the other.
The $3-$manifold itself is a Lens space $L(p,q)$.
Remember that lens spaces are defined as quotients of ${\bf S}^3$ by
a ${\bf Z}_p$ action. The space $L(p,q)$ is given by
\eqn\lsp{|x|^2+|u|^2=1\quad\;\;(x,u) \sim (\exp(2i\pi/p) x,
\exp(2 i\pi q/p)u).}
To see that, consider an ${\bf S}^3$ which, as explained above, is a $T^2$
fibration over an interval, where the cycles of the $T^2$
are generated by phases of $x,u$.
If the complex structure of the $T^2$ corresponding to ${\bf S}^3$
it is $\tau$, then an ${\rm SL}(2,{\bf Z})$ transformation that takes
this $T^2$ to a $T^2$ with $(1,0)$ and $(q,p)$ cycles vanishing
over the endpoints will take $\tau$ to $\tau'=\frac{\tau+q}{p}$.
But the $T^2$ with the new complex structure is precisely
a quotient of the original one by the ${\bf Z}_p$ action specified in
\lsp.
Note that $L(p,q)$ is homeomorphic to $L(p,1)$.
In the present context this corresponds to the fact that
global ${\rm SL}(2,{\bf Z})$ transformations preserving the
$(1,0)$ cycle of the $T^2$ can be used to set $q$ to one.

For our later considerations in this paper it is important
to have another view on this construction of a three-manifold
$M$ as a $T^2$ fiber over
interval. The construction is as follows:
we are gluing two solid tori over (say) the midpoint of
the interval, up to an ${\rm SL}(2,{\bf Z})$ transformation $V_M$
that corresponds to a diffeomorphism identification of their boundaries.
Let us call the two tori on each
side of the midpoint by $T^2_L$ and $T^2_R$.
The embedding of this in the Calabi-Yau geometry
provides a canonical choice of $V_M$.
In the Calabi-Yau geometry, there is a natural choice of basis
of cycles $\alpha,\beta$ of the $T^2$
that fibers $X$, which is provided by the choice of complex structure on $X$.
We can identify the one-cycles of the $T^2$ fiber
that shrink over the left and the right sides of the interval with
the shrinking 1-cycles of $T_L$ and $T_R$.
The diffeomorphism map $V_M$ is the ${\rm SL}(2,{\bf Z})$ transformation
that relates one of the shrinking cycles
of the fiber of $X$ to the other one.

Let us now explain the construction of the gluing matrices that will
suit our purpose.
Let $(p_L,q_L)$ be the cycle of the $T^2$ fiber
that degenerates over the left half on $M$, and let $(p_R,q_R)$
be the cycle that degenerates over the right half.
The gluing matrix $V_M$ can be written as
\eqn\glue{
V_M = V_L^{-1} V_R,}
where $V_{L,R} = \left(\matrix{p_{L,R} & s_{L,R}\cr
q_{L,R} &t_{L,R}}\right)\in SL(2,\bf{Z})$
Clearly, $V_M$ is unique up to a homeomorphism that changes the ``framing''
of three-manifold \jones\ and takes
$$V_{L,R}\rightarrow V_{L,R}\; T^{n_{L,R}}$$
where $T$ is a generator of $SL(2,\bf{Z})$,
$T=\left(\matrix{1&1\cr 0&1}\right)$.
This is a consequence of the fact that there is no natural choice of
the cycle that is finite on the solid torus.

In the case of $M={\bf S}^3$ above,
since $(1,0)$ degenerates in the left
half of $M$ and $(0,1)$ in the right half,
$V_{M} = S,$ where $S=\left(\matrix{0&-1\cr 1&0}\right)$.
As a small modification, we could make $(p,1)$ degenerate over
the left half instead, so that $V_L=T^{p}S$  is a lens space $L(p,1)$ and
$V$ is $S^{-1}T^{-p}S$.
For most considerations in this paper we will be considering the cases
$L(1,1)$ or $L(1,0)$, which are homeomorphic to ${\bf S}^3$.

\newsec{Open String Theory}

We are interested in the topological A-model
on the Calabi-Yau geometries described above, with D-branes wrapping
special Lagrangian three-spheres.
The local geometry in some neighborhood of a Lagrangian
three-manifold $M$ is $T^*M$ and it was shown in \wittcs\
that the topological A-model corresponding to $N$ D-branes
on $M$ is a $U(N)$ Chern-Simons theory on three-manifold $M$,
$$Z= \int {\cal D} A e^{S_{\rm CS}(A)}$$
where
$$S_{\rm CS}(A) = \frac{i k}{4 \pi}\int_{M} {\rm Tr} (A\wedge dA
+ \frac{2}{3} A\wedge A\wedge A)$$
is the Chern-Simons action. The basic
idea of this equivalence is as follows:
the path-integral of the topological A-model
localizes on holomorphic curves. When there are D-branes, this
means holomorphic curves with boundaries
ending on them. In the $T^*M$ geometry
with D-branes wrapping $M$ there are no honest holomorphic curves,
however there are degenerate holomorphic curves that look like trivalent
ribbon graphs and come from the boundaries of the moduli space.
This leads to a field theory
description in target space, which is equivalent to topological
Chern-Simons theory (as the abstract open string field theory
formulation demonstrates \sft).
In this map, the level $k$ would be naively related
to the inverse of the string coupling constant $g_s$.
However, quantum corrections \jones\ shift this identification to
$$\frac{2\pi i}{k+N} = g_s.$$

More globally, however,
the geometry is generally not that of the cotangent space to any
manifold, and there can be D-branes wrapping other minimal three-spheres
in $X$. In this case the topological open strings will have contributions
from degenerate holomorphic curves, which are captured
by Chern-Simons theories, as well as some honest holomorphic curves,
which lead to insertion of some
Wilson loop observables for the Chern-Simons theory \wittcs .
If we have a number of $M_i$'s distributed in some way inside
a Calabi-Yau, with $N_i$ D-branes wrapped over $M_i$, then
we can trade the degenerate holomorphic curves
by including the corresponding Chern-Simons theories
$S_i=S_{\rm CS}(A_i)$ coupled in an appropriate way with the honest holomorphic
curves.  Namely, we have
\eqn\impo{e^{F_{\rm all}}=\int \prod_i DA_i e^{S_i+F_{\rm ndg}(U_i(\gamma_i))}}
where $F_{\rm all}$ denotes the full topological
A-model amplitude, and $F_{\rm ndg}$ denotes the contribution
of the non-degenerate holomorphic curves to the topological
amplitudes. These holomorphic curves give rise to Wilson loops
on the D-branes: each holomorphic curve with area
$A$ ending on $M_i$ over the knot ${\gamma}_i$
leads to the contribution $e^{-A} \prod_i {\rm Tr}U_i({\gamma}_i)$
to $F_{\rm ndg}$, where $U_i(\gamma_i)$ denotes the holonomy of the
Chern-Simons gauge connection around the knot $\gamma_i$. Notice
that all these Chern-Simons theories have the same coupling constant.
More precisely,
$${2\pi i\over k_i+N_i}=g_s$$
In the toric examples we will consider in this
paper it turns out that only holomorphic
annuli contribute to $F_{\rm ndg}$ and thus
this connection
with Chern-Simons theory is a useful way to compute the topological A-model
amplitudes
as some particular correlation function in a system of coupled
Chern-Simons theories.

\ifig\twospheres{Calabi-Yau geometry with $b_2=1,b_3=2$ and
two minimal ${\bf S}^3$'s as the dashed lines.}
{\epsfxsize 3.0truein\epsfbox{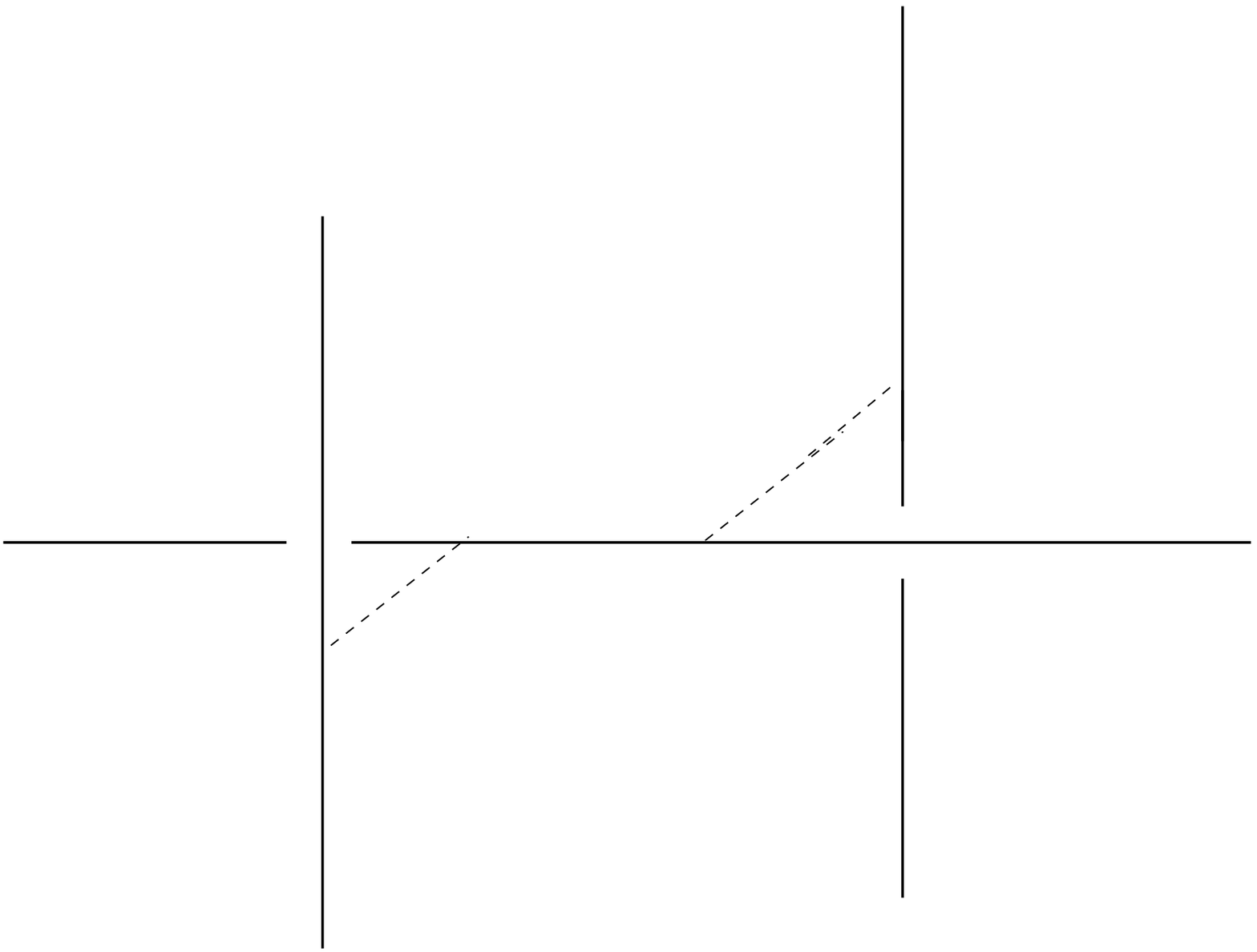}}

\subsec{The Annulus Amplitude}
As an example, let us consider the Calabi-Yau manifold $X$
with $b_3 = 2, b_2=1$ in \twospheres, whose
complex structure is described by
\eqn\two{\eqalign{xy&=z\cr
                 x'y'&=(z-\mu_1)(z-\mu_2).}}
and was
studied in a physics context in \DD.
In $X$, the
${\alpha}$ cycle of the $T^2$ degenerates over the point $z=0$, but
the $\beta$ cycle degenerates twice, over $z={\mu_{1}}$ and ${\mu_2}$.
The cycles over $[\mu_1,0]$ and $[0,\mu_2]$ are three--spheres
$M_{1,2}$ that generate $H_3(X)$ .
The base space of $X$, with loci with degenerate fibers -- is pictured in
\twospheres, where we have taken $\mu$'s
to be real, and  $\mu_1<0<\mu_2$ (the reader should keep in mind that only
one of the dimensions of the $z$ plane is visible in the base).

As is clear from the picture, there is an additional parameter visible
in the base: the relative distance of two $\beta$-branes.
This is a K\"ahler parameter corresponding to the one compact
2-cycle in $X$ (since $X$ contains an ${\bf S}^2\times {\bf S}^1$,
as we discussed
above, it certainly contains an ${\bf S}^2$ that cannot be contracted --
we simply pick a point on the ${\bf S^1}$ in ${\bf S}^2\times {\bf S}^1$)
\foot{The interpretation of this K\"ahler parameter is obvious in
the type IIB dual, as it is
the scalar field of the six dimensional ${\cal N}=(1,1)$ supersymmetric
theory on the two parallel $(0,1)$ five-branes. Duality relates this
to a Calabi-Yau three-fold in M-theory containing a curve
of $A_1$ singularities \DD.}.

As we discussed above, at the level of topological strings
the theory is a $U(N_1)\times U(N_2)$ Chern-Simons theory,
but since there are two stacks of D-branes,
there is a new open string sector where one end of the string
is on the D-branes wrapping $M_1$ and the other on $M_2$.
The ground states of this string correspond to constant maps to the
${\bf S}^1$ that the three-manifolds ``intersect'' over. Correspondingly,
there are two states in the Ramond sector of the topological string,
a real scalar and a one form,
with $U(1)_R$ charges $-1/2$ and $1/2$. Only the scalar is physical,
and taking into
account both orientations of the string, we get a complex scalar $\phi$ in
$(N_1,\overline N_2)$. This complex scalar is generically massive, and
its mass is proportional to
the ``distance'' between $M_1$ and $M_2$ given by the
complexified K\"ahler parameter $r$.
We will show below that the only modification of the
topological string we need to make in this geometry is to include
the minimally coupled complex scalar in this sector.
Because of the topological invariance
of the theory the action of a charged scalar with minimal coupling
is of the form ${\cal L_{\phi}} \sim\; \oint_{\gamma}\;
{\rm Tr} \,\bar{\phi}\,(d-A_1+A_2)\, \phi $.
Note that the scalar field gets a mass from turning on
a Wilson line on the ${\bf S}^1$ it propagates on. We will pick
its ``background'' value which we denote by $r$ below.
The path integral involving
$\phi$ is Gaussian so it can be easily evaluated \ov\ and gives:
\eqn\ove{\eqalign{
{\cal O}(U_1, U_2; r) =&\exp\Bigl[-{\rm Tr} \; \log(
{\rm e}^{r/2} U_1^{-1/2}\otimes
U_2^{1/2} - {\rm e}^{-r/2} U_1^{1/2}\otimes  U_2^{-1/2})\Bigr]\cr
=&\exp\Bigl\{\sum_{n=1}^{\infty}\; \frac{{\rm e}^{-nr}}{n}
\;{\rm Tr} U_1^{n}\; {\rm Tr}U_2^{-n}\Bigr\},}}
where $U_{1,2}$ are the holonomies of the corresponding
gauge fields around a loop
\foot{In going from the first to the second line in
\ove, we have dropped
a factor of
${\rm det}(U_1^{1/2}){\rm det}(U_2^{-1/2})$ in ${\cal O}$.
This factor, which equals $\exp({\sqrt{N_1}\over 2}
\oint_{\gamma} {\rm Tr} A_1 - {\sqrt{N_2}\over 2}
\oint_{\gamma} {\rm Tr} A_2)$, can be absorbed away in a
redefinition of $r$. It is likely that this is related to the holomorphic
anomaly of topological strings \BCOV, and this clearly
deserves further investigation.}
$$\quad U_i={\rm P}\exp \oint_{\gamma}A_i\;\; \in U(N_i),\;\; i=1,2.$$

\ifig\oneannulus{There is one holomorphic annulus connecting
the two ${\bf S}^3$'s. This corresponds to a one-loop computation with
a bifundamental string running around the loop.}
{\epsfxsize 4.0truein\epsfbox{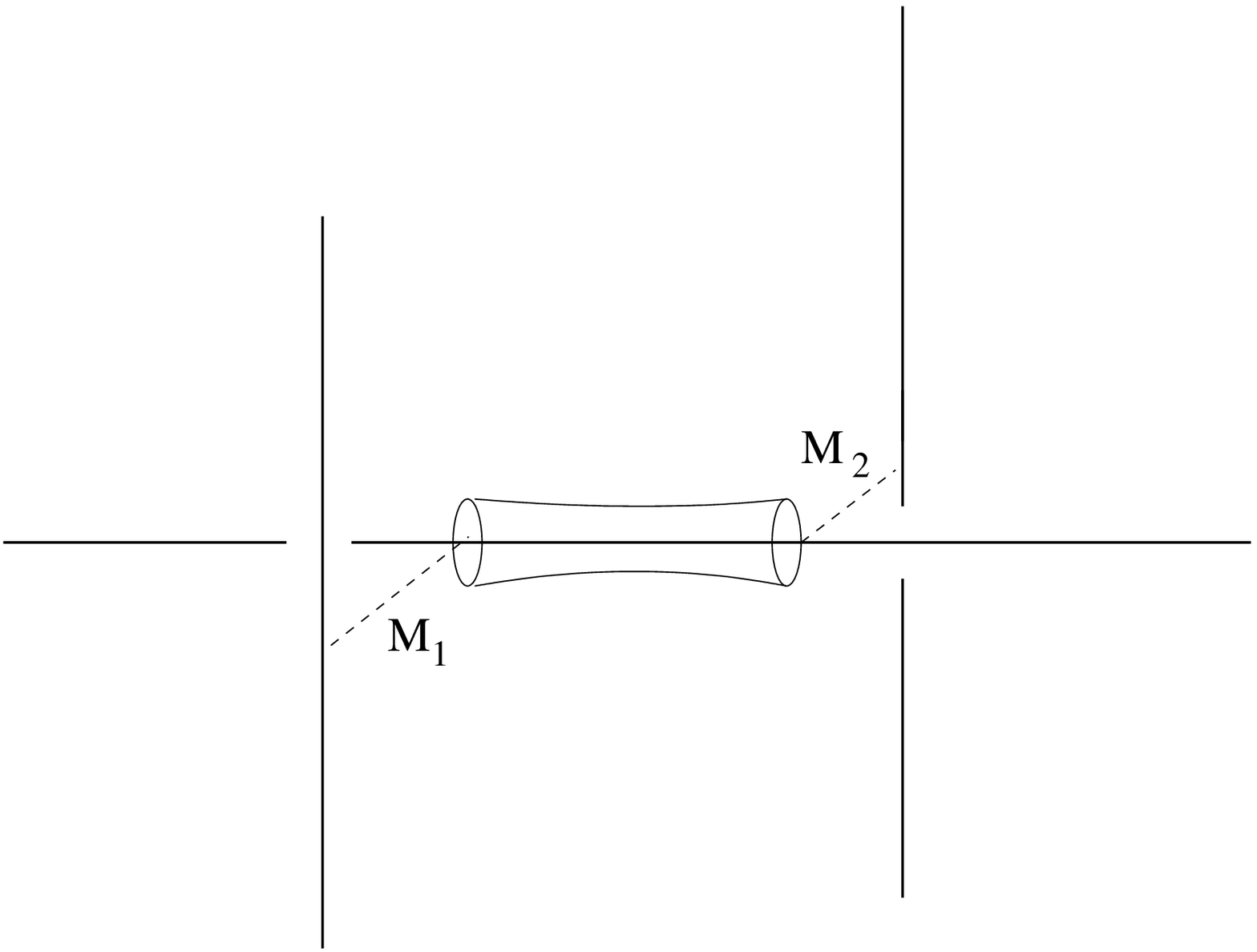}}

Note that the operator ${\cal O}$ is
the amplitude for a primitive annulus of size $r$
with boundaries on $M_1$ and $M_2$, together with its
multicovers \ov\lmv\mv. This annulus is depicted in \oneannulus, and
it is a piece of the holomorphic curve that is wrapped
by the $(1,0)$ brane. This curve is obtained by setting  $x=0=y=z$ in
\two\ and is given by $x'y'=\mu_1\mu_2.$

The Chern-Simons path integral in this geometry
is therefore defined with the insertion of the above operator.
The path integral of the A-model string field theory
in this background is therefore given by:
\eqn\tp{Z= \int {\cal D} A_1 {\cal D}A_2
e^{S_{\rm CS}(A_1)+S_{\rm CS}(A_2)}
\exp \Bigl\{ - \sum_{n=1}^{\infty} \frac{{\rm e}^{-nr}}{n}
{\rm Tr} U_1^{n} {\rm Tr}U_2^{-n}\Bigr\}}
To recapitulate,
we have Chern Simons theory on two three-manifolds
$M_1$ and $M_2$ connected via an annulus. The boundaries of
the annulus look like ${\bf S}^1$'s in both of them, {\it i.e.}
we have one knot in each $M_i$. The
topological theory is computing
the expectation value of the operator ${\cal O}(U_1,U_2;r)$,
which involves Wilson loop operators around the two knots.
This obviously extends to more general configurations: for every pair of
three-manifolds $M_1$ and $M_2$ that are connected by a holomorphic
annulus, we
will get a bifundamental complex scalar. Integrating
this field out, we find we need to insert an operator \ove,
where $r$ is the size of the corresponding annulus diagram
in spacetime.
\ifig\move{In the left figure, it looks like there is a family of
holomorphic annuli between $M_1$ and $M_2$, and a holomorphic annulus
connecting $M_1$ with $M_3$. However, by moving in the complex structure
moduli space we get to the figure in the right, where it is clear that there
is an isolated holomorphic annulus connecting $M_1$ and $M_2$ and no
holomorphic curve between $M_1$ and $M_3$.}
{\epsfxsize 5.0truein\epsfbox{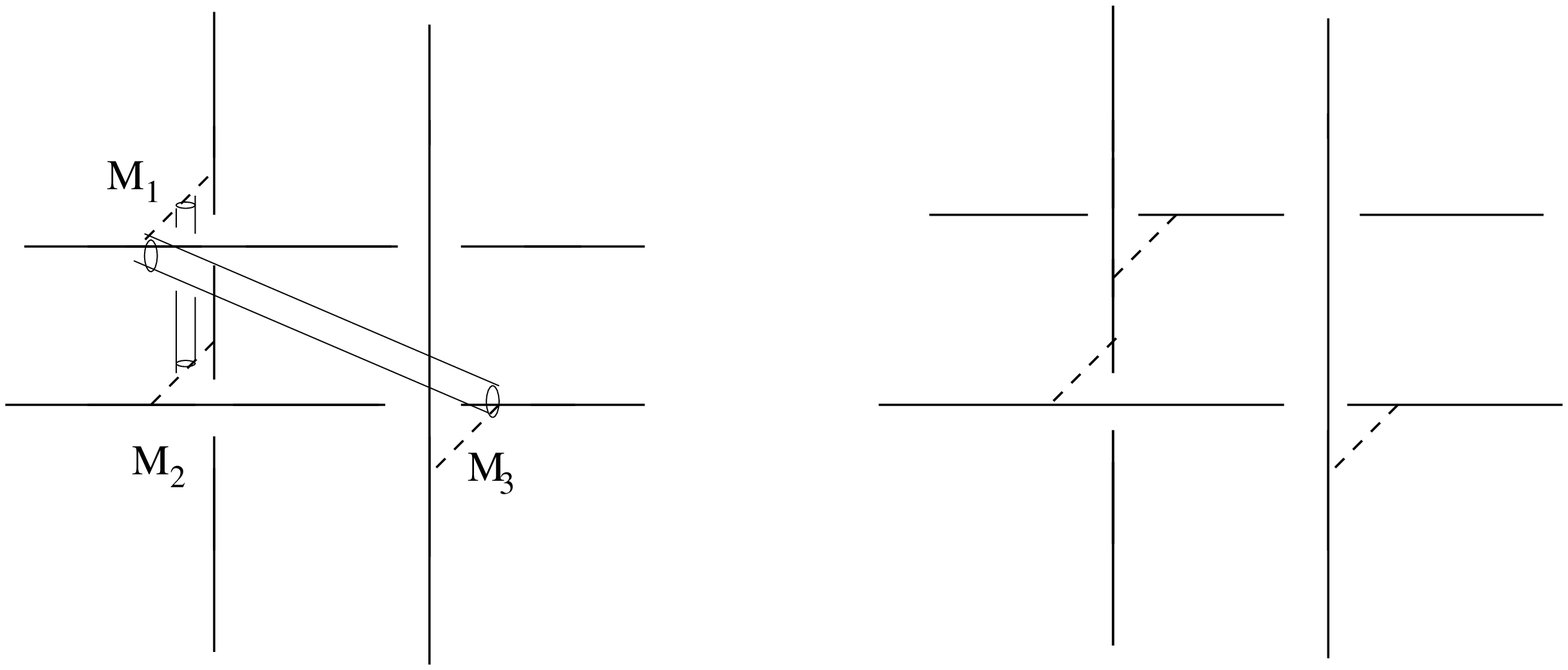}}

As an example, consider \move.  There are
$N_i$ D-branes wrapping a chain of four minimal spheres $M_i$, $i=1,\ldots
4$ connecting two $(1,0)$ branes and two $(0,1)$ branes.
For every pair of spheres intersecting
over an ${\bf S}^1$ we get a bifundamental scalar field, so we have matter in
representations $(N_i,{\overline N}_{i+1})$,
where $i=5$ corresponds to the first
sphere again. Note that in the $(N_1,{\overline N}_2)$ and $(N_3,{\overline
N}_4)$
sector, the bifundamental scalar is not localized, and
correspondingly in \move\ there is a family of annuli.
In fact, a careful reader has probably noticed
that this could have happened in the two-sphere case
as well, had we not chosen judiciously the
ordering of $(1,0)$ and $(0,1)$ branes in the $z$ direction
of the base.  In other words, we could have picked $\mu_1,\mu_2>0$ in
\two\ , and we would have found a family of annuli.  See \move.
This objection is in fact its own cure. Namely,
by changing the complex structure of $X$ we could
go from one configuration to the other.
In fact, using the other direction in the $z$-plane
we can do this in a smooth way, as the $\mu$'s
are complex, without passing through a singularity
of the three manifold.
On the other hand, the topological A-model
amplitudes cannot depend on the complex structure moduli.
As a consequence, the value of operators ${\cal O}(U_i,U_{i+1};r_i)$
cannot change in passing between the two configurations,
and they are given by the annulus computation
we already outlined.

This idea is rather powerful and it leads to the fact that, in all
the toric cases, the only holomorphic curves are annuli connecting
pairs of ${\bf S}^3$'s along lines on the toric base ({\it i.e.} along loci
of $(p,q)$ 5-branes). The argument for this is extremely simple:
as we explained before,
we can deform the theory to a generic point in the complex
structure moduli space, where it is manifest that the only holomorphic
curves are annuli. By the fact that topological A-model amplitudes do not
depend on complex structure moduli, we can immediately conclude
that only annuli contribute to topological string amplitudes.
Let us now explain why this preserves only annuli.
At a generic point of the complex structure moduli space
(which we can always choose),
the $T^2$ fibers degenerate over a
set of points in the $z$ plane, and the three-cycles $M_i$ in $X$ project to
lines connecting them, which are generically not aligned.
Recall that the holomorphic curves project to points in the $z$ plane.
This means that the ``large'' holomorphic curves must project to
points in the $z$-plane
(as discussed in section 2) where different $M_i$'s intersect,
and for a generic choice of complex structure these are the points
where the $T^{2}$ fibers degenerate.
In other words, we are left only with annuli over loci
where the $T^{2}$ fibration degenerates, as we wished
to show\foot{
Note that at a generic point in complex structure moduli space,
$M_i$ are not mutually supersymmetric, and when we add D-branes
supersymmetry is broken. However, for topological A-model amplitudes
to make sense we do not
require supersymmetry, and D-branes need only be Lagrangian,
which holds for any complex structure.}.

\newsec{D-branes on Chains of Three-Manifolds and Knot Invariants}

In order to evaluate the A-model partition functions in these
backgrounds we need a few additional pieces of data.
Namely, we need to know how the different knots
are linked, in particular their linking numbers ${\rm lk}(\gamma_i,
\gamma_j)$, and also what is their framing -- the self linking
number of each of ${\gamma}_i$'s. As it is explained in \jones\gmm, the
framing is a rather subtle effect from the point of view of Chern-Simons
theory, having to do with the fact that in evaluating
expectation values of Wilson loop operator associated to the knot,
one encounters certain ambiguities in the calculation.
These are akin to a choice of point-splitting regularisation,
since to calculate the self-linking number
in a way that is consistent with topological invariance one must
choose a ``framing'' by thickening
a knot into a ribbon. Different framings differ by adding twists
to the ribbon, the framing itself being
defined as the linking number of the two edges of the ribbon.
The Wilson loop operators are not invariant under the change
of framing. We will show below that different
choices of framing correspond
in the present context to different target space geometries.
The role of framing in topological string theory was discovered in
\AKV\ in a closely related context and
studied subsequently in \refs{\mv,\kl,\zasgr}.

\subsec{Rewriting ${\cal O}$}

Before we proceed, it is key to note
that there is an illuminating way to write the operator
${\cal O}(U_1,U_2;r)$, by
using the techniques of \lm.
If we expand the exponential explicitly, we get:
\eqn\expans{
1+ \sum_{h=1}^{\infty} \sum_{n_1, \cdots, n_h}
{1 \over h!} {{\rm e}^{-r\sum_{i=1}^h n_i} \over n_1 \cdots n_h}
{\rm Tr}U_1^{n_1}
\cdots {\rm Tr}U_1^{n_h} {\rm Tr}U_2^{-n_1} \cdots {\rm Tr}U_2^{-n_h}.}
Now we write the $h$-uples $(n_1, \cdots, n_h)$ in terms of a
vector $\vec k$, as in \lm: $k_i$ is the number of $i$'s in
$(n_1, \cdots, n_h)$. Taking into account that there are
$h!/\prod_j k_j!$ $h$-uples that give the same vector $\vec k$, and
that $n_1 \cdots n_h =\prod_j j^{k_j}$, we find that \expans\ equals
\eqn\expanstwo{
1+ \sum_{\vec k} {{\rm e}^{-\ell r}\over z_{\vec k}} \Upsilon_{\vec k} (U_1)
\Upsilon_{\vec k}(U^{-1}_2),}
where $z_{\vec k} = \prod_j k_j! j^{k_j}$,
\eqn\basis{
\Upsilon_{\vec k}(U)= \prod_{j=1}^\infty \Big( {\rm Tr}\,U^j \Big)^{k_j},}
and $\ell = \sum_j j k_j$.
Now, Frobenius formula tells us that
\eqn\frob{
{\rm Tr}_R (U)=\sum_{\vec k} {1 \over z_{\vec k}} \chi_R ( C(\vec k))
\Upsilon_{\vec k} (U).} Using this together with orthonormality of the
characters gives immediately that
\eqn\transp{
{\cal O}(U_1,U_2;r)=\sum_R {\rm Tr}_R
U_1 {\rm e}^{-\ell r}{\rm Tr}_R U^{-1}_2,}
where $\ell$ is the number of boxes in the Young tableau of $R$
and the sum is a sum over all representations,
including the trivial one. We remind the reader that
$U_i$ is a Wilson line in the three-manifold $M_i$. Notice
that this operator is the cylinder propagator for a two-dimensional
gauge theory \qgt\cmr, in which $r$ plays the role of time and the Hamiltonian
is given by the first Casimir of $U(N)$ (which counts precisely the number
of boxes $\ell$ of a representation).

\subsec{Framing}

One of the key ideas used in \jones\ is  that one can cut the manifold
up into pieces on which one can solve the theory, and then glue them back
together. Central to the story is
also the relation of the Hilbert space of Chern-Simons theory
with the space of conformal blocks of WZW models.
Recall that all manifolds
$M_i$ in our geometry can be obtained by gluing together
solid tori with a diffeomorphism identification of the boundary.
Associated to the boundary $T^2$ we have a finite dimensional
Hilbert space, and a basis of states is labeled by
the representations of the affine Lie
algebra \jones. We will denote this basis by $|R\rangle$.
The dual Hilbert space has a basis $\langle {\overline R}|$,
where ${\overline R}$ denotes the representation conjugate to
$R$. The dual pairing is simply $\langle {\overline R}_1 | R_2\rangle =
\delta_{R_1 R_2}$. Notice that $|R \rangle$ can be computed
by the path integral on a solid torus
with insertion of a Wilson line in representation
$R$ around the cycle that is non-trivial in homology.
The corresponding state in the dual Hilbert space $\langle \overline R|$
is obtained by doing the same path integral but
over the manifold with opposite orientation.

In the context of Chern-Simons theory with no insertions,
because the diffeomorphism of the boundary induces a linear
transformation of the Hilbert space,
one can think about the path integral on $M$
in terms of the path integrals on two solid tori,
that are then glued together with
an ${\rm SL}(2,{\bf Z})$ matrix $V_{M}$ that specifies $M$.
Since we are making no insertions, the state associated to each
of the solid tori is the vacuum $|0\rangle$ (corresponding to the
trivial representation), and the partition function
of Chern-Simons theory on $M$ is $Z(M) =\langle 0|V_M|0\rangle$.

In the problem at hand, we are interested
in the Chern-Simons amplitude not in the vacuum but
in the presence of Wilson lines.
The gluing procedure that gives the partition function
can be generalized to this setting, since the role of the insertions
will simply be that the solid tori give rise to states $|R \rangle$ with
arbitrary $R$. In our problem we have insertions of
operators ${\cal O}(U_i,U_j;r)$ corresponding to annuli connecting the two
manifolds. Each annulus is attached to the $M_i$'s
either on its left or the right ``half'',
and by \transp\ we can regard it as carrying a Wilson line in
an arbitrary representation $R$ of the gauge group on the right half, and
the conjugate representation ${\overline R}$ in the left half. We also
have to sum over all representations.
For example, in the two-sphere case, there is a knot on the right half
of $M_1$ and the left half of $M_2$. We thus have
$$Z(M_1,{\rm Tr}_R U)=\langle 0|V_{M_1}|R\rangle,$$
and
$$Z(M_2,{\rm Tr}_R U^{-1})=\langle {\overline R}|V_{M_2}|0\rangle,$$
where the ${\rm Tr}_R U$, ${\rm Tr}_R U^{-1}$ mean that we do the
path integral with the insertion of these operators.
Thus, by using \transp\ and the above gluing techniques, the full
amplitude \tp\ can be written as:
\eqn\twop{
Z=\sum_R \langle 0|V_{M_1}|R\rangle e^{-\ell r}\langle
{\overline R}|V_{M_2}|0\rangle,}

Note that were it not for the weight of $e^{-\ell r}$, we could use
the resolution of the identity $\sum_R |R \rangle \langle {\overline R}|=
{\bf 1}$ and the operator insertion in \twop\
would correspond to a surgery operation that
glues together $M_1$ and $M_2$. The resulting manifold $M_1\# M_2$ would
have gluing matrix
$$V_{M_1\# M_2}=V_{M_1} V_{M_2}.$$
This corresponds to the geometric fact that, when $r=0$,
the two special Lagrangian three-spheres that we called
$M_1$ and $M_2$ are exactly degenerate with $M_1\# M_2 ={\bf S}^{2}\times
{\bf S}^1$
that is their sum in homology.
But instead we have a finite $r$-time
propagation with a Hamiltonian
that counts the numbers of boxes.
Namely, insertion of the operator ${\cal O}$
corresponds to cutting off the right half of $M_1$ in the vacuum
and the left half of $M_2$, and gluing in instead
$${\cal O}(U_1,U_2;r) = \sum_R |R\rangle e^{-\ell r}\langle
{\overline R}|.$$

In the case of \oneannulus, $M_1$ and $M_2$
are ${\bf S}^3$'s with canonical framing.
Since $\alpha=(1,0)$ degenerates in the left
half of $M_1$ and $\beta=(0,1)$ in its right half,
and $\alpha$ and $\beta$ are exchanged for $M_2$,
the gluing matrices are
$V_{M_1} = S = V_{M_2}^{-1},$
where $S=\left(\matrix{0&-1\cr 1&0}\right)$.
Note that standard surgery gives ${\bf S}^2\times {\bf S}^{1}$ with
partition function equal to
one, as expected.
The amplitude \twop\ receives contributions from
unknots on $M_1$ and $M_2$ in representation $R$.

Note that the transformation that changes the framing of the three-manifold
affects the Wilson loop amplitudes. The
diffeomorphism by $T^n$ on the boundary of the solid torus
with a Wilson loop in representation $R$ in the center, adds
$n$ twists to the ``ribbon'' that frames the knot. The change of
framing acts on the Wilson loop amplitude by
$$T|R\rangle = {\rm e}^{2 \pi i (h_R-c/24)}|R\rangle,$$
where $c$ is the central charge of the current algebra, and
$h_R$ is the conformal weight of
the WZW primary field in representation $R$. Recall
that $h_R$ is given by
$$
h_R = {\Lambda_R \cdot (\Lambda_R + \rho) \over 2(k+N)}
$$
where $\Lambda_R$ is the highest weight of $R$ and $\rho$ is the
Weyl vector. The numerator $C_R= \Lambda_R \cdot (\Lambda_R + \rho)$
is the quadratic
Casimir of the representation.
While there is no natural choice of framing, there is a canonical
choice at least on ${\bf S}^3$ and this corresponds to zero self-linking
number. In the above example, both unknots on ${\bf S}^3$
were framed canonically. Below, we will see that other choices of framing
arise as well.

\subsec{Linking}

The considerations above completely determine the gluing
matrices, up to irrelevant framings that do not affect
the amplitude by other than renormalization of $r$.
Therefore, the linking of the different knots should be
determined as well. Let us discuss this issue with a
concrete example.

\ifig\threespherewanuli{
In the figure there is a
chain of three minimal spheres connecting $(1,0)$,
$(0,1)$ and $(1,-1)$ branes, with branes wrapped on each sphere. }
{\epsfxsize 4.0truein\epsfbox{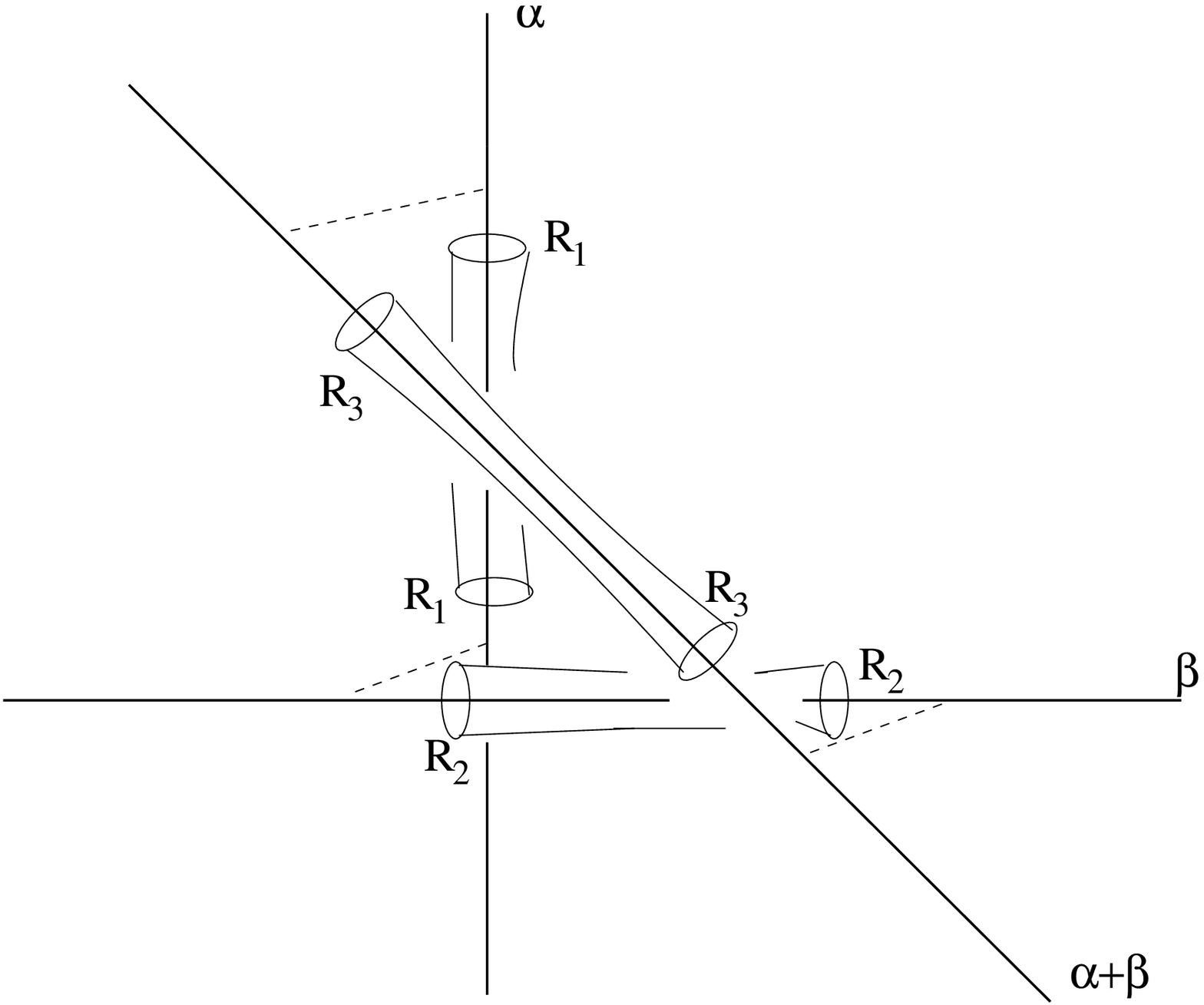}}
The three-spheres in \threespherewanuli\ form a
necklace where each is intersecting the other two
over an ${\bf S}^1$,
so we get scalar fields in the bifundamental,
and integrating them out leaves us with the annuli shown in the figure.
The path integral of the A-model in this background
involves four Chern-Simons theories on a chain
of four three-spheres connected
with annuli:
\eqn\fourp{Z=\int \prod_{i=1}^3 {\cal D}A_i e^{S_{\rm CS}(A_i)}
{\cal O}(U_1,U_2;r_1){\cal O}(U_2,U_3;r_2){\cal O}(U_3,U_1;r_3)}

There are two unknots on
each three-sphere, and the amplitude will depend
on their linking, in addition to their framing.
Just as above, we can use \transp\ to write this in a more transparent form
\eqn\fourg{Z= \sum_{R_1,R_2,R_3}
\langle {\overline R}_1|V_{M_3}R_3\rangle e^{-\ell_3 r_3}\langle
{\overline R}_3|V_{M_2}R_2\rangle e^{-\ell_2 r_2}
\langle {\overline R}_2|V_{M_1} R_1\rangle e^{-\ell_1 r_1}.}
Since the D-branes go around the loop from one degeneration locus to the
other
and then back to the first one, we must have
$$V_{M_3}V_{M_2}V_{M_1}=1.$$
This will hold generally whenever there are closed loops with D-branes
in the toric diagram.

\ifig\hopf{The Hopf link with linking number ${\rm lk}=+1$
. }
{\epsfxsize 2.0truein\epsfbox{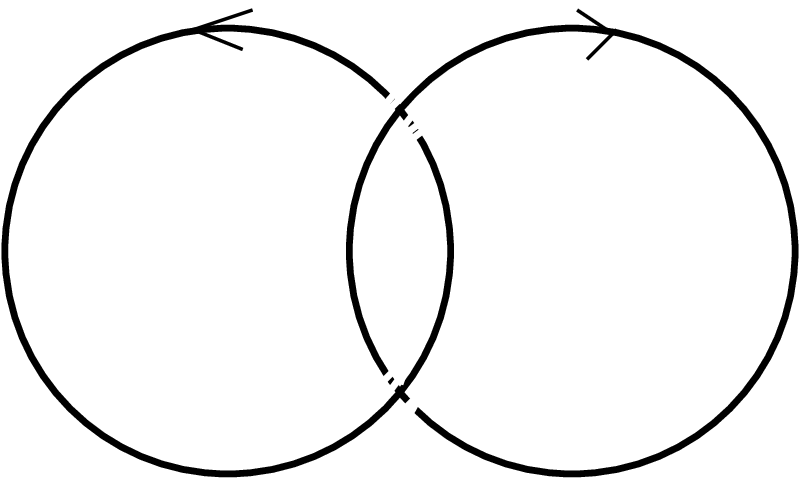}}

Looking at \threespherewanuli\ we can read off,
\eqn\matrices{V_{M_1}=S^{-1}, \quad V_{M_2} = ST^{-1}S, \quad V_{M_3}= TS^{-1},}
so that the last factor of \fourg\ is given by
\eqn\firstlink{
Z(M_{1},{\cal L}(R_2,R_1))=\langle{\overline R}_2|S^{-1}|R_1\rangle
=S^{-1}_{R_1 R_2}.}
This means that
$M_{1}$ is a three-sphere
with two unknots that are linked into a
Hopf link ${\cal L}(R_2,R_1)$ of linking number ${\rm
lk}= 1$ (see \hopf).
Namely, as we have seen previously, ${\bf S}^3$ is obtained
by identifying two solid tori up to an $S$ transformation that
exchanges the $\alpha$ and the $\beta$ cycles of the $T^2$.
If $\beta$ is the nontrivial cycle of the solid $T^2$, and
along this cycle one has a knot in representation $R_1$
and another one in representation $R_2$, then the $S^{-1}$ transformation
results in two unknots with zero framing (this does
not add any twists to the ribbons that frame the knots),
but which are linked in a Hopf link with linking number ${\rm lk}=1$
(an $S$ transformation would give a Hopf link with
linking number $-1$).
Similarly, using $ST^{-1}S=TS^{-1}T$,
we see that $M_{2}$ has a Hopf link
with two knots of framing $+1$.
\eqn\thirdlink{
Z(M_{2},{\cal L}(R_3,R_2))=\langle {\overline R}_3|TS^{-1}T|R_2
\rangle.}
Finally, $M_{3}$ is a three-sphere with a Hopf link
whose components are an unknot carrying representation $R_1$ and
with framing $+1$, and an unknot carrying representation $R_3$ with
canonical framing:
\eqn\secondlink{
Z(M_{3}, {\cal L}(R_1,R_3))=\langle{\overline R}_1|TS^{-1}|R_3\rangle.}
\ifig\arbitrary{The amplitude associated to this geometry can be
interpreted in terms
of a lattice model. The annuli correspond
to states of the lattice. The 3-manifolds correspond to the
interaction vertices. The figure shows the annuli on the
``primitive'' edges and some ``non-primitive'' ones.}
{\epsfxsize 3.5truein\epsfbox{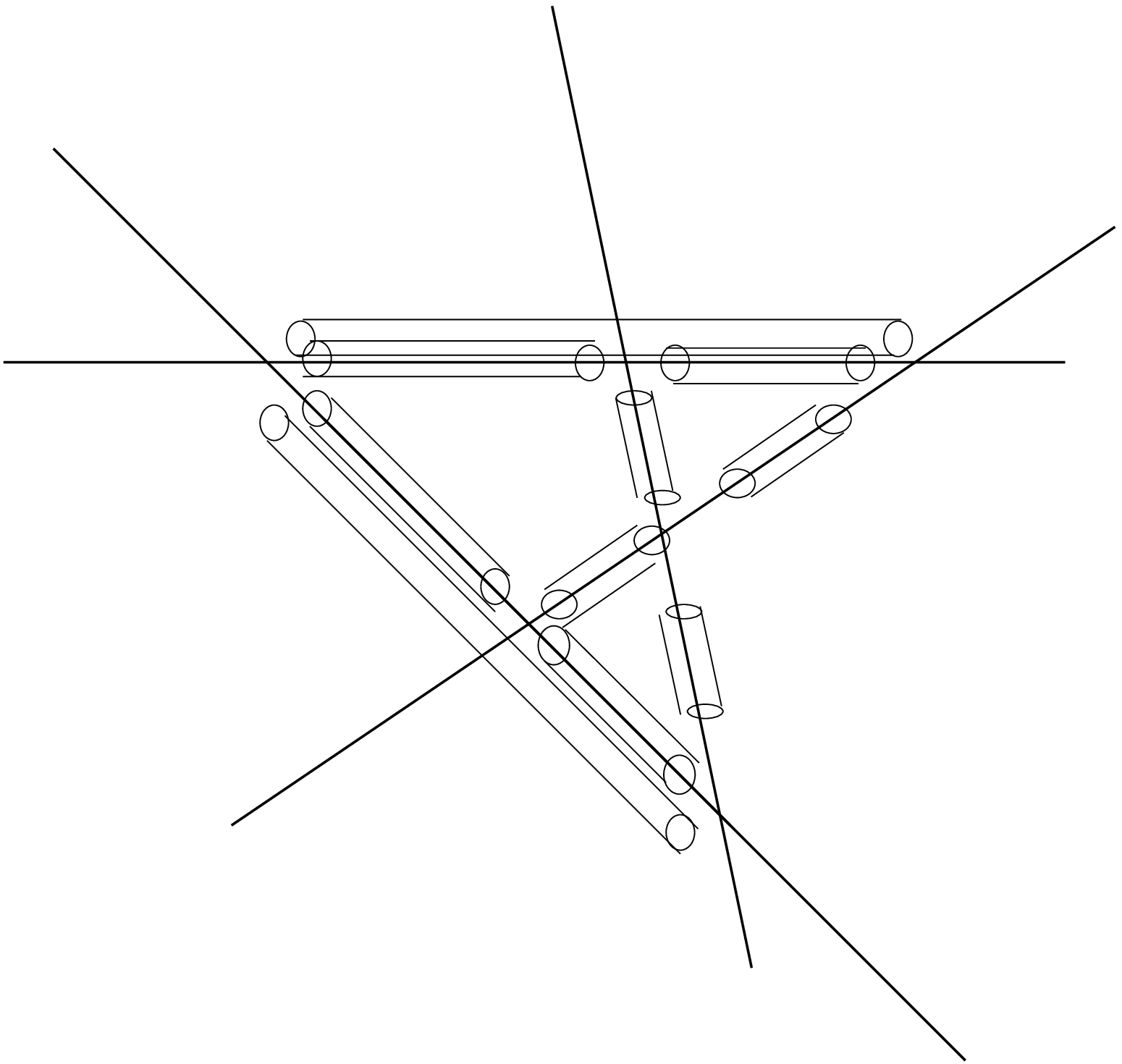}}
\subsec{Lattice model interpretation}
The models discussed above clearly generalize to more complicated
geometries
like the one depicted in \arbitrary, where we have suppressed
one direction of the base. The rules for computing the amplitudes
should be clear
from the previous discussion:
\medskip
1) The model has  states associated
to all the edges of the lattice.  Some of the edges
are ``primitive'' (connecting nearest neighbor nodes)
and some are ``non-primitive'' (connecting other nodes
but always along the straight lines of the lattice)
The states are labeled by representations of
the affine Lie algebra ({\it i.e.} a state in the
Hilbert space of $T^2$, ${\cal H}$). To each state on the $i$-th edge
we associate a weight ${\rm e}^{-\ell r_i}$, where $r_i$ is the length of
the corresponding edge and the $\ell$ is the number
of boxes in the corresponding representation.
\medskip
2) To every vertex we associate a linear operator.
This linear operator is obtained by
computing matrix elements like the ones depicted below:
\ifig\amp{The four-point vertex.}
{\epsfxsize 3.5truein\epsfbox{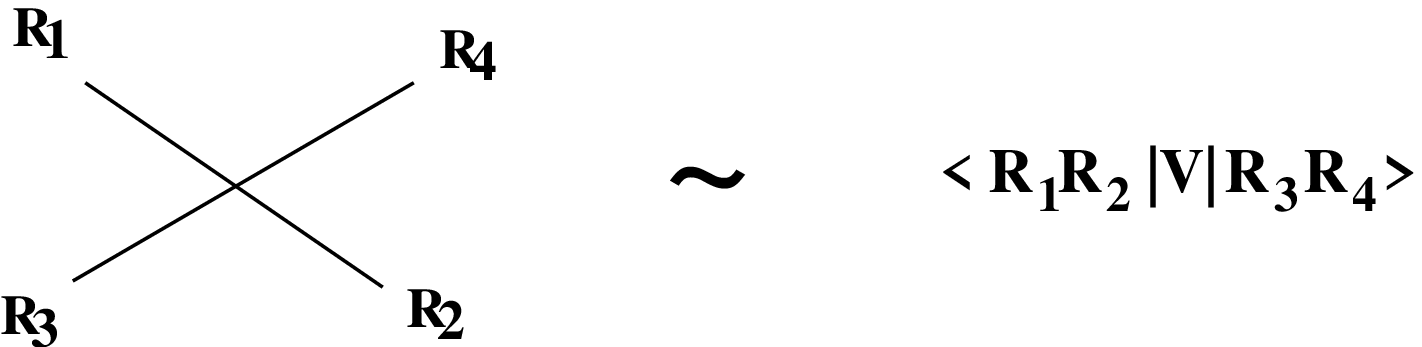}}
Here, $R_1, R_2$ and $R_3, R_4$ are the
two pairs of representations corresponding to the
collinear edges. A state $|R, R'\rangle$
is obtained by doing the Chern-Simons path integral over the solid torus
with two parallel Wilson lines inserted along its nontrivial
cycle, in representations $R$, $R'$. $V$ is the gluing matrix, as
explained before. As is well known,
using the fusion rules of the WZW theory, we can write
$$ |R, R'\rangle = \sum_{R''} N_{R R'}^{R''} |R''\rangle$$
where the fusion coefficients are given by the Verlinde formula \verlinde\
\eqn\verlindef{
N_{R R'}^{R''}= \sum_{Q} { S_{R Q} S_{R' Q} S^{-1}_{R'' Q}
\over S_{0 Q}}, }
\medskip
Using this we can write the four-point vertex as
\eqn\matel{
\langle \overline{R}_1, \overline{R}_2 | V | R_3, R_4 \rangle=
\sum_{Q,Q'}N_{R_1 R_2}^{Q}V_{QQ'}N_{R_3R_4}^{Q'},
}
where $V$ denotes the corresponding modular
transformation matrix.
Notice that, although there are four primitive edges ending on each vertex,
one can have many non-primitive edges ending
on the same vertex. In that case, we will have matrix elements of the
form
\eqn\genover{
\langle \overline{R}_1, \cdots, \overline{R}_n | V | R'_1, \cdots, R'_m
\rangle
}
where the in- and out-states can be evaluated by a repeated use of the
fusion rules. In \genover, $\overline{R}_1, \cdots, \overline{R}_n$
correspond to a set of collinear edges, and
$R'_1, \cdots, R'_m$ to the other set.
As explained before the solid tori are glued together
by an ${\rm SL}(2,{\bf Z})$ matrix $V$ that is computed as in \glue.

\medskip
3) Since there are edges that go off to infinity, there are
boundary conditions: the state on these edges is always
 the trivial representation\foot{It would also be interesting
to put periodic boundary conditions and interpret
them as partially compact Calabi-Yau models.}.
\medskip
4) The amplitude is the product of the linear operators over all the
vertices, together with the weights associated to the connecting edges.
At the end we sum over all representations on each link.

\newsec{Large $N$ Duality}
As was recently demonstrated \proof,
one can derive the large $N$ duality conjecture of
Chern-Simons on ${\bf S}^3$ with topological strings \gv.
In this derivation one starts with the linear sigma model description
of the closed string side and finds that in some limit the theory
develops Coulomb and Higgs branch. The Coulomb branch plays the
role of holes in the dual Chern-Simons description.

The models we are considering here all admit a linear sigma model
description \phases , as discussed in \kmv .
Thus one can start from the gravity side, and go to the point
on moduli for each $U(1)$ gauge factor and repeat
the analysis of \proof, which should lead to topological open
string description with $N_i$ D-branes wrapped around ${\bf S}^3_i$.
The analysis we did for the open string demonstrated that this
open string can in turn be written in terms of some link observables
in the product of $U(N_i)$ Chern-Simons theories.  Thus we find the general
prediction that
$$F_{\rm closed}(t_i,r'_a)=F_{\rm open}(N_ig_s, r_a)$$
where by $F_{\rm open}$ we mean the open string amplitude
with link observables inserted, and the $r_a$ correspond
to sizes of the annuli. In this equation, the $t_i$ on the closed
string side are the K\"ahler moduli of the blowups
corresponding to where the ${\bf S}^3_i$ were, and $t_i=N_i g_s$.
As we will see later, $r'_a=r_a-{1\over 2}(t_{a1}+t_{a2})$ where $t_{ai}$
denote the K\"ahler moduli associated to the two ends of the annulus
$a$. It would be interesting to repeat in detail the analysis of \proof\
for the case at hand and thus obtain these shifts directly.

\newsec{Closed String Localization}

In this section we argue that the large $N$
duality proposal given in the previous section
is in accord with localization ideas in computation
of the closed string invariants.

It was suggested in \kon\ that one can use circle
actions to localize closed topological string amplitudes.
The final answer takes the form of sum over certain graphs
with nodes corresponding to genus $g$ Riemann surfaces.
This idea has been further developed \gp\
and applied to some concrete examples in \ckyz\ (at genus zero) and
\kz\ (for higher genus).
The geometry of the localization is very much related to the
$(p,q)$ 5-brane graphs we have in our setup. Basically one
ends up with sums over graphs whose links correspond
to intervals in the $(p,q)$ 5-brane web connecting
adjacent vertices.  These correspond to rational curves
in the closed string side, and in the gauge theory
setup they correspond to annuli.  Moreover one is instructed
to consider all genera computation on each node, which corresponds
to mapping the whole Riemann surface to that point on the toric geometry.
This ends up with a particular computation of a characteristic class
on the moduli space of Riemann surfaces, that
in particular depends on which links
have been used in the graph. This seems to match naturally
with the Chern-Simons computation, where each node is replaced by open
string Riemann surfaces captured by Chern-Simons, coupled
to each other through the Wilson loop expectation values coming
from annuli. It is as if the Chern-Simons theory is computing
directly the relevant characteristic classes on moduli of Riemann surfaces.
This is not at all surprising, in light of the observations
in \mv\ where one can use the framing dependence
of unknot in Chern-Simons theory to compute all intersection numbers
of Mumford classes with up to three Hodge classes, which is what
the closed string side computes \kl.
It would be extremely interesting to make this connection
with Kontsevich integral more precise and reduce
the statement of the equivalence to some concrete computation at each node,
which is being done using Chern-Simons gauge theory.  Incidentally
this is in the same spirit of the current methods of computation
of these invariants where one uses Kontsevich's results on
matrix realization of Mumford classes \konts\ together with certain
results of Faber \faber. However, the Chern-Simon gauge theory
is a more natural
realization of this computation.

\newsec{Closed String Invariants from Chern-Simons Theory}

In this section we will show that the large $N$ duality
proposed in section 5 is a powerful way to compute
closed string topological A-model amplitudes for local
Calabi-Yau manifolds, in terms of Chern-Simons amplitudes.
In particular we will consider examples of $\IP^2$ blown
up at three points ($\IB_3$ del Pezzo), and $\IP^1\times
\IP^1$ blown up at four points.  Since the size
of the blow ups are proportional to the rank of the
corresponding dual gauge group, we can also consider
the limit where the blown up $\IP^1$'s have infinite size
by considering the $N_i\rightarrow \infty$ limit. This
in particular leads to computation of topological
strings for $\IP^2$ and $\IP^1\times \IP^1$ inside a Calabi-Yau
threefold.

\subsec{Chern-Simons Invariants of Unknots and Hopf Links}

The toric geometries that we have described involve framed unknots and
Hopf links, therefore in the evaluation of the Chern-Simons
amplitudes we will need the
invariants of the unknot and the Hopf link in arbitrary representations
of $SU(N)$. In this section we give precise formulae for these invariants.
Our notation is as follows: $W_{R_1, R_2}({\cal L})$ denotes the vacuum
expectation value in Chern-Simons theory corresponding to the link ${\cal
L}$ with components ${\cal K}$, ${\cal K}'$:
\eqn\linknot{
W_{R_1, R_2}({\cal L}) =\langle {\rm Tr}_{R_1}(U_1) {\rm Tr}_{R_2}(U_2)
\rangle, }
where $U_1$, $U_2$ are the holonomies of the gauge field around the knots
${\cal K}$ and ${\cal K}'$, respectively. If, say $R_2=\cdot$ is the
trivial representation, the vev \linknot\ becomes the vev of the knot
${\cal K}$ (the second knot disappears), and we will denote this vev by
$W_R ({\cal K})$. The vacuum
expectation values denoted by $\langle \cdot \rangle$ are normalized, so
that they
denote the path integral with insertions and divided by the
partition function (in other words, the vev
of the identity operator is one).  Of course the duality
also cares for the overall normalization (i.e., the vacuum
energy) and this we will
put in at the end of the computation.
 We also recall
our notation for the Chern-Simons
variables:
\eqn\csvar{
q=\exp \biggl( { 2\pi i \over k+N} \biggr), \,\,\,\,
\lambda = q^N.}

It is well-known that the Chern-Simons invariant of the unknot in an
arbitrary representation $R$ is given by the quantum dimension of $R$:
\eqn\qdimvev{
W_R = {S_{0R} \over S_{00}}={\rm dim}_q R.
}
The explicit expression for ${\rm dim}_q R$ is as
follows. Let $R$ be a representation corresponding to a Young tableau
with row lengths $\{ \mu_i \}_{i=1, \cdots, d(\mu)}$, with
$\mu_1 \ge \mu_2 \ge \cdots$, and where $d(\mu)$
denotes the number
of rows. Define the following $q$-numbers:
\eqn\qnumbers{
\eqalign{
[x] =& q^{x\over2} - q^{-{x\over2}},\cr
[x]_{\lambda} =& \lambda^{1\over2} q^{x\over2} - \lambda^{-{1\over2}}
q^{-{x\over2}}.\cr}}
Then, the quantum dimension of $R$ is given by
\eqn\qdim{
{\rm dim}_q R=\prod_{1\le i<j\le d(\mu)} { [\mu_i -\mu_j + j-i]
\over [j-i]} \prod_{i=1}^{d(\mu)}  {\prod_{v=-i+1}^{\mu_i-i}
[v]_{\lambda} \over \prod_{v=1}^{\mu_i} [v-i+d(\mu)]}.}
The quantum dimension is a Laurent polynomial in
$\lambda^{\pm {1\over2}}$ whose
coefficients are rational functions of $q^{\pm {1 \over 2}}$.
In what follows in some cases we will also be interested in
the leading power of $\lambda$ in the above expression. It is easy to see
that this power is $\ell/2$, where $\ell=\sum_i \mu_i$ is the total
number of boxes in the representation $R$, and the coefficient of this
power is the rational function of $q^{\pm {1 \over 2}}$
\eqn\leadingqdim{
q^{\kappa_R/4} \prod_{1\le i<j\le d(\mu)} { [\mu_i -\mu_j + j-i]
\over [j-i]} \prod_{i=1}^{d(\mu)}
\prod_{v=1}^{\mu_i} {1 \over [v-i+d(\mu)]},
}
where
\eqn\cas{
\kappa_R = \ell+ \sum_{i=1}^{d(\mu)} \mu_i (\mu_i -2 i).}
This quantity is related to the quadratic Casimir of the representation.
In fact, one has $\Lambda_R \cdot (\Lambda_R + \rho) = \kappa_R + N\ell -
\ell^2/N$.

Let us now consider the Hopf link with linking number $1$. Its invariant
for representations
$R_1$, $R_2$ is given by
\eqn\hopfinv{
W_{R_1, R_2} = q^{\ell_1 \ell_2/N}{S^{-1}_{R_1 R_2} \over S_{0 0}},}
where $\ell_i$ is the total number of boxes in the Young tableau of $R_i$,
$i=1,2$. The prefactor $q^{\ell_1 \ell_2/N}$ in \hopfinv\ is a
correction which was pointed out in \mv, and is
due to the fact that the vev $W_{R_1, R_2}$ has
to be computed in the theory with gauge group $U(N)$.
Although the expression for the $S$-matrix is explicitly known, it is not
straightforward to write it in terms of $q$ and $\lambda$, which is
what we need.
We can use the Verlinde formula \verlindef\ giving the fusion coefficients
in terms of the $S$-matrix elements,
as well as the well-known identity $(ST)^3=S^2=C$, to obtain the
expression
\eqn\hopflr{
W_{R_1, R_2} =\sum_R N_{R_1, R_2}^R
q^{{1 \over 2}(\kappa_R-\kappa_{R_1}-\kappa_{R_2})} {\rm dim}_q R.}
This can be also derived by using
the formalism of knot operators \lllr\ilr, and it was used in \lmv\ to
obtain the integral invariants associated to the Hopf link (we must observe,
however, that in \lmv\ the Hopf link with linking number $-1$ was
considered).
Notice that the fusion coefficients $N_{R_1, R_2}^R$ become, in the
large $k$ limit, the Littlewood-Richardson coefficients
for the tensor product $R_1\otimes R_2 =\sum_R N_{R_1 R_2}^R
R$, and since we are evaluating the invariants at large $k$, $N$, to
compute \hopflr\ we
have to use these tensor product coefficients.

Another expression for the Chern-Simons invariant
of the Hopf link in arbitrary
representations has been recently obtained by Morton and Lukac by using
skein theory \ml\lukac. Let us briefly describe their result, which turns
out to be very useful in order to compute the invariants. Let $\mu$
be a Young tableau, and let $\mu^\vee$ denote its transposed
 tableau (remember that this tableau is obtained from $\mu$ by
exchanging rows and columns). The Schur polynomial
in the variables $(x_1,\cdots,x_N)$ corresponding to $\mu$
(which is the character of the diagonal
$SU(N)$ matrix $(x_1,\cdots,x_N)$ in the representation
corresponding to $\mu$), will be denoted by  $s_{\mu}$.
They can be written in terms of elementary
symmetric polynomials $e_i(x_1,\cdots,x_N)$, $i\ge 1$, as follows \macdonald:
\eqn\jt{
s_{\mu} = {\rm det} M_\mu}
where
$$M_\mu^{ij}=(e_{\mu^\vee_i +j-i})$$
 $M_\mu$ is an $r \times r$ matrix, with
$r= d(\mu^\vee)$. To evaluate $s_{\mu}$ we put $e_0=1$, $e_k=0$
for $k<0$. The expression \jt, known sometimes
as the Jacobi-Trudy identity, can be formally extended to give the
Schur polynomial $s_{\mu} (E(t))$ associated to any formal power series
$E(t)=1 + \sum_{n=1}^{\infty} a_i t^i$. To obtain this, we simply use the
Jacobi-Trudy formula \jt, but where $e_i$ denote now the coefficients
of the series $E(t)$, i.e. $e_i=a_i$. Morton and Lukac define the series
$E_{\emptyset}(t)$ as follows:
\eqn\trive{
E_{\emptyset}(t)=1+ \sum_{n=1}^{\infty} c_n t^n, }
where the coefficients $c_n$ are defined by
\eqn\coeffs{
c_n =\prod_{i=1}^n { {1 - \lambda^{-{1\over 2}} q^{i-1}}
\over {q^i-1}}.} They also define a formal power series associated to
a tableau $\mu$, $E_{\mu}(t)$, as follows:
\eqn\eser{
E_{\mu}(t)= E_{\emptyset}(t)
\prod_{j=1}^{d(\mu)} { {1 + q^{\mu_j -j} t}
\over {1 + q^{-j} t}}.} One can then consider the Schur function of the
power series \eser, $s_{\mu}(E_{\mu'}(t))$, for any pair of tableaux
$\mu$, $\mu'$, by expanding
$E_{\mu'}(t)$ and substituting its coefficients in the Jacobi-Trudy
formula \jt. It turns out that this Schur function is essentially the
invariant we were looking for. More precisely, one has
\eqn\mlfor{
W_{R_1, R_2} = ({\rm dim}_q R_1) (\lambda q)^{\ell_2 \over 2}
s_{\mu_2}(E_{\mu_1}(t)),} where $\mu_{1,2}$ are the tableaux
corresponding to $R_{1,2}$, and
$\ell_2$ is the number of boxes of $\mu_2$. More details and examples
can be found in \ml. It is easy to see from
\mlfor\ that the leading power in $\lambda$ of $W_{R_1, R_2}$ is $(\ell_1 +
\ell_2)/2$, and its coefficient is given by the leading coefficient of the
quantum dimension, \leadingqdim, times a rational function of $q^{\pm {1
\over 2}}$ that can
be easily computed by taking $\lambda \rightarrow \infty$ in
$E_{\emptyset}(t)$.

As a simple example of the Morton-Lukac formula, we can compute
$W_{(\tableau{1}, \tableau{1})}$. In this case, $s_{\{1\}}=e_1$, and
it is enough to expand $E_{\{1\}}(t)$ at first order,
\eqn\firstor{
E_{\{ 1 \}}(t)=1+ \Bigl\{ 1-q^{-1} + {{ 1-\lambda^{-1}} \over q
-1} \Bigr\} t + \cdots,}
so that we obtain
\eqn\homfly{
W_{(\tableau{1}, \tableau{1})}=\biggl( {{\lambda^{1\over2}
-\lambda^{-{1\over2}} }\over  q^{1\over2} -q^{-{1\over2}}} \biggr)^2
+ \lambda -1.}
In the same way, one can easily find:
\eqn\morevalues{
\eqalign{
W_{(\tableau{2}, \tableau{1})}= &\lambda^3 {1 - q^2 + q^3
\over (q^{1 \over 2} - q^{-{1 \over 2}})^3 (q+1)}
- \lambda{q^{-1} + 1 + q^3
\over (q^{1 \over 2} - q^{-{1 \over 2}})^3 (q+1)} \cr &+
\lambda^{-1}{q^{-1} + 1 + q^2
\over (q^{1 \over 2} - q^{-{1 \over 2}})^3 (q+1)}-
\lambda^{-3}  {1 \over (q^{1 \over 2} - q^{-{1 \over 2}})^3 (q+1)}, \cr
W_{(\tableau{1 1}, \tableau{1})}=&\lambda^3 {q^{-2} - q^{-1} + q
\over (q^{1 \over 2} - q^{-{1 \over 2}})^3 (q+1)}
- \lambda{q^{-2} + q + q^2
\over (q^{1 \over 2} - q^{-{1 \over 2}})^3 (q+1)} \cr &+
\lambda^{-1}{q^{-1} + q + q^2
\over (q^{1 \over 2} - q^{-{1 \over 2}})^3 (q+1)}-
\lambda^{-3}  {q \over (q^{1 \over 2} - q^{-{1 \over 2}})^3 (q+1)},  \cr}}
and so on. In the computations that give the invariants of $\IP^2$, we
only need the rational function of $q^{\pm {1\over 2}}$ which
multiplies the highest power in $\lambda$.

The above results are for knots and links in the standard framing. The
framing can be incorporated as in \mv, by simply multiplying the
Chern-Simons invariant of a link with components
in the representations $R_1, \cdots, R_L$, by the factor
\eqn\framefactor{
(-1)^{\sum_{\alpha =1}^L p_{\alpha}\ell_{\alpha}} q^{{1 \over
2}\sum_{\alpha=1}^L p_{\alpha} \kappa_{R_{\alpha}}},}
where $p_{\alpha}$, $\alpha=1, \cdots, L$ are integers labeling
the choice of framing for each component.

\subsec{Evaluation of the Two-Sphere Example \foot{The results in this
section have been obtained in collaboration with P. Ramadevi}}

The simplest example of how to compute a closed string
amplitude from Chern-Simons theory comes from the geometry depicted in
 \oneannulus.
As explained there, this involves computing the vacuum expectation
value of the operator \transp:
\eqn\finalres{
{\cal O}(U_1,U_2;r)=\sum_R {\rm Tr}_R U_1 {\rm e}^{-\ell r}{\rm Tr}_R U^{-1}_2,}
where $U_1$ and $U_2$ are the holonomies of dynamical gauge fields
(in other words, we are computing the vev in a $U(N_1) \times U(N_2)$
theory, but with the same coupling constant).
We are going to discuss the operator \finalres\ in a more general
setting, so that $U_1$ is the holonomy around an arbitrary knot.
We are going to assume however that $U_2$ is the holonomy around an unframed
unknot.
We now have to take the vev of this expression by doing the functional
integral over both the $U(N_1)$ field $A_1$ and the $U(N_2)$ field $A_2$.
Since we are assuming that $U_2$ is the holonomy around an unknot with
zero framing, we have that \ov\
\eqn\uzero{
\langle {\rm Tr}_R U^{-1}_2 \rangle_{A_2} =  {\rm Tr}_R U^{-1}_0
,} where $U_0$ is the element in the Cartan subalgebra that
corresponds to $\exp( 2 \pi i \rho/(k_2+N_2))$. Therefore, the vev with
respect to the field $A_2$ gives
\eqn\atildevev{
\langle {\cal O}(U_1,U_2;r) \rangle_{A_2}=\sum_R {\rm Tr}_R U_1 {\rm e}^{-\ell
r}{\rm Tr}_R U^{-1}_0.}
Notice that we can regard this vev as the generating functional
considered in \ov, where the source takes the particular value $U^{-1}_0$.
We can now take the vev with respect to the $A_1$ field, and use the
results of \ov\ to write:
\eqn\conj{
\langle {\cal O}(U_1,U_2;r) \rangle_{A_1,A_2}=
\exp\Big(\sum_{d=1}^\infty {1 \over d} \sum_{R}
f_R (q^d, \lambda_1^d){\rm e}^{-d\ell r} {\rm Tr}_R
{U_0^{-d}} \Big),}
This expression is valid for any framed knot along which we
take the holonomy $U_1$ of the gauge field $A_1$. In this equation,
$\lambda_1 =q^{N_1}$ is the exponential of the 't Hooft
coupling for the $U(N_1)$ Chern-Simons theory. We can write the
exponent of the right hand side as follows,
\eqn\kexp{
\sum_{R} f_R (q^d, \lambda_1^d){\rm e}^{-\ell r} {\rm Tr}_R
{U_0^{-d}}=\sum_{\vec k} {1 \over z_{\vec k}} f_{\vec k} (q^d, \lambda_1^d)
{\rm e}^{-\ell r} \Upsilon_{\vec k} (U_0^{-d}),}
where $f_{\vec k}$ was introduced in \lmv\ and is simply the character
transform of the $f_R$. The last factor can be easily computed to be
\eqn\generalvev{
\Upsilon_{\vec k} (U_0^{-d})=
\prod_j \biggl( { \lambda_2^{dj\over 2} - \lambda_2^{-{dj\over 2}}
\over q^{dj\over 2} - q^{-{dj\over 2}} } \biggr)^{k_j},}
where $\lambda_2 = q^{N_2}$.

The vev \conj\ can be written in a very suggestive way by using the
results of \lmv\ on Chern-Simons vevs. In particular,
$f_{\vec k}$ has the structure:
\eqn\strucfk{
f_{\vec k} (q, \lambda_1) =
\prod_j (q^{-{j\over 2}} - q^{j\over 2} )^{k_j}
\sum_{g, Q} n_{\vec k, g, Q}
(q^{-{1\over 2}} - q^{{1\over 2}} )^{2g-2} \lambda_1^{Q}.}
Therefore, one finds for \conj:
\eqn\closed{
\eqalign{
&\log \langle {\cal O}(U_1,U_2;r) \rangle_{A_1, A_2}
=\cr
&\sum_{d=1}^{\infty} \sum_{g,Q} {1 \over d}
\bigl(q^{-{d\over 2}} - q^{d\over2} \bigr)^{2g-2} \sum_{\vec k}
{n_{\vec k, g, Q} \over z_{\vec k}} {\rm e}^{-d\ell r } \prod_j
(  \lambda_2^{-{dj\over 2}} - \lambda_2^{{dj\over 2}})^{k_j}
\lambda_1^{dQ},\cr}}
This has the structure of the free energy for a {\it closed} string \gvtwo:
\eqn\gvmulti{
\sum_{d=1}^{\infty} \sum_{g,m} n^g_{\vec m} {1 \over d} \biggl(
2 \sin {d g_s \over 2} \biggr)^{2g-2} {\rm e}^{-d\vec m \cdot \vec t }, }
provided one finds the appropriate relation between the
closed string K\"ahler parameters $\vec t$, and the K\"ahler parameters
for the open string appearing in \closed.
We will find the precise relation
in the examples below. We also have to show that the expansion in \closed\
involves integers in a manifest way, since in \closed\ we are dividing the
integers $n_{\vec k, g, Q}$ by $z_{\vec k}$. The way to fix that is
to recall that, as shown in \lmv, the ``primitive'' integer invariants are
not $n_{\vec k, g, Q}$, but ${\widehat N}_{R,g,Q}$. They are related by
a linear transformation involving the characters of the symmetric group,
\eqn\relns{
n_{\vec k, g, Q} = \sum_R \chi_R(C(\vec k)) {\widehat N}_{R,g,Q}.}
Using again
the results of \lmv, one can show that
\eqn\rearlam{ \prod_j
(  \lambda^{-{j\over 2}} - \lambda^{{j\over 2}})^{k_j} =
(\lambda^{-{1\over 2}} - \lambda^{1\over 2}) \sum_R \chi_R (C(\vec k)) S_R
(\lambda^{-1}),}
where $S_R (\lambda)$ is the monomial defined in \lmv: if $R$ is not a
hook representation, it is zero, and if $R$ is a hook of $\ell$ boxes
with $\ell-s$ boxes in the first row, then
$$S_R (\lambda)=
(-1)^s \lambda^{-{\ell -1\over 2} +s}$$
Using this, we find
\eqn\manifestint{
\sum_{\vec k}
{n_{\vec k, g, Q} \over z_{\vec k}} \prod_j
(  \lambda_2^{-{dj\over 2}} - \lambda_2^{{dj\over 2}})^{k_j} =
(\lambda_2^{-{d\over 2}} - \lambda_2^{d\over 2}) \sum_R S_R (\lambda_2^{-d})
{\widehat N}_{R,g,Q},}
and this would lead to the identification
$$\sum_{m} n^g_{\vec m} {\rm e}^{-\vec m \cdot \vec t }=
(\lambda_2^{-{1\over 2}} - \lambda_2^{1\over 2})
\sum_{R,Q} e^{-\ell r} S_R (\lambda_2^{-1})\lambda_1^Q
{\widehat N}_{R,g,Q}$$
{}From this expression, together with a suitable linear map between K\"ahler
parameters $\vec t$ and $(r, N_1 g_s,N_2 g_s)$ (which will
lead to only {\it negative} exponents for $\vec t$ and which will
depend on the choice of knot),
the integral structure on both sides is
compatible and one can express the
closed string integral invariants $n_d^g$ in terms of the open string
integral invariants ${\widehat N}_{R,g,Q}$. In the examples
we will encounter in this paper, both knots will be unknots
and the relation will be rather simple.
It is interesting to notice that, when $N=M$, i.e. both gauge groups
coincide, and $r=0$, \finalres\ is the partition function of the
three-manifold obtained after performing surgery on the knot where $U$ is
supported (for finite $k$ and $N$) \jones\kaul\kr.

The above result can be easily generalized to more complicated
situations. For example, one can consider $L$ arbitrary knots, ${\cal
K}_{\alpha}$, where $\alpha=1,\cdots, L$, and suppose that in each of the
components we have a $U(N_{\alpha})$ Chern-Simons theory. Let us
also consider $L$ unknots at zero framing ${\cal K}_{\alpha}$ with a
$U(M_{\alpha})$ Chern-Simons theory in each of them. If we denote by
$U_{\alpha}$, $V_{\alpha}$ the holonomies of the $U(N_{\alpha})$,
$U(M_{\alpha})$ fields, respectively, one can construct the operator
\eqn\linkgen{
{\cal O}(U_1, \cdots, U_{L};V_1, \cdots, V_L) = \exp \biggl[
\sum_{\alpha=1}^L \sum_{n=1}^{\infty} { {\rm e}^{-n r_{\alpha}} \over n}
{\rm Tr} U^n_{\alpha} {\rm Tr}V^{-n}_{\alpha} \biggr].}
Again it can be easily shown that $\log Z(U_1, \cdots, U_{L};V_1, \cdots,
V_L)$ has the structure of the free energy for a closed topological
string.

As an application of the above computation, we can evaluate
\twop\ when $p=0$, corresponding to \oneannulus.
In this case, $U_1$ is
the holonomy around an unknot with trivial framing, and the only nontrivial
$f_R$ corresponds to the fundamental representation. We then find,
$$Z(M_1,M_2) = \exp(-F(M_1)-F(M_2) - F(M_1,M_2;r))$$
where $F(M_i)$ is the free energy of the three-sphere $M_i$,
and
\eqn\closcs{
F(M_1,M_2;r)= \log \langle {\cal O}(U_1,U_2;r) \rangle_{A_1,A_2}=
\sum_{d=1}^{\infty}
\frac{{\rm e}^{-dr'}(1-{\rm e}^{-dt_1})(1-{\rm e}^{-dt_2})}
{d(2 \sin(dg_s/2))^2}.}
Notice that, in writing \closcs, we have defined:
\eqn\redef{
r'= r-{t_1 + t_2 \over 2},}
{\it i.e.} the parameter $r$ that appears in \finalres\ has to be
renormalized in order to match the closed string K\"ahler parameter $r'$.
We will see below other examples of this, in which the same structure
\redef\ appears. This shift was first observed in \dfg.
\smallskip

\ifig\twotransition{After the transition,
the two ${\bf S}^3$'s that we
have wrapped D-branes on disappear, and with them
all of $H_3(X)$, so we are
left with $b_2(X) =3$.}
{\epsfxsize 5.5truein\epsfbox{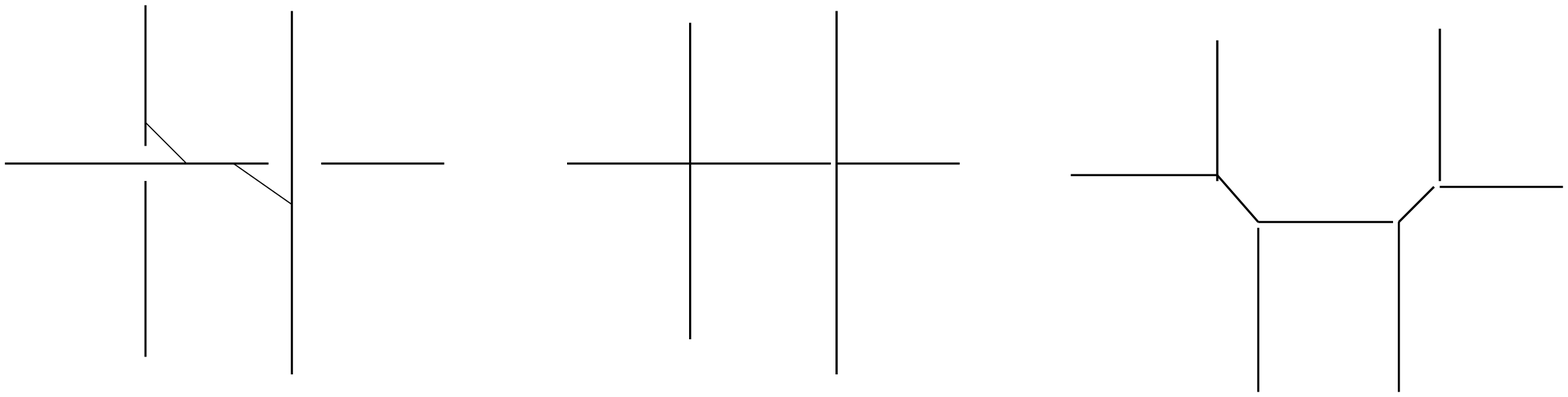}}
The dual closed string geometry is depicted in \twotransition.
The two ${\bf S}^3$'s, whose local neighborhood are $T^{*}{\bf S}^3$'s
are replaced by two $\IP^1$'s with normal bundle ${\cal O}(-1)\oplus
{\cal O}(-1)$. In the total geometry, each of them intersects a $\IP^1$
with ${\cal O}\oplus {\cal O}(-2)$ normal bundle.

It is not difficult to calculate the genus zero amplitude of the closed
string background using mirror symmetry, as in \AVG,
and we find
$$F_{g=0}(r',t_1,t_2) = \sum_{i=1,2}
\sum_{d=1}^{\infty}  \frac{{\rm e}^{-dt_i}}{d^3} +
\sum_{d=1}^{\infty}\frac{{\rm e}^{-dr'}(1-{\rm e}^{-dt_1})
(1-{\rm e}^{-dt_2})}{d^3},$$
This has contribution from a single primitive curve in each of the classes
$[r'],[t_i],[r+t_i],[r+t_1+t_2]$.
In fact, the only primitive curves in this geometry are
rational curves, which are enumerated by the genus zero amplitude.
One easy way to see this is as follows.

Note that under the duality of M-theory on $X$/
type IIB theory on $B\times S^1$ that we discussed in the previous
sections, M2 branes wrapping holomorphic curves in $X$ that have
components along the $T^2$ directions map to $(p,q)$ string web in type IIB
string theory ending on the five-brane web (recall that M2 brane wrapping
$(p,q)$ cycle of the $T^2$ maps to a $(p,q)$ string in type
IIB string theory). The requirement for
supersymmetry is that the $(p,q)$ string must be parallel
to $(p,q)$ five brane, and that stings must be parallel to the plane
defined by the five-branes.
Compact curves in $X$ correspond to string webs ending
on the five-branes, but not any string can
end on any five brane -- a $(p,q)$ string can only end on the
$(p,q)$ five brane. These are conditions for holomorphic curves
in $X$, rephrased in the IIB language (to be complete, there is also
the zero force condition and string string charge conservation
that must be conserved at each
vertex, in addition to the junctions with five-branes).

Stated this way, it is clear from \twotransition\ that there are no
BPS strings of finite length in the IIB dual, and correspondingly, no
curves in $X$ other than the ones counted above.
This, together with the prediction for integrality properties of the
amplitude \gvmulti,
implies that the all genus partition function of the closed string theory
is given by
\eqn\gz{F=\sum_{d=1}^{\infty}\frac{
e^{-d t_1}+e^{-d t_2}+e^{-dr'}(1-e^{-dt_1})(1-e^{-dt_2})}{d(2 \sin(d g_s/2))^2}}
This agrees exactly with the Chern-Simons answer \closcs.
Note that $$t_i = N_i g_s$$
are two new K\"ahler parameters, the sizes of two-spheres that grow with $N$,
replacing the ${\bf S}^3$'s. The K\"ahler
parameter $r'$ was already present in the open string
geometry as the size of the holomorphic annulus with boundaries on the two
spheres, but we have seen that their precise relation is given by \redef.

\subsec{$ {\cal O}(-3) \rightarrow \IP^2$}

The geometric transition in the three-sphere case is similar,
and we have depicted it in \threetrans.
The dual closed string geometry contains a $\IP^2$ with
three exceptional $\IP^1$'s touching it at three points.
We can send the size of these $\IP^1$'s to infinity by
taking $N_i\rightarrow \infty$
in our calculation to recover the
$\IP^2$ amplitude.
The novelty in this case is that the closed string geometry has
curves of arbitrarily high genus
contributing to the topological A-model amplitudes,
and infinitely many integer invariants are non-zero, as we will see.
\smallskip

\ifig\threelinks{The figure shows the Hopf links in
the manifolds $M_1$, $M_2$ and $M_3$ respectively. The numbers
indicate the framing of each knot.}
{\epsfxsize 5.0truein\epsfbox{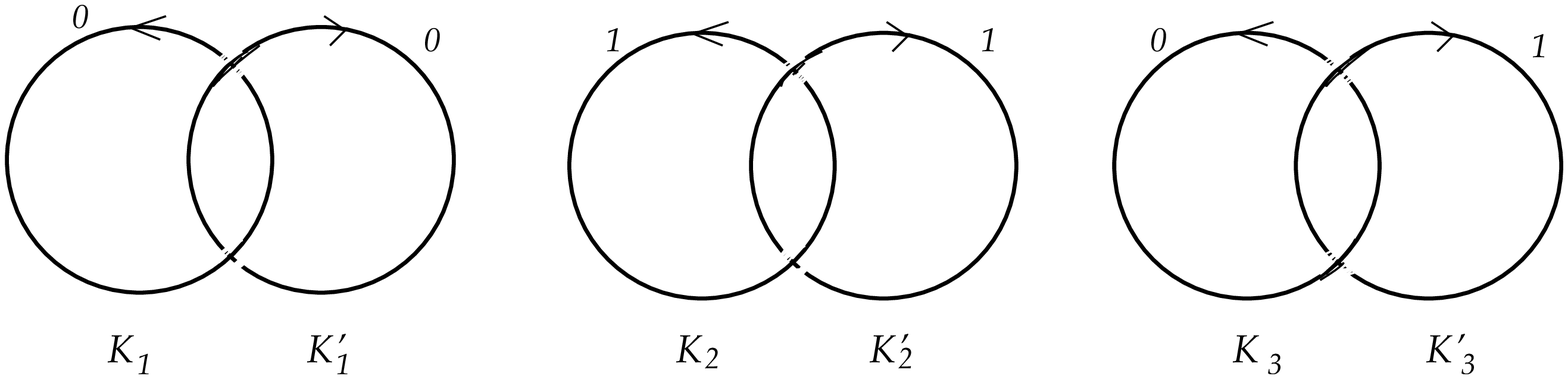}}

We now focus on the Chern-Simons computation that gives
the invariants for $ {\cal O}(-3) \rightarrow \IP^2$. According to what
we discussed before, the Chern-Simons scenario involves
three different gauge groups, with 't Hooft parameters $t_1$, $t_2$, $t_3$,
and with the same coupling constant $g_s$. Accordingly, the quantum
invariants will be a rational function of $q$ and
$\lambda_i ={\rm e}^{t_i}$, $i=1,2,3$. As we saw in
section 5, for each Chern-Simons theory
we have a Hopf link, and we will denote them by
${\cal L}_i$, with components ${\cal K}_i$ and ${\cal
K}_i'$, $i=1,2,3$. The framings can be read from \firstlink, \secondlink\
and \thirdlink, and are as follows: ${\cal K}_1$, ${\cal K}'_1$
and ${\cal K}_3$ have framing zero, while the remaining knots have framing
$p=1$. This means that ${\cal L}_1$ is in the canonical framing, in ${\cal
L}_2$ both components are framed, while in ${\cal L}_3$ only one of the
components, ${\cal K}'_3$, is framed.
The free energy at all genus for the topological closed string is given by
\eqn\ptwof{
F=\log \biggr\{ \sum_{R_1,R_2, R_3} {\rm e}^{-\sum_{i=1}^3 \ell_i r_i }
W_{R_1, R_2} ({\cal L}_1)  W_{R_2, R_3} ({\cal L}_2)
W_{R_3, R_1} ({\cal L}_3) \biggl\}. }
 The sum in \ptwof\ is over
all possible representations,
including the trivial one (and that will be denoted by $\cdot$).
In this equation,
$r_i$ are ``bare'' K\"ahler
parameters that will lead to a ``renormalized'' K\"ahler parameter $r$ for the
K\"ahler class of $ {\cal O}(-3) \rightarrow \IP^2$. The relation between
$r_i$ and $r$ can be obtained by requiring a consistent limit  $t_i
\rightarrow \infty$ (which corresponds to the local $\IP^2$ limit of the
original, more complicated geometry). We will discuss this relation in a
moment. Of course, there is also a
piece $F(M_{1})+ F(M_{2})+ F(M_{3})$ given by
the sum of the free energies of the spheres that should be added to
\ptwof.

Once the ``renormalized'' parameter has been restored, and the
limit $t_i \rightarrow \infty$ taken, the free energy has the
structure:
\eqn\logf{
F=\log \biggr\{ 1 + \sum_{\ell=1}^{\infty} a_{\ell} (q) {\rm e}^{-\ell
r}\biggr\}= \sum_{\ell=1}^\infty a_{\ell}^{(c)}(q) {\rm e}^{-\ell r},}
and we will refer to $a_{\ell}^{(c)} (q)$ as to the connected
coefficients. If we keep the $t_i$ finite, we find the
free energy for a closed string propagating in $\IP^2$ blown
up at 3 points, which is known as $\IB_3$ (a particular case
of del Pezzo).
The $1$ inside the brackets in \logf\
corresponds to $R_1=R_2=R_3=\cdot$, {\it i.e.}
all the representations being the trivial one. In order to
 extract the integral invariants from this expression, we have
to recall the general structure of the closed string topological
amplitudes \gvmulti .  One can use this formula
to write the integral invariants in terms of closed
string amplitudes
by using the M\"obius function $\mu (n)$ \bp.
Recall that $\mu
(n)=0$ if $n$ is not square-free, and it is $(-1)^f$ otherwise, where $f$
is the number of factors in the prime decomposition of $n$. One finds:
\eqn\finalgv{
\sum_{g\ge 0} (-1)^{g-1}n_d^g (q^{-{1\over2}} -q^{1\over2})^{2g-2}
=\sum_{k|d} {\mu (k) \over k} a_{d/k}^{(c)}(q^k).}
Therefore, if we know the coefficients $a_{\ell}^{(c)}$ for $\ell=1,
\cdots, d$, we can extract the integral invariants $n_{\ell}^g$ for
degrees $\ell=1, \cdots, d$, and {\it for all genera}. Of course, from the
point of view of Chern-Simons theory, it is highly nontrivial that the
coefficients $a_{\ell}^{(c)}$ extracted from \ptwof\ have the structure
required by \finalgv. In the examples discussed in the previous section,
the properties
of the Chern-Simons invariants derived in \ov\lmv\ guaranteed that one
obtained a closed string expansion at all degrees and genera. Here we do not
have a general proof, but we will explicitly show at low degrees that
again, the rather constraining structural properties of the invariants of
knots and links guarantee that one finds the right structure \finalgv.

Let us now look at the expansion of \logf\ at low degrees, which will also
fix the relation between $r_i$ and $r$. First notice that, at every given
degree $d$, we have a combinatorial problem of finding all possible
representations $R_1, R_2, R_3$ such that $\ell_1 + \ell_2 + \ell_3 =d$.
At degree one, we have three possibilities ($\tableau{1} \cdot \cdot$
and two permutations), and we find:
\eqn\firstor{
{\rm e}^{-r} a_1 =
{\rm e}^{-r_1}W_{\tableau{1}}({\cal K}_1) W_{\tableau{1}}({\cal K}_3') +
{\rm e}^{-r_2} W_{\tableau{1}}({\cal K}_2)W_{\tableau{1}}({\cal K}_1') +
{\rm e}^{-r_3}W_{\tableau{1}}({\cal K}_3) W_{\tableau{1}}({\cal K}_2').}
The knots involved in this computation are just framed unknots.
Since the invariant of an unknot with framing $p$ in the fundamental
representation is
\eqn\framedun{
(-1)^p {{\lambda^{-{1\over2}} -\lambda^{1\over2}}\over  q^{-{1\over2}} -q^{1\over2}}}
we see that in order to have a finite limit as
$t_i \rightarrow \infty$, we must have:
\eqn\renorm{
r=r_1 -{ {t_1 + t_3} \over 2}=r_2 -{ {t_1 + t_2} \over 2}=
r_3 -{ {t_2 + t_3} \over 2},}
so we take the limit $t_i \rightarrow \infty$ and at the same time
$r_i \rightarrow \infty$ in such a way that the above combinations
remain finite and equal to the closed string K\"ahler parameter $r$. The
relation \renorm\ is identical to the one we found before, in \redef. Notice
that, from the point of view of the Chern-Simons computation, this means
that we have to renormalize every vev
in the representations $R$, $R'$ as follows,
\eqn\normfactor{
W_{R, R'}({\cal L}_i) \rightarrow
\lambda_i^{-{{\ell + \ell'} \over 2}}W_{R, R'}({\cal L}_i),
}
where $\ell, \ell'$ are the number of boxes in the representations $R$,
$R'$. We have assumed that the renormalized K\"ahler parameters
corresponding to the different annuli are all equal to the single K\"ahler
parameter of the local $\IP^2$ geometry. As we will see later, one can
take the renormalized parameters to be different to obtain a refined
version of the invariants.

The conclusion of the above analysis is that,
in order to recover the local $\IP^2$
geometry, we have to take the limit $\lambda_i \rightarrow \infty$ after the
renormalization factor has been introduced. We then find that, for the
local $\IP^2$ geometry,
\eqn\degreeone{
a_1(q)= - { 3 \over (q^{-{1\over2}} -q^{1\over2})^2},}
which gives immediately the integral invariants at degree one:
\eqn\firstinv{
n_1^0 = 3, \,\,\,\,\,\,\,\,\,\,\,\,\,\,\,\,\,\,\,  n_1^g =0 \,\,\, {\rm for}\, g>0,}
indeed the right result \kz\kkv.

Let us do the computation at degree two. There are nine possible
choices of representations that lead to this degree:
$\tableau{2} \cdot \cdot$, $\tableau{1 1} \cdot \cdot$,
$\tableau{1} \tableau{1} \cdot$, and their permutations. This
gives
\eqn\degreetwo{
\eqalign{
a_2 &= \lambda_1^{-1}(\lambda_2\lambda_3)^{-{1\over 2}}
W_{(\tableau{1}, \tableau{1})}({\cal L}_1) W_{\tableau{1}}({\cal K}_3')
W_{\tableau{1}}({\cal K}_2) \cr &+ (\lambda_1\lambda_3)^{-1}
 \bigl( W_{\tableau{2}}({\cal K}_1)W_{\tableau{2}}({\cal
K}_3')
 +
W_{\tableau{1 1}}({\cal K}_1)W_{\tableau{1 1}}({\cal K}_3') \bigr)
+ {\rm perms},\cr}
}
where the permutations act cyclically as follows: ${\cal L}_1 \rightarrow
{\cal L}_2 \rightarrow {\cal L}_3$, ${\cal K}_1 \rightarrow
{\cal K}_2 \rightarrow {\cal K}_3$, ${\cal K}'_1 \rightarrow
{\cal K}'_2 \rightarrow {\cal K}'_3$. The connected coefficient
can be easily computed to be,
\eqn\degtwocon{
\eqalign{
 a_2^{(c)} &=\lambda_1^{-1}(\lambda_2\lambda_3)^{-{1\over 2}}
f_{(\tableau{1}, \tableau{1})}({\cal L}_1) W_{\tableau{1}}({\cal K}_3')
W_{\tableau{1}}({\cal K}_2) + {\rm perms} \cr
&+ \sum_{R=\tableau{2}, \tableau{1 1}} \bigl( (\lambda_2\lambda_3)^{-1}
f_R({\cal K}'_2 )W_R({\cal K}_3) +(\lambda_1\lambda_3)^{-1}
 f_R({\cal K}'_3 )W_R({\cal K}_1)\cr &+ (\lambda_1\lambda_2)^{-1}
f_R({\cal K}_2 )W_R({\cal K}'_1) \bigr)
\cr &+
{1 \over 2} (\lambda_1\lambda_3)^{-1} W^{(2)}_{\tableau{1}}({\cal K}_3')
W^{(2)}_{\tableau{1}}({\cal K}_2) + {\rm perms}. \cr}}
We have denoted $W^{(2)}_{R}(q, \lambda)=W_R(q^2, \lambda^2)$. The
invariant $f_{(R_1, R_2)}({\cal L})$ was introduced in \lmv, and we
recall that for $R_1 =R_2 =\tableau{1}$, one has
\eqn\foneone{
f_{(\tableau{1}, \tableau{1})}({\cal L}) =
W_{(\tableau{1}, \tableau{1})}({\cal L}) - W_{\tableau{1}}({\cal K}_1)
W_{\tableau{1}}({\cal K}_2) ,}
where ${\cal K}_{1,2}$ are the components of ${\cal L}$. One can also
prove \lmv\ that $f_{(\tableau{1}, \tableau{1})}({\cal L})$ has the
structure:
\eqn\fstr{
f_{(\tableau{1}, \tableau{1})}({\cal L}) =\sum_{g,Q}
{\widehat N}_{(\tableau{1}, \tableau{1}), g, Q} (q^{-{1 \over 2}} - q^{1
\over 2})^{2g}\lambda^Q,}
so it is an even polynomial in $q^{-{1 \over 2}} - q^{1
\over 2}$. We can now show that the structure of \degtwocon\ is compatible
with \finalgv: the first three lines give
a Laurent polynomial in $q^{-{1 \over 2}} - q^{1
\over 2}$ with even powers $\ge -2$. For the first line, this is guaranteed
by \fstr, while for the second and third lines we just notice that they
have the
structure of \kexp\ (with $d=1$), therefore we can use \strucfk\rearlam\
to write it in the desired form. In fact, the last three lines of \degtwocon\
are precisely what we would obtain if the
knots were all unlinked, and therefore the arguments of the
last section guarantee that they have the right structure. Indeed,
the very last line gives very
precisely the two-cover of the degree one contribution,
in agreement with \finalgv.

In the above computation we have considered the most general geometry
with four K\"ahler classes. In order to recover the local $\IP^2$ case,
we have to take $t_i \rightarrow \infty$. It is easy to see that in this
limit the only relevant integral invariants of the links are
\eqn\nlinks{
{\widehat N}_{(\tableau{1}, \tableau{1}), g=0, Q=1}({\cal L}_1)=
{\widehat N}_{(\tableau{1}, \tableau{1}), g=0, Q=1}({\cal L}_2)=
-{\widehat N}_{(\tableau{1}, \tableau{1}), g=0, Q=1}({\cal L}_3)=1.}
These invariants can be computed from \foneone\ and \homfly, after
including the framing corrections\foot{In fact, many of the Chern-Simons
invariants of the Hopf link with trivial framing can be read
from section 6.2 of \lmv, after changing $\lambda, q \rightarrow
\lambda^{-1},
q^{-1}$ due to the fact that the Hopf link considered there has
the opposite linking number to the one considered here.},
while for the framed unknots we only need, in this limit, the invariant \mv\
\eqn\framedn{
{\widehat N}_{\tableau{1 1}, g=0, Q=1}(p=1)=1.}
The relevant integral invariants with $g>0$ all vanish. We then find,
\eqn\connac{
a_2^{(c)}(q)=
{ 6 \over (q^{-{1\over2}} -q^{1\over2})^2} + {1 \over 2}a_1 (q^2),}
therefore:
\eqn\secondinv{
n_2^0 = -6, \,\,\,\,\,\,\,\,\,\,\,\,\,\,\,\,\,\,\,  n_2^g =0 \,\,\, {\rm for}\, g>0,}
again in agreement with the A and B-model computations \kz\kkv.

The procedure is now clear: to any given degree $d$, one has to
compute the coefficient $a_d$ as a sum of different contributions given by
the combinatorics of Young tableaux, compute the connected piece, and
finally extract the multicovering contributions. Notice that, if one is just
interested in the local $\IP^2$ results,  one can take
the limit $t_i \rightarrow \infty$ at the beginning
of the computation. In this limit we only have to keep the leading term in
$\lambda$ in the Chern-Simons invariant of the Hopf link presented in
section 5.1. In this way we have a very powerful method to compute the
integral invariants of the local $\IP^2$ geometry that can be easily implemented in
a symbolic manipulation program. The results, up to degree 12 and
at all genera, are presented in the following tables:
\bigskip

{\vbox{\ninepoint{
$$
\vbox{\offinterlineskip\tabskip=0pt
\halign{\strut
\vrule#
&
&\hfil ~$#$
&\hfil ~$#$
&\hfil ~$#$
&\hfil ~$#$
&\hfil ~$#$
&\hfil ~$#$
&\hfil ~$#$
&\hfil ~$#$
&\hfil ~$#$
&\hfil ~$#$
&\vrule#\cr
\noalign{\hrule}
&g
&d=1
&2
&3
&4
&5
&6
&7
&8
&9
&
\cr
\noalign{\hrule}
&0
&3
&-6
&27
&-192
&1695
&-17064
&188454
&-2228160
&27748899
&
\cr
\noalign{\hrule}
&1
&0
&0
&-10
&231
&-4452
&80948
&-1438086
&25301295
&-443384578
&
\cr
\noalign{\hrule}
&2
&0
&0
&0
&-102
&5430
&-194022
&5784837
&-155322234
&3894455457
&
\cr
\noalign{\hrule}
&3
&0
&0
&0
&15
&-3672
&290853
&-15363990
&649358826
&-23769907110
&
\cr
\noalign{\hrule}
&4
&0
&0
&0
&0
&1386
&-290400
&29056614
&-2003386626
&109496290149
&
\cr
\noalign{\hrule}
&5
&0
&0
&0
&0
&-270
&196857
&-40492272
&4741754985
&-396521732268
&
\cr
\noalign{\hrule}
&6
&0
&0
&0
&0
&21
&-90390
&42297741
&-8802201084
&1156156082181
&
\cr
\noalign{\hrule}
&7
&0
&0
&0
&0
&0
&27538
&-33388020
&12991744968
&-2756768768616
&
\cr
\noalign{\hrule}
&8
&0
&0
&0
&0
&0
&-5310
&19956294
&-15382690248
&5434042220973
&
\cr
\noalign{\hrule}
&9
&0
&0
&0
&0
&0
&585
&-9001908
&14696175789
&-8925467876838
&
\cr
\noalign{\hrule}
&10
&0
&0
&0
&0
&0
&-28
&3035271
&-11368277886
&12289618988434
&\cr
\noalign{\hrule}
&11
&0
&0
&0
&0
&0
&0
&-751218
&7130565654
&-14251504205448
&
\cr
\noalign{\hrule}
&12
&0
&0
&0
&0
&0
&0
&132201
&-3624105918
&13968129299517
&\cr
\noalign{\hrule}
&13
&0
&0
&0
&0
&0
&0
&-15636
&1487970738
&-11600960414160
&
\cr
\noalign{\hrule}
&14
&0
&0
&0
&0
&0
&0
&1113
&-490564242
&8178041540439
&
\cr
\noalign{\hrule}
&15
&0
&0
&0
&0
&0
&0
&-36
&128595720
&-4896802729542
&\cr
\noalign{\hrule}
&16
&0
&0
&0
&0
&0
&0
&0
&-26398788
&2489687953666
&\cr
\noalign{\hrule}
&17
&0
&0
&0
&0
&0
&0
&0
&4146627
&-1073258752968
&\cr
\noalign{\hrule}
&18
&0
&0
&0
&0
&0
&0
&0
&-480636
&391168899747
&\cr
\noalign{\hrule}
&19
&0
&0
&0
&0
&0
&0
&0
&38703
&-120003463932
&\cr
\noalign{\hrule}
&20
&0
&0
&0
&0
&0
&0
&0
&-1932
&30788199027
&\cr
\noalign{\hrule}
&21
&0
&0
&0
&0
&0
&0
&0
&45
&-6546191256
&\cr
\noalign{\hrule}
&22
&0
&0
&0
&0
&0
&0
&0
&0
&1138978170
&\cr
\noalign{\hrule}
&23
&0
&0
&0
&0
&0
&0
&0
&0
&-159318126
&\cr
\noalign{\hrule}
&24
&0
&0
&0
&0
&0
&0
&0
&0
&17465232
&\cr
\noalign{\hrule}
&25
&0
&0
&0
&0
&0
&0
&0
&0
&-1444132
&\cr
\noalign{\hrule}
&26
&0
&0
&0
&0
&0
&0
&0
&0
&84636
&\cr
\noalign{\hrule}
&27
&0
&0
&0
&0
&0
&0
&0
&0
&-3132
&\cr
\noalign{\hrule}
&28
&0
&0
&0
&0
&0
&0
&0
&0
&55
&
\cr
\noalign{\hrule}
}\hrule}$$}
\vskip - 7 mm
\centerline{{\bf Table 1:} The integral invariants $n_d^g$ for the local
$\IP^2$ case.}
\vskip7pt}
\noindent
\smallskip

{\vbox{\ninepoint{
$$
\vbox{\offinterlineskip\tabskip=0pt
\halign{\strut
\vrule#
&
&\hfil ~$#$
&\hfil ~$#$
&\hfil ~$#$
&\hfil ~$#$
&\vrule#\cr
\noalign{\hrule}
&g
&d=10
&11
&12
&
\cr
\noalign{\hrule}
&0
&-360012150
&4827935937
&-66537713520
&
\cr
\noalign{\hrule}
&1
&7760515332
&-135854179422
&2380305803719
&
\cr
\noalign{\hrule}
&2
&-93050366010
&2145146041119
&-48109281322212
&
\cr
\noalign{\hrule}
&3
&786400843911
&-24130293606924
&698473748830878
&
\cr
\noalign{\hrule}
&4
&-5094944994204
&210503102300868
&-7935125096754762
&
\cr
\noalign{\hrule}
&5
&26383404443193
&-1485630816648252
&73613315148586317
&
\cr
\noalign{\hrule}
&6
&-111935744536416
&8698748079113310
&-572001241783007370
&
\cr
\noalign{\hrule}
&7
&395499033672279
&-42968546119317066
&3786284014554551293
&
\cr
\noalign{\hrule}
&8
&-1177301126712306
&181202644392392127
&-21609631514881755756
&
\cr
\noalign{\hrule}
&9
&2978210177817558
&-658244675887405242
&107311593188998164015
&
\cr
\noalign{\hrule}
&10
&-6445913624274390
&2074294284130247058
&-466990545532708577390
&\cr
\noalign{\hrule}
&11
&12001782164043306
&-5702866358492557440
&1791208287019324701495
&
\cr
\noalign{\hrule}
&12
&-19310842755095748
&13744538465609779287
&-6085017394087513680618
&\cr
\noalign{\hrule}
&13
&26952467292328782
&-29157942375100015002
&18384612378910358924791
&
\cr
\noalign{\hrule}
&14
&-32736035592797946
&54641056077839878893
&-49578782776769125835658
&
\cr
\noalign{\hrule}
&15
&34693175820656421
&-90735478019244786786
&119723947998685791289164
&\cr
\noalign{\hrule}
&16
&-32151370513161966
&133885726253316075984
&-259634731498425150837576
&\cr
\noalign{\hrule}
&17
&26099440805196660
&-175976406401479949154
&506961721474582218552270
&\cr
\noalign{\hrule}
&18
&-18580932613650624
&206477591201198965488
&-893407075206205808615238
&\cr
\noalign{\hrule}
&19
&11609627766170547
&-216671841840838260606
&1424048002136300951108030
&\cr
\noalign{\hrule}
&20
&-6367395873587820
&203674311322868998065
&-2057099617415644933602618
&\cr
\noalign{\hrule}
&21
&3064262549419899
&-171730940091766865658
&2697839037217627321703085
&\cr
\noalign{\hrule}
&22
&-1292593922494452
&130015073789764141299
&-3217397468483821476968358
&\cr
\noalign{\hrule}
&23
&477101143946277
&-88451172530198637924
&3494176460021369389735746
&\cr
\noalign{\hrule}
&24
&-153692555590206
&54098277648908454123
&-3460084190968494003073062
&\cr
\noalign{\hrule}
&25
&43057471189239
&-29751302949160261398
&3127576636374963802648718
&\cr
\noalign{\hrule}
&26
&-10441089412308
&14709694749741501501
&-2582938330708242629937150
&\cr
\noalign{\hrule}
&27
&2177999212647
&-6535189635435373326
&1950461493734929553600580
&\cr
\noalign{\hrule}
&28
&-387688567518
&2606677300588276035
&-1347524558332336039964082
&
\cr
\noalign{\hrule}
&29
&58269383541
&-932238829973577348
&852109374825775079556606
&
\cr
\noalign{\hrule}
&30
&-7292193288
&298408032566091294
&-493309207337589509893062
&
\cr
\noalign{\hrule}
&31
&745600245
&-85297647759486510
&261477149328500781917776
&
\cr
\noalign{\hrule}
&32
&-60650490
&21708810999461607
&-126876156355185161374314
&
\cr
\noalign{\hrule}
&33
&3773652
&-4901354114590566
&56339101711825399890960
&
\cr
\noalign{\hrule}
}\hrule}$$}
\vskip - 7 mm
\centerline{{\bf Table 2:} The integral invariants $n_d^g$ for the local
$\IP^2$ case (continuation).}
\vskip7pt}

\noindent
\smallskip
\bigskip{\vbox{\ninepoint{
$$
\vbox{\offinterlineskip\tabskip=0pt
\halign{\strut
\vrule#
&
&\hfil ~$#$
&\hfil ~$#$
&\hfil ~$#$
&\hfil ~$#$
&\vrule#\cr
\noalign{\hrule}
&g
&d=10
&11
&12
&
\cr
\noalign{\hrule}
&34
&-168606
&977233475777499
&-22881258328195868502320
&
\cr
\noalign{\hrule}
&35
&4815
&171090302865948
&8492649924309368930964
&
\cr
\noalign{\hrule}
&36
&-66
&26117674453665
&-2877665040430021956492
&
\cr
\noalign{\hrule}
&37
&0
&-3445690553358
&888968505074075552261
&
\cr
\noalign{\hrule}
&38
&0
&388460380746
&-249952226921825722236
&
\cr
\noalign{\hrule}
&39
&0
&-36878620320
&63836429603183934921
&
\cr
\noalign{\hrule}
&40
&0
&2891025822
&-14772524364719546808
&
\cr
\noalign{\hrule}
&41
&0
&-182125500
&3088415413809592461
&
\cr
\noalign{\hrule}
&42
&0
&8859513
&-581271967556317272
&
\cr
\noalign{\hrule}
&43
&0
&-312270
&98073062075574517
&
\cr
\noalign{\hrule}
&44
&0
&7095
&-14758388168491098
&\cr
\noalign{\hrule}
&45
&0
&-78
&1968679573589997
&
\cr
\noalign{\hrule}
&46
&0
&0
&-231043750764510
&\cr
\noalign{\hrule}
&47
&0
&0
&23635158339861
&\cr
\noalign{\hrule}
&48
&0
&0
&-2082988758060
&\cr
\noalign{\hrule}
&49
&0
&0
&155790863415
&\cr
\noalign{\hrule}
&50
&0
&0
&-9693024822
&\cr
\noalign{\hrule}
&51
&0
&0
&488072208
&\cr
\noalign{\hrule}
&52
&0
&0
&-19105426
&\cr
\noalign{\hrule}
&53
&0
&0
&545391
&\cr
\noalign{\hrule}
&54
&0
&0
&-10098
&\cr
\noalign{\hrule}
&55
&0
&0
&91
&\cr
\noalign{\hrule}
}\hrule}$$}
\vskip - 7 mm
\centerline{{\bf Table 3:} The integral invariants $n_d^g$ for the local
$\IP^2$ case (continuation).}
\vskip7pt}
\noindent
\smallskip
\bigskip
It is interesting to compare this procedure to obtain the integer invariants
with the ones based in the A and the B-model. As in the A-model
computations based on localization, our procedure has to proceed degree by
degree, and as the degree is increased the number of terms that contribute
to $a_d$ grows very rapidly: to evaluate the integer invariants
up to degree $12$, one has to find $a_1, \cdots, a_{12}$, and this
involves evaluating $18239$ terms in total. Degree $20$ involves $943304$
terms (there are
$341649$ terms contributing just to $a_{20}$).
However, the number of terms seems to be
substantially lower than in a localization computation (compare for
example with \kz), and of course
the crucial advantage of the Chern-Simons approach is that one gets
the invariants for all genera. This is also its main advantage
with respect to the B-model computations, which also become more and more
difficult as the genus is increased. The B-model results for higher genera
are in fact determined only up to some unknown constants, due to the
holomorphic ambiguity \BCOV , and in order to find the actual value of
the invariants one has to provide the value of the integral invariants coming
from A-model computations. Therefore, the computation via Chern-Simons provides
another way of fixing the holomorphic ambiguity of the B-model.

Some comments on the results listed in Tables 1-3 are in order. First observe
that, for a given degree $d$, $n_d^g$ vanishes for $g> (d-1)(d-2)/2$.
Indeed, $(d-1)(d-2)/2$ is the genus of a
nondegenerate curve of degree $d$ in $\IP^2$. As shown in \kkv, one has in
this case
\eqn\bpsone{
n_d^{(d-1)(d-2)/2}={(-1)^{d(d+3)/2} \over 2} (d+1)(d+2),}
in full agreement with the corresponding entries in Table 1 for $d=1,
\cdots, 12$. For $d>2$, we have contributions from curves
with one node (therefore $g=d(d-3)/2$), and the arguments of \kkv\ give
\eqn\bpstwo{
n_d^{d(d-3)/2}=-(-1)^{d(d+3)/2} {d \choose 2} (d^2 + d-3),}
again in full agreement with the results that we have obtained. Curves with
two nodes start contributing at $d>3$, and one finds:
\eqn\bpsthree{
n_d^{(d^2-3d -2)/2}={(-1)^{d(d+3)/2}\over 4}(d-1)
(d^5 -2 d^4 - 6d^3 + 9d^2 +36),}
which reproduces our results for $4\le d\le 12$.
For curves with three nodes, the integral invariant is given by
\eqn\bpsfour{
n_d^{(d^2-3d-4)/2}=-{(-1)^{d(d+3)/2}\over 12}
(-96 + 222\,d - 323\,d^2 + 54\,d^3 -
34\,d^4 + 36\,d^5 + 2\,d^6 - 6\,d^7 + d^8),}
which reproduces our results for $5\le d\le 12$.
For $d=4$ there are reducible curves with three nodes,
and in order to reproduce $n_4^0$ one has to introduce a correction, as
explained in detail in \kkv. We then see that the results
obtained from Chern-Simons theory are in full agreement
with what is expected from the geometric interpretation of the
integral invariants. Notice that we have been able to check results
for very high genus, which is not easy to do in the A or B model computations.

\subsec{$\IB_3$}

In the previous subsection we have seen how to recover the
integer invariants for $\IP^2$ by taking the limit $t_i \rightarrow
\infty$. Keeping the blow up parameters $t_i$ finite we obtain the integer
invariants of the local del Pezzo $\IB_3$ (we remind that
$\IB_3$ is the rational surface obtained from
$\IP^2$ by blowing up three points). We will write the generating functional
for the integer invariants at genus $g$ as
\eqn\gendp{
{\cal F}_g (r, t_1, t_2, t_3)=\sum_\ell {\rm e}^{-\ell r}{\cal F}_\ell^g
(t_1, t_2, t_3)}
where
\eqn\gendpp{
{\cal F}_d^g(t_1, t_2, t_3)=\sum_{d_1, d_2 d_3}
n^g_{d_1, d_2, d_3} q_1^{d_1} q_2^{d_2} q_3^{d_3},}
and we have written $q_i={\rm e}^{-t_i}$ (these shouldn't be confused
with the Chern-Simons variable introduced before).
We present the results for these generating functionals
up to degree four in $\IP^2$:
\eqn\resdp{
\eqalign{
{\cal F}_1^0=& 3 -2(q_1 + q_2 + q_3)+ q_1q_2 + q_1 q_3 + q_2 q_3,\cr
{\cal F}_2^0=&-6 +5(q_1 + q_2 + q_3)-4( q_1q_2 + q_1 q_3 + q_2 q_3)
+ 3\, q_1 q_2 q_3, \cr
{\cal F}_3^0=&27 -32(q_1 + q_2 + q_3)+35( q_1q_2 + q_1 q_3 + q_2 q_3)
+7(q_1^2 + q_2^2 + q_3^2) \cr
&-6 (q_1 q_2^2 + q_1 q_3^2 + q_2 q_1^2 +
q_2 q_3^2+q_3 q_1^4 + q_3 q_2^2)-36\,  q_1 q_2 q_3 \cr &
+ 5(q_1 q_2^2 q_3^2 + q_2 q_1^2 q_3^2 +q_3 q_1^2 q_2^2),\cr
{\cal F}_3^1=&-10 + 9(q_1 + q_2 + q_3) -8(q_1q_2 + q_1 q_3 + q_2 q_3)
+ 7 \, q_1 q_2 q_3,\cr
{\cal F}_4^0=&-192 + 286(q_1 + q_2 + q_3)-400(q_1q_2 + q_1 q_3 + q_2 q_3)
-110  (q_1^2 + q_2^2 + q_3^2)\cr
& +135 (q_1 q_2^2 + q_1 q_3^2 + q_2 q_1^2 +
q_2 q_3^2+q_3 q_1^4 + q_3 q_2^2) + 531 \, q_1 q_2 q_3
-9(q_1^3 + q_2^3 + q_3^2) \cr
& -8 (q_1 q_2^3 + q_1 q_3^3 + q_2 q_1^3 +
q_2 q_3^3+q_3 q_1^3 + q_3 q_2^3)-32 (q_1^2q_2^2 + q_1^2 q_3^2 + q_2^2
q_3^2)
\cr
&-160(q_1 q_2 q_3^2 +q_1 q_3 q_2^2 + q_2 q_3 q_1^2)
+ 35(q_1 q_2^2 q_3^2 +q_2 q_3^3 q_2^2 + q_3 q_1^2 q_2^2)\cr&
+ 7( q_1 q_2 q_3^3 +q_1 q_3 q_2^3 + q_2 q_3 q_1^3) -6\, q_1^2 q_2^2
q_3^2,\cr}}
and finally,
\eqn\findp{
\eqalign{
{\cal F}_4^1=&231 -  288(q_1 + q_2 + q_3)+ 344(q_1q_2 + q_1 q_3 + q_2 q_3)
+68  (q_1^2 + q_2^2 + q_3^2)\cr
& -72(q_1 q_2^2 + q_1 q_3^2 + q_2 q_1^2 +
q_2 q_3^2+q_3 q_1^4 + q_3 q_2^2) + 396 \, q_1 q_2 q_3 \cr
& +9  (q_1^2q_2^2 + q_1^2 q_3^2 + q_2^2
q_3^2)
+74(q_1 q_2 q_3^2 +q_1 q_3 q_2^2 + q_2 q_3 q_1^2) \cr &
-8(q_1 q_2^2 q_3^2 +q_2 q_3^3 q_2^2 + q_3 q_1^2 q_2^2),
\cr
{\cal F}_4^2=&-102 +108(q_1 + q_2 + q_3)-112( q_1q_2 + q_1 q_3 + q_2 q_3)
-12(q_1^2 + q_2^2 + q_3^2) \cr
&+11 (q_1 q_2^2 + q_1 q_3^2 + q_2 q_1^2 +
q_2 q_3^2+q_3 q_1^4 + q_3 q_2^2) + 114\, q_1 q_2 q_3 \cr
& -10(q_1 q_2 q_3^2 +q_1 q_3 q_2^2 + q_2 q_3 q_1^2) \cr
{\cal F}_4^3=&15 - 14(q_1 + q_2 + q_3) +13(q_1q_2 + q_1 q_3 + q_2 q_3)
-12 \, q_1 q_2 q_3.\cr}}
We can take the limit in which one of the $q_i$'s, say $q_3$, goes to
zero. The corresponding results for the $g=0$ amplitudes agree with
those presented in \ckyz\ for the $\IB_2$ local del Pezzo, after
relabeling $t_{1,2} \rightarrow -t_{1,2}$, $r\rightarrow r+t_1 + t_2$.

\ifig\fourspheres{The figure depicts four ${\bf S}^3$'s connected with annuli.}
{\epsfxsize 4.0truein\epsfbox{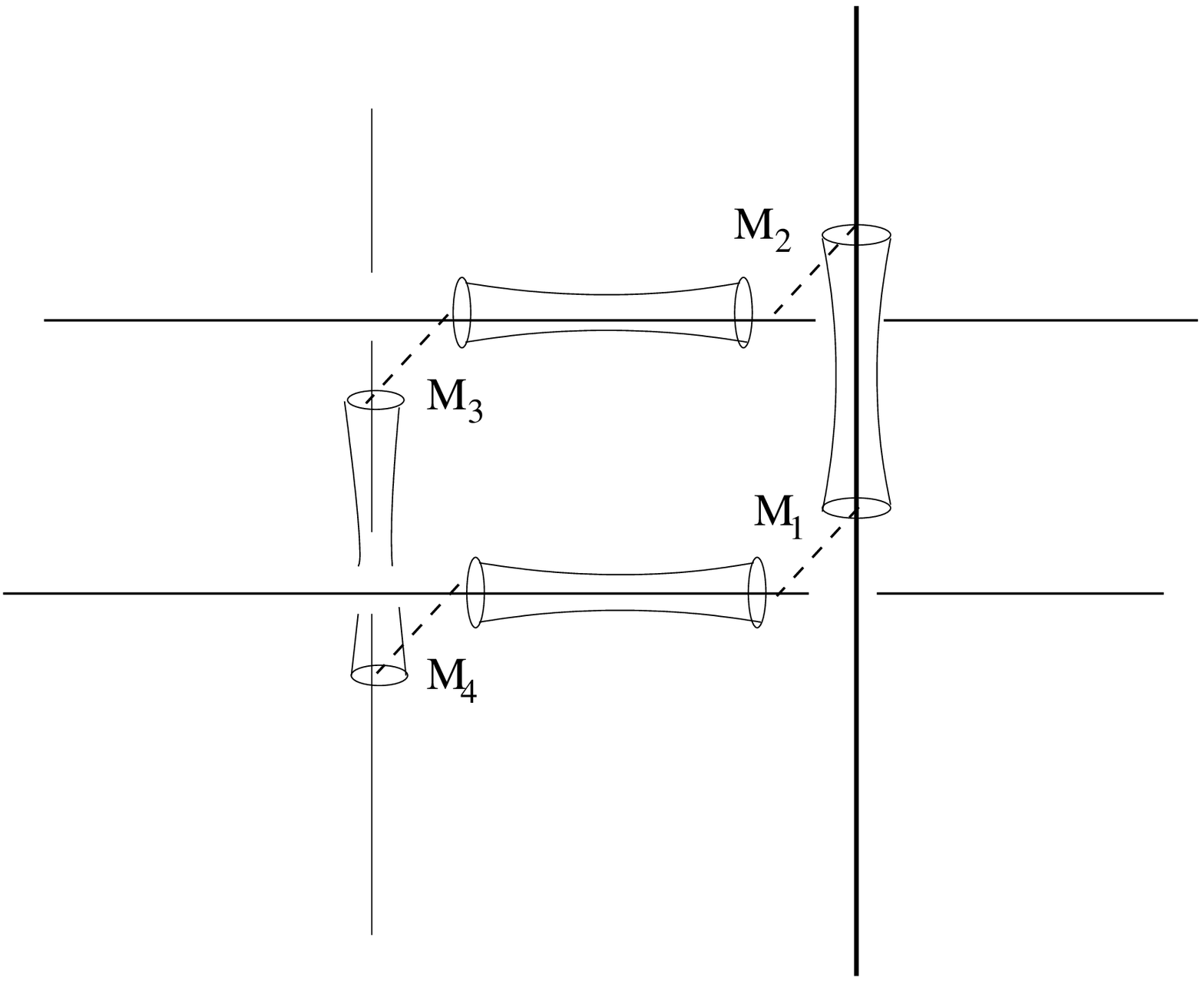}}

\subsec{$ {\cal O}(K) \rightarrow \IP^1 \times \IP^1$}

We now consider the geometry that leads to local $\IP^1 \times \IP^1$.
In \fourspheres\ there are $N_i$
D-branes, $i=1,\ldots,4$, wrapping a chain of four minimal spheres connecting two $(1,0)$
branes and two $(0,1)$ branes. For every pair of spheres ``intersecting''
over an ${\bf S}^1$ we get a bifundamental scalar field, so we have matter in
representation $(N_i,{\overline N}_{i+1})$, where $i=5$ corresponds to the first
sphere again.
The path integral of the A-model in this background
can be written as:
\eqn\fourp{Z=\int \prod_{i=1}^4{\cal D}A_i e^{S_{\rm CS}(A_i)}
{\cal O}(U_1,U_2){\cal O}(U_2,U_3){\cal O}(U_3,U_4){\cal O}(U_4,U_1)}

There are two unknots on
each three-sphere and the amplitude will depend
on their linking numbers, in addition to framing.
As before, we can use \transp\ to write this in a more transparent form
\eqn\fourg{\eqalign{Z= \sum_{R_1,R_2,R_3,R_4}
&\langle V_1R_1|V_{4}R_4\rangle {\rm e}^{-\ell_4 r_4}
\langle V_4 R_4|V_3 R_3\rangle {\rm e}^{-\ell_3 r_3}\cr
&\cdot \langle V_3 R_3|V_2 R_2\rangle {\rm e}^{-\ell_2 r_2}
\langle V_2 R_2|V_1R_1\rangle {\rm e}^{-\ell_1 r_1}.}}
As in the previous case, the requisite diffeomorphism are
determined by the geometry. From the figure, we have,
$$V_1=S,\quad V_2=C, \quad V_3=SC,\quad V_4={\bf 1}.$$
This gives four ${\bf S}^3$'s, each of which has a Hopf link with
linking number $+1$ and whose
components have zero framing.

\ifig\ponefi{The figure shows a geometric transition
of four ${\bf S}^3$ in the previous figure. The dual geometry
is related by four flops of the external $\IP^1$'s
to a non-generic del Pezzo $\IB_5$.}
{\epsfxsize 4.0truein\epsfbox{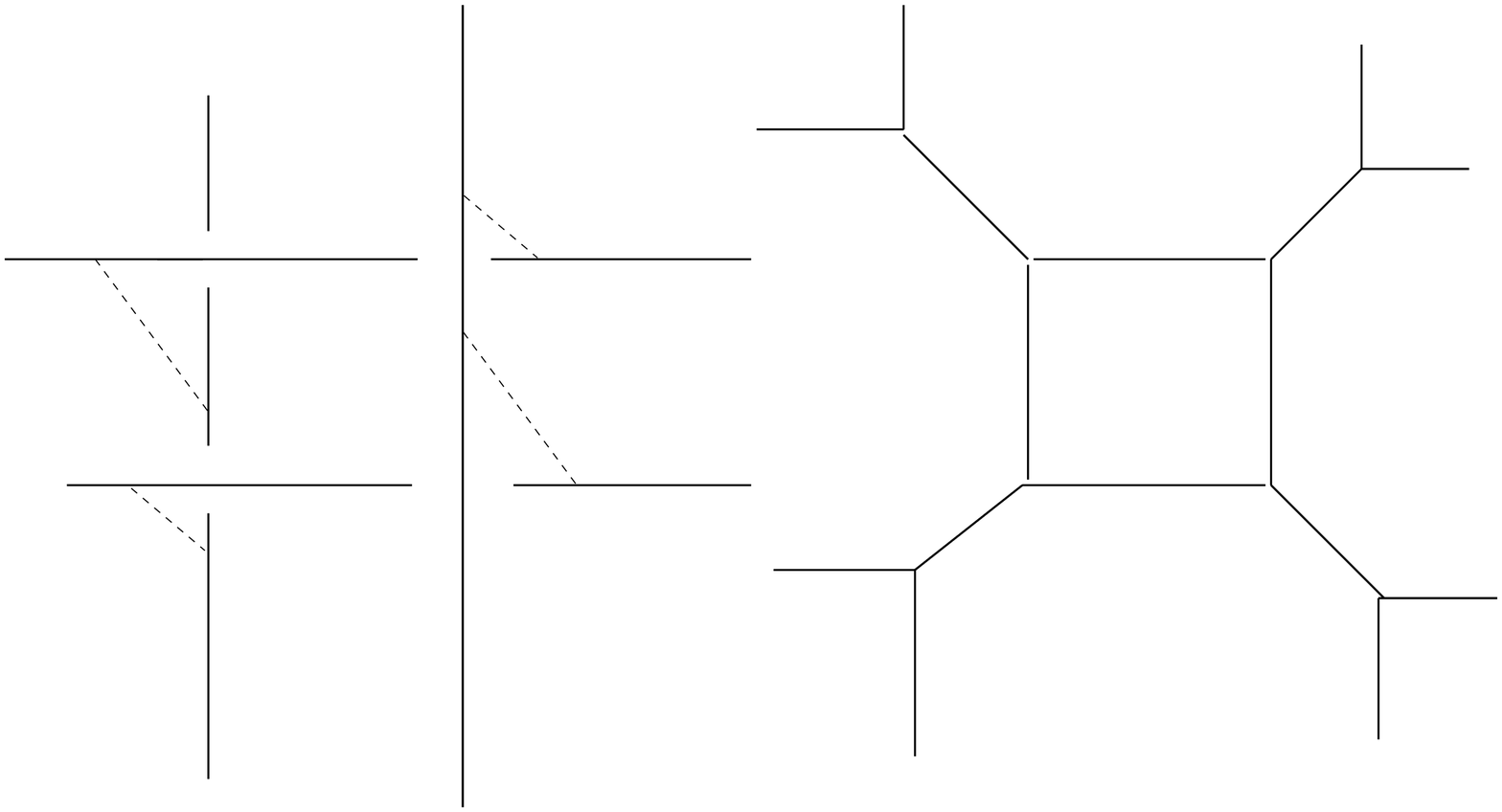}}

The geometric transition is represented in \ponefi. The resulting
dual closed string geometry contains a $\IP^1 \times \IP^1$, together
with four exceptional $\IP^1$'s. As in the previous case, we can
take the limit $t_i \rightarrow \infty$ in order to extract the integer
invariants of the local $\IP^1 \times \IP^1$ geometry.
The Chern-Simons computation that gives the invariants
is very similar to the one we discussed in the
previous section, so we won't give all the details. According to the
geometric picture, we have four Chern-Simons theories with 't Hooft
parameters $t_1$, $t_2$, $t_3$, and $t_4$,
and with the same coupling constant $g_s$. Since
all knots and links are identical, it is sufficient to label the vevs
by indicating explicitly the corresponding 't Hooft parameter.
The free energy at all genus for the topological
closed string is then given by
\eqn\ptwof{
F=\log \biggr\{ \sum_{R_1,R_2, R_3, R_4} {\rm e}^{-\ell_1 r_1 -\ell_3 r_2
-\ell_2 s_1 -\ell_4 s_2 }
W_{R_1, R_2}(t_1)  W_{R_2, R_3} (t_2)
W_{R_3, R_4} (t_3) W_{R_3, R_4} (t_4) \biggl\}.}
Again,
$r_i$ and $s_i$  are ``bare'' K\"ahler
parameters that will lead to two ``renormalized'' K\"ahler parameter $r$,
$s$. The relation between them can be obtained as in the previous case, and
one easily finds:
\eqn\renormpone{
\eqalign{
r&=r_1 -{ {t_1 + t_4} \over 2}=r_2 -{ {t_2 + t_3} \over 2},\cr
s&=s_1 -{ {t_1 + t_2} \over 2}=s_2 -{ {t_3 + t_4} \over 2},\cr} }
and we have to rescale the Chern-Simons vevs as before,
\eqn\multi{
W_{R, R'}(t_i) \rightarrow
\lambda_i^{-{{\ell + \ell'} \over 2}}W_{R, R'}(t_i).
}
Once we have done that, the free energy is given by:
\eqn\logf{
F=\log \biggr\{ 1 + \sum_{\ell_1, \ell_2=1}^{\infty} a_{\ell_1, \ell_2} (q)
{\rm e}^{-\ell_1 r -\ell_2 s} \biggr\}=
\sum_{\ell_1, \ell_2=1}^\infty a_{\ell_1, \ell_2}^{(c)}(q) {\rm e}^{-\ell_1 r -\ell_2 s},}
and from here we can again extract the integral invariants $n^g_{\ell_1,
\ell_2}$ by subtracting multicovering effects.

Let us present some explicit results at lower degree. For degrees $(\ell_1,
\ell_2)=(1,0)$ and $(0,1)$, we find:
\eqn\aonezero{
\eqalign{
a_{1,0} &=(\lambda_1\lambda_4)^{-{1 \over 2}}W_{\tableau{1}}(t_1) W_{\tableau{1}}(t_4)+ (\lambda_2\lambda_3)^{-{1 \over
2}}
W_{\tableau{1}}(t_2) W_{\tableau{1}}(t_3),\cr
a_{0,1} &=(\lambda_1\lambda_2)^{-{1 \over 2}}W_{\tableau{1}}(t_1) W_{\tableau{1}}(t_2)+ (\lambda_3\lambda_4)^{-{1 \over
2}}
W_{\tableau{1}}(t_3) W_{\tableau{1}}(t_4).\cr}}
In general, the coefficients $a_{n,m}$ and $a_{m,n}$ are related by
exchanging $t_2 \leftrightarrow t_4$. By taking the limit $t_i \rightarrow
\infty$, we find
\eqn\responeone{
a_{1,0}=a_{0,1}= {2 \over (q^{-{1\over2}} -q^{1\over2})^2},}
therefore
\eqn\gvdegone{
n_{1,0}^0=n_{0,1}^0=-2,}
and the invariants for higher genus all vanish. This is indeed the right
result \geomen\ckyz. For $a_{1,1}^{(c)}$ we find:
\eqn\aoneone{
a_{1,1}^{(c)}=\lambda_1^{-1} (\lambda_2\lambda_4)^{-{1 \over 2}} f_{(\tableau{1}, \tableau{1})}(t_1) W_{\tableau{1}}(t_2)
W_{\tableau{1}}(t_4) + {\rm
perms},}
where perms stands for three terms that are obtained from the first one by
permuting $t_i \rightarrow t_{i+1}$. Due to \fstr, this has the structure of the degree $(1,1)$ term
in a closed string free energy. After taking the limit $t_i \rightarrow
\infty$, one finds
\eqn\oneoneq{
a_{1,1}^{(c)}= {4\over (q^{-{1\over2}} -q^{1\over2})^2},}
therefore
\eqn\gvdegone{
n_{1,1}^0=-4,}
while the invariants for higher genera vanish. Again this is the right
value for the invariant.

We can again easily implement the computation of these invariants. In the
following tables we present most
of the results up to total degree $10$ and genus $8$
(the non-trivial invariants for total degree $10$ go
all the way to genus $16$, which we have obtained, but
have not included here for the economy of space):
\bigskip
{\vbox{\ninepoint{
$$
\vbox{\offinterlineskip\tabskip=0pt
\halign{\strut
\vrule#
&
&\hfil ~$#$
&\hfil ~$#$
&\hfil ~$#$
&\hfil ~$#$
&\hfil ~$#$
&\hfil ~$#$
&\hfil ~$#$
&\hfil ~$#$
&\vrule#\cr
\noalign{\hrule}
&d_2
&d_1=0
&1
&2
&3
&4
&5
&6
&
\cr
\noalign{\hrule}
&0
&
&-2
&0
&0
&0
&0
&0
&
\cr
\noalign{\hrule}
&1
&-2
&-4
&-6
&-8
&-10
&-12
&-14
&
\cr
\noalign{\hrule}
&2
&0
&-6
&-32
&-110
&-288
&-644
&-1280
&
\cr
\noalign{\hrule}
&3
&0
&-8
&-110
&-756
&-3556
&-13072
&-40338
&
\cr
\noalign{\hrule}
&4
&0
&-10
&-288
&-3556
&-27264
&-153324
&-690400
&
\cr
\noalign{\hrule}
&5
&0
&-12
&-644
&-13072
&-153324
&-1252040
&
&
\cr
\noalign{\hrule}
&6
&0
&-14
&-1280
&-40338
&-690400
&
&
&
\cr
\noalign{\hrule}
}\hrule}$$}
\vskip - 7 mm
\centerline{{\bf Table 4:} The integral invariants $n_d^0$ for the local
$\IP^1\times\IP^1 $ case.}
\vskip7pt}
\noindent
\smallskip

{\vbox{\ninepoint{
$$
\vbox{\offinterlineskip\tabskip=0pt
\halign{\strut
\vrule#
&
&\hfil ~$#$
&\hfil ~$#$
&\hfil ~$#$
&\hfil ~$#$
&\hfil ~$#$
&\hfil ~$#$
&\vrule#\cr
\noalign{\hrule}
&d_2
&d_1=2
&3
&4
&5
&6
&
\cr
\noalign{\hrule}
&2
&9
&68
&300
&988
&2698
&
\cr
\noalign{\hrule}
&3
&68
&1016
&7792
&41376
&172124
&
\cr
\noalign{\hrule}
&4
&300
&7792
&95313
&760764
&4552692
&
\cr
\noalign{\hrule}
&5
&988
&41736
&760764
&8695048
&
&
\cr
\noalign{\hrule}
&6
&2698
&172124
&4552692
&
&
&
\cr
\noalign{\hrule}
}\hrule}$$}
\vskip - 7 mm
\centerline{{\bf Table 5:} The integral invariants $n_d^1$ for the local
$\IP^1\times\IP^1 $ case.}
\vskip7pt}
\noindent
\smallskip

{\vbox{\ninepoint{
$$
\vbox{\offinterlineskip\tabskip=0pt
\halign{\strut
\vrule#
&
&\hfil ~$#$
&\hfil ~$#$
&\hfil ~$#$
&\hfil ~$#$
&\hfil ~$#$
&\hfil ~$#$
&\vrule#\cr
\noalign{\hrule}
&d_2
&d_1=2
&3
&4
&5
&6
&
\cr
\noalign{\hrule}
&2
&0
&-12
&-116
&-628
&-2488
&
\cr
\noalign{\hrule}
&3
&-12
&-580
&-8042
&-64624
&-371980
&
\cr
\noalign{\hrule}
&4
&-116
&-8042
&-167936
&-1964440
&-15913228
&
\cr
\noalign{\hrule}
&5
&-628
&-64624
&-1964440
&-32242268
&
&
\cr
\noalign{\hrule}
&6
&-2488
&-371980
&-15913228
&
&
&
\cr
\noalign{\hrule}
}\hrule}$$}
\vskip - 7 mm
\centerline{{\bf Table 6:} The integral invariants $n_d^2$ for the local
$\IP^1\times\IP^1 $ case.}
\vskip7pt}
\noindent
\smallskip

{\vbox{\ninepoint{
$$
\vbox{\offinterlineskip\tabskip=0pt
\halign{\strut
\vrule#
&
&\hfil ~$#$
&\hfil ~$#$
&\hfil ~$#$
&\hfil ~$#$
&\hfil ~$#$
&\hfil ~$#$
&\vrule#\cr
\noalign{\hrule}
&d_2
&d_1=2
&3
&4
&5
&6
&
\cr
\noalign{\hrule}
&2
&0
&0
&15
&176
&1130
&
\cr
\noalign{\hrule}
&3
&0
&156
&4680
&60840
&501440
&
\cr
\noalign{\hrule}
&4
&15
&4680
&184056
&3288688
&36882969
&
\cr
\noalign{\hrule}
&5
&176
&60840
&3288688
&80072160
&
&
\cr
\noalign{\hrule}
&6
&1130
&501440
&36882969
&
&
&
\cr
\noalign{\hrule}
}\hrule}$$}
\vskip - 7 mm
\centerline{{\bf Table 7:} The integral invariants $n_d^3$ for the local
$\IP^1\times\IP^1 $ case.}
\vskip7pt}
\noindent
\smallskip

{\vbox{\ninepoint{
$$
\vbox{\offinterlineskip\tabskip=0pt
\halign{\strut
\vrule#
&
&\hfil ~$#$
&\hfil ~$#$
&\hfil ~$#$
&\hfil ~$#$
&\hfil ~$#$
&\hfil ~$#$
&\vrule#\cr
\noalign{\hrule}
&d_2
&d_1=2
&3
&4
&5
&6
&
\cr
\noalign{\hrule}
&2
&0
&0
&0
&-18
&-248
&
\cr
\noalign{\hrule}
&3
&0
&-16
&-1560
&-36048
&-450438
&
\cr
\noalign{\hrule}
&4
&0
&-1560
&-133464
&-3839632
&-61250176
&
\cr
\noalign{\hrule}
&5
&-18
&-36048
&-3839632
&-144085372
&
&
\cr
\noalign{\hrule}
&6
&-248
&-450438
&-61250176
&
&
&
\cr
\noalign{\hrule}
}\hrule}$$}
\vskip - 7 mm
\centerline{{\bf Table 8:} The integral invariants $n_d^4$ for the local
$\IP^1\times\IP^1 $ case.}
\vskip7pt}
\noindent
\smallskip

{\vbox{\ninepoint{
$$
\vbox{\offinterlineskip\tabskip=0pt
\halign{\strut
\vrule#
&
&\hfil ~$#$
&\hfil ~$#$
&\hfil ~$#$
&\hfil ~$#$
&\hfil ~$#$
&\hfil ~$#$
&\vrule#\cr
\noalign{\hrule}
&d_2
&d_1=2
&3
&4
&5
&6
&
\cr
\noalign{\hrule}
&2
&0
&0
&0
&0
&21
&
\cr
\noalign{\hrule}
&3
&0
&0
&276
&13888
&276144
&
\cr
\noalign{\hrule}
&4
&0
&276
&64973
&3224340
&75592238
&
\cr
\noalign{\hrule}
&5
&0
&13888
&3224340
&195035824
&
&
\cr
\noalign{\hrule}
&6
&21
&276144
&75592238
&
&
&
\cr
\noalign{\hrule}
}\hrule}$$}
\vskip - 7 mm
\centerline{{\bf Table 9:} The integral invariants $n_d^5$ for the local
$\IP^1\times\IP^1 $ case.}
\vskip7pt}
\noindent
\smallskip

{\vbox{\ninepoint{
$$
\vbox{\offinterlineskip\tabskip=0pt
\halign{\strut
\vrule#
&
&\hfil ~$#$
&\hfil ~$#$
&\hfil ~$#$
&\hfil ~$#$
&\hfil ~$#$
&\vrule#\cr
\noalign{\hrule}
&d_2
&d_1=3
&4
&5
&6
&
\cr
\noalign{\hrule}
&3
&0
&-20
&-3260
&-115744
&
\cr
\noalign{\hrule}
&4
&-20
&-20936
&-1969710
&-70665312
&
\cr
\noalign{\hrule}
&5
&-3260
&-1969710
&-202598268
&
&
\cr
\noalign{\hrule}
&6
&-115744
&-70665312
&
&
&
\cr
\noalign{\hrule}
}\hrule}$$}
\vskip - 7 mm
\centerline{{\bf Table 10:} The integral invariants $n_d^6$ for the local
$\IP^1\times\IP^1 $ case.}
\vskip7pt}
\noindent
\smallskip
{\vbox{\ninepoint{
$$
\vbox{\offinterlineskip\tabskip=0pt
\halign{\strut
\vrule#
&
&\hfil ~$#$
&\hfil ~$#$
&\hfil ~$#$
&\hfil ~$#$
&\hfil ~$#$
&\vrule#\cr
\noalign{\hrule}
&d_2
&d_1=3
&4
&5
&6
&
\cr
\noalign{\hrule}
&3
&0
&0
&428
&32568
&
\cr
\noalign{\hrule}
&4
&0
&4266
&873972
&50501308
&
\cr
\noalign{\hrule}
&5
&428
&873972
&163185964
&
&
\cr
\noalign{\hrule}
&6
&32568
&50501308
&
&
&
\cr
\noalign{\hrule}
}\hrule}$$}
\vskip - 7 mm
\centerline{{\bf Table 11:} The integral invariants $n_d^7$ for the local
$\IP^1\times\IP^1 $ case.}
\vskip7pt}
\noindent
\smallskip

{\vbox{\ninepoint{
$$
\vbox{\offinterlineskip\tabskip=0pt
\halign{\strut
\vrule#
&
&\hfil ~$#$
&\hfil ~$#$
&\hfil ~$#$
&\hfil ~$#$
&\hfil ~$#$
&\vrule#\cr
\noalign{\hrule}
&d_2
&d_1=3
&4
&5
&6
&
\cr
\noalign{\hrule}
&3
&0
&0
&-24
&-5872
&
\cr
\noalign{\hrule}
&4
&0
&-496
&-277880
&-27655024
&
\cr
\noalign{\hrule}
&5
&-24
&-277880
&-102321184
&
&
\cr
\noalign{\hrule}
&6
&-5872
&-27655024
&
&
&
\cr
\noalign{\hrule}
}\hrule}$$}
\vskip - 7 mm
\centerline{{\bf Table 12:} The integral invariants $n_d^8$ for the local
$\IP^1\times\IP^1 $ case.}
\vskip7pt}
\noindent
\smallskip

These results are in full agreement with the ones presented in
\geomen\ckyz\kkv. Again, we can verify many of these
numbers with the geometric formulae
of \kkv. For a given bidegree $(a,b)$, $n^g_{(a,b)}$ vanishes for
$g>(a-1)(b-1)$, which is indeed the arithmetic genus of a curve of bidegree
$(a, b)$ in $\IP^1 \times \IP^1$. One finds,
\eqn\nopone{
n_{(a,b)}^{(a-1)(b-1)}= -(-1)^{(a+1)(b+1)} (a+1)(b+1),}
which reproduces the corresponding results listed in the tables above.
For curves with one node, one finds:
\eqn\onepone{
n_{(a,b)}^{(a-1)(b-1)-1}= 2 (-1)^{(a+1)(b+1)}(a+b + ab -a^2 -b^2 + a^2
b^2),}
again in full agreement with the tables.
For curves with two nodes (extending
the derivation in \kkv ) we have:
\eqn\twopone{
\eqalign{
n_{(a,b)}^{(a-1)(b-1)-2}=&-(-1)^{(a+1)(b+1)}
\Bigl( -14 + 9(a+b)-3 ab -3(a^2
+ b^2)+ 3a^2 b^2 \cr
& \,\,\,\,\,\,\,\,\,+  2(a^3 + b^3 +a^2 b +b^2 a)
 -2 (a^3 b + b^3 a)
-2 (a^3 b^2 + b^3 a^2) + 2a^3 b^3 \Bigr),\cr}}
which reproduces for example $n_{(3,3)}^2=-580$, $n_{(3,6)}^8=-5872$ and
$n_{(4,4)}^7=4266$. For the invariants corresponding to bidegrees $(2,n)$,
where $3\le n \le 6$, and curves with two nodes,
one has to introduce corrections associated to
reducible curves. For example, for bidegrees $(2,6)$, \twopone\ gives the
value $1166$, but there are reducible curves of type  $(5,2) \cup
(1,0)$ with two nodes. Since $n^0_{(1,0)}=-2$, and $n_{(5,2)}^4=-18$, the
subtraction scheme proposed in \kkv\ gives $n_{(2,6)}^3=1166- (-2)(-18) =
1130$, in agreement with the result of table 7.

\subsec{Refined Integral Invariants}

The integral invariants $n_{d}^g$ defined in \gvtwo\ denote the (net)
number of wrapped M2 branes in $4+1$ dimensional effective theory, obtained
by compactification of M-theory on the corresponding Calabi-Yau,
where $d \in H_2(X)$ denotes the class the M2 brane is wrapped
and $g$ denotes a basis for the $SU(2)_L$ rotation subgroup of
$SO(4)$ (see \gvtwo\ for details). If the Calabi-Yau space has
global symmetries,
then these states also form representations of this group.
Compact Calabi-Yau manifolds do not admit global symmetries,
so this does not arise in that context. However for local
toric 3-folds there always are extra global symmetries and
one can ask how the $n_d^g$ decompose in representations
of this symmetry algebra. Thus it is natural to ask whether
we can use our techniques to also compute these refined invariants.

For example, consider the linear sigma
model describing
${\cal O}(-3) \rightarrow \IP^2$, which
contains three matter fields of charge $+1$.  In this case there
are two extra $U(1)$ global symmetries, which
for some metric in $\IP^2$ could give rise to the Cartan
of $SU(3)$.  This can be implemented in terms of the toric diagram, by
assigning different sizes to the different edges. In the local
$\IP^2$ case, we should assign different sizes to the triangle
describing the base of the $\IP^2$, {\it i.e.} we should introduce
three K\"ahler parameters instead of one. Notice that this is perfectly
natural from the point of view of the Chern-Simons description,
because in the limit where we took the $N_i\rightarrow \infty$
we had to tune $r$'s.  Nothing prevents us from
tuning the three edges to different values by considering a
suitable limit.

Let us consider the computation of the refined integral invariants
in some detail, in the case of local $\IP^2$.
We have to introduce three parameters associated to the three different
edges, and we will denote them by $r_i'$, with $i=1,2,3$,
where we view $e^{-r_i'}$ as forming a Cartan torus of $U(3)$.
Notice that, if we write $U(3)=U(1)
\times SU(3)$, the overall $U(1)$ quantum number is precisely the
degree $d$. We then have to further decompose the spectrum with respect to
the $SU(3)$. This goes as follows: due to the underlying symmetry,
the closed string free energy will be now of the form:
\eqn\globalsymgv{
\sum_{m=1}^{\infty} \sum_{g,d}
n^g_{d}(x^m_1, x^m_2, x^m_3)  {1 \over m} \biggl(
2 \sin {m g_s \over 2} \biggr)^{2g-2},}
where $x_i={\rm e}^{-r'_i}$, $i=1,2,3$, and
$n^g_{d}(x_1, x_2, x_3)$ is now a symmetric polynomial of degree $d$
in the $x_i$, with integer coefficients. Therefore, we can expand it in terms
of Schur polynomials $s_R$ in three variables and of degree $d$, which are
labeled by representations $R$ of $SU(3)$ with $d$ boxes. We then write:
\eqn\refin{
n^g_{d}(x_1, x_2, x_3) =\sum_{R} n^g_{d, R}\,  s_{R}(x_1, x_2, x_3),}
where the sum is over representations of $SU(3)$ with $d$ boxes,
and $n^g_{d, R}$ denote
the number of M2 branes of degree $d$
with $SU(2)_L$ representation $g$ and transforming
as representation
$R$ of the $SU(3)$ global symmetry. These are the refined invariants of
the local $\IP^2$. Notice that, if we put $x_1=x_2=x_3=1$ in \refin, we
recover the usual integer invariants, therefore one has
\eqn\relaref{
n^g_{d}=\sum_R ({\rm dim} R)\,  n^g_{d, R}.}

The computation of the refined invariants can be
easily done in the Chern-Simons setting, by taking the renormalized
sizes of the annuli to be different. The renormalized sizes
will give in this way the parameters $r_i'$ appearing in the closed
string side, in other words:
$$
r'_1=r -{t_1 + t_2 \over 2}
$$ and so on. The refined invariants for the first few degrees
can be easily computed.
At degree one we find,
\eqn\dgone{
n^0_1 (x_1, x_2, x_3)=x_1 + x_2 + x_3, }
therefore $n^0_{1, \tableau{1}}=1$. At degree two, one has:
\eqn\degtwo{
n^0_{2, \tableau{2}}=-1,\,\,\,\,\,\,\,\,\,\,\,\,\,\,\,\,\,\,
n^0_{2, \tableau{1 1}}=0.}
At degree three, we find:
\eqn\degthree{
\eqalign{
n^0_{3, \tableau{3}}=& 2\,\,\,\,\,\,\,\,\,\,\,\,\,\,\,\,\,\,
n^0_{3, \tableau{2 1}}=1, \,\,\,\,\,\,\,\,\,\,\,\,\,\,\,\,\,\,
n^0_{3, \tableau{1 1 1}}=-1,\cr
n^1_{3, \tableau{3}}=& -1\,\,\,\,\,\,\,\,\,\,\,\,\,\,\,\,\,\,
n^1_{3, \tableau{2 1}}=0, \,\,\,\,\,\,\,\,\,\,\,\,\,\,\,\,\,\,
n^0_{3, \tableau{1 1 1}}=0,\cr}}
We finally list the results for degree four:
\eqn\degthree{
\eqalign{
n^0_{4, \tableau{4}}=& -7\,\,\,\,\,\,\,\,\,\,\,\,\,\,\,\,\,\,
n^0_{4, \tableau{3 1}}=-6, \,\,\,\,\,\,\,\,\,\,\,\,\,\,\,\,\,\,
n^0_{4, \tableau{2 2}}=-2,\,\,\,\,\,\,\,\,\,\,\,\,\,\,\,\,\,\,
n^0_{4, \tableau{2 1 1}}=5,\cr
n^1_{4, \tableau{4}}=& 11\,\,\,\,\,\,\,\,\,\,\,\,\,\,\,\,\,\,
n^1_{4, \tableau{3 1}}=5, \,\,\,\,\,\,\,\,\,\,\,\,\,\,\,\,\,\,
n^1_{4, \tableau{2 2}}=1,\,\,\,\,\,\,\,\,\,\,\,\,\,\,\,\,\,\,
n^1_{4, \tableau{2 1 1}}=-5,\cr
n^2_{4, \tableau{4}}=& -6\,\,\,\,\,\,\,\,\,\,\,\,\,\,\,\,\,\,
n^2_{4, \tableau{3 1}}=-1, \,\,\,\,\,\,\,\,\,\,\,\,\,\,\,\,\,\,
n^2_{4, \tableau{2 2}}=0,\,\,\,\,\,\,\,\,\,\,\,\,\,\,\,\,\,\,
n^2_{4, \tableau{2 1 1}}=-1,\cr
n^3_{4, \tableau{4}}=&1\,\,\,\,\,\,\,\,\,\,\,\,\,\,\,\,\,\,
n^3_{4, \tableau{3 1}}=0, \,\,\,\,\,\,\,\,\,\,\,\,\,\,\,\,\,\,
n^3_{4, \tableau{2 2}}=0,\,\,\,\,\,\,\,\,\,\,\,\,\,\,\,\,\,\,
n^3_{4, \tableau{2 1 1}}=0.\cr}}

For $\IP^1 \times \IP^1$ one can similarly decompose the
invariants with respect to the $SU(2) \times SU(2)$ global symmetry of
the model. Note that from a mathematical point of view,
these refined integer invariants should be
related to the equivariant Gromov-Witten invariants associated
to the group action on the manifold, as was studied
in the Fano case in \HV.  Moreover, one could use the techniques
of \HV\ to obtain the mirror of these deformations for the
toric Calabi-Yau manifolds we have discussed and check
the results obtained here against the predictions of mirror symmetry
(at least for genus 0).

\newsec{Embedding in Superstrings}

It is natural to ask what kind of dualities these
geometric transitions lead to, once we embed them in superstrings, as was
done in \vaaug\ for the original Chern-Simons
duality \gv.

Embedding these dualities for topological
strings in type IIA strings is  easily
done
by replacing the branes with D6 branes wrapping ${\bf S}^3$'s
and filling 4 dimensional spacetime.  Thus we end up, at low
energy with a system involving ${\cal N}=1$ $U(N_i)$ gauge symmetry.
Moreover for each annulus
contribution we end up with a bifundamental matter ``hypermultiplet''
in the superstring context.  Of course this is only the low
energy limit of the brane system.  The high energy aspects of this
theory differ from that of pure Yang-Mills.  This can be deduced
by considering the superpotential for this theory, as was
done in \vaaug.  In the IR the gaugino condensation will
take place where the ${\bf S}^3$'s are replaced by blown up
${\bf S}^2$'s with RR fluxes through them.
There is no RR flux through ${\bf S}^2$'s
which come from matter bifundamentals. In the applications we have looked at,
we have also considered the interesting limit where
the sizes of blown up ${\bf S}^2$'s go to infinity, while keeping
the effective masses of the bifundamental fields finite.
This was, for example, how we got the full answer for $\IP^2$
in topological strings.  In the gauge theory setup the size
of the blown up ${\bf S}^2$'s correspond to the size of the gaugino
condensate getting large, which can be adjusted by increasing
the corresponding gauge coupling.  Note that in this limit
we will have no RR flux left in the type IIA superstring theory.
Since we have fixed total RR flux through the ${\bf S}^2$'s
which get infinitely large,
in the limit we are considering the flux per unit volume
goes to zero. Moreover,
the finite ${\bf S}^2$'s in this limit correspond to where the bifundamental
matter came from, and there is no flux through them. Thus we end
up with a novel large $N$ duality, were the bifundamental matter
structure dictates the geometry of the dual and this geometry
has no RR flux in it.

The statement of the above dualities correspond to gauge theories
with all the string interactions on them.  One would naturally ask
if there are any large $N$ dualities along these lines
for pure gauge theories.  For this purpose
it is convenient to go to the type IIB mirror setup.

To illustrate the idea let us first consider a simple
example.  Consider the ${\cal N}=2$, $U(2N)$ gauge theory deformed
by the addition of superpotential
$$W=g{\rm tr}[{1\over 3}\Phi^3-m^2\Phi]$$
where $\Phi$ is the adjoint field.  There are two classical values
for the eigenvalues of $\Phi$, namely $\Phi=\pm m$.
Let us choose $N$ eigenvalues of $\Phi$ to be at $+m$ and $N$
to be at $-N$.  Then the large $N$ dual of this system
in type IIB is proposed in \FI\
(and further elaborated recently in \cv) to be
given by propagation in the non-compact Calabi-Yau given by
the hypersurface in $\IC^4$:
$$uv+y^2+g^2(x^2-m^2)^2+g^2\Lambda^4=0$$
where $\Lambda$ is related to the scale of the original ${\cal N}=2$
theory.  Note that for small $\Lambda$ we have two conifold
points centered near $x=m$ and $x=-m$.  In this dual
gravitational geometry, there is RR flux of $N$ units through
each of the corresponding ${\bf S}^3$'s.  However there is
no RR flux through the compact ${\bf S}^3$ which runs between
these two ${\bf S}^3$'s (and intersects both at 1 point).
It is convenient to rewrite the above geometry as
$$uv+y^2+g^2P(x)=0$$
where
$$P(x)=(x^2-M^2)(x^2-a^2)$$
and we identify
$$m^2={1\over 2}(M^2+a^2)$$
$$\Lambda^4= {-(M^2-a^2)^2\over 4}$$
In this parameterization the two ${\bf S}^3$'s with
RR flux project in the x-plane to the intervals
$-M\leq x\leq -a$ and $a\leq x\leq M$.
In particular there is no flux through the ${\bf S}^3$
which projects to the interval $-a\leq x\leq a$.
We consider the situation where $(M/a)>>1$.
In this limit the two ${\bf S}^3$'s have become big.  In particular
as $M\rightarrow \infty$, keeping $a$ and $\alpha =-g^2 M^2$ fixed the
geometry reduces to
$$uv+y^2+\alpha (x^2-a^2)=0$$
which is the ordinary conifold.  Moreover, in this limit
the RR fields per unit volume go to zero everywhere.  Thus
we have found a gauge theory/gravity duality where the geometry
is free of RR flux.  To be precise we have to note that we need
to complete the duality by going farther in the UV of gauge system,
which forces a cascade structure \FIII\
generalizing the construction of \klebstra\ to the case
at hand.  Namely, we will end up with a $U(2N+M)\times U(M)$ gauge
system, with two bifundamental hypermultiplets, as $M\rightarrow \infty$.
Moreover we have superpotentials $W_1(\Phi_1)$ and
$W_2(\Phi_2)$
which have the same functional form as the superpotential $W(\Phi)$
discussed before, namely $W_1=-W_2=W$, where the coefficients of $W$
are carefully tuned, as discussed above.
Thus we have a proposal for a gauge dual description of
the standard conifold
with no flux through it.

Clearly this example can be generalized.  In fact a large
class of local Calabi-Yau threefolds were constructed
in \FII\ as duals
to gauge systems, which were analyzed in \FIII.
Applying a similar kind of reasoning as the above example
we end up describing a rather large class of local threefold without
fluxes, as duals to some limits of ${\cal N}=2$
gauge systems deformed to ${\cal N}=1$ by superpotential terms.
It would be very interesting to study the physical implications of these
dualities.

\bigskip
\centerline{\bf Acknowledgements}
\medskip
We would like to thank E. Diaconescu, A. Grassi, S. Katz and P. Ramadevi
for valuable discussions.

This research is supported in part by NSF grants PHY-9802709
and DMS-0074329.

\listrefs

\bye